\documentclass{desyprocA4}
\usepackage{color}
\usepackage{parskip}               % suppress hanging indentation
\usepackage{graphicx}				% Use pdf, png, jpg, or eps with pdflatex; use eps in DVI mode
\usepackage{multirow}								% TeX will automatically convert eps --> pdf in pdflatex		
\usepackage{amssymb}
\usepackage{subfigure}
\usepackage[ , ,bf,it]{caption}    % nice captions
\usepackage{url}
\newcommand\pubdate{\today}

\def\to{\rightarrow}

\def\bi{\begin{itemize}}
 \def\ei{\end{itemize}}
\def\te{\tilde e}

\def\c1p{C1^\prime}
\def\msq3{\overline{m}_{\tilde{q}}(3)}

\def\tb{\tilde b}

\def\tst{\tilde t}
\def\ttau{\tilde \tau}
\def\tmu{\tilde \mu}

\def\tnu{\tilde\nu}

\def\be{\begin{equation}}  
\def\ee{\end{equation}}  
\def\bea{\begin{eqnarray}}  
\def\eea{\end{eqnarray}}  
\def\tw{\tilde\chi}
\def\twp{\tilde\chi^+}
\def\twm{\tilde\chi^-}
\def\twpm{\tilde\chi^\pm}
\def\tz{\tilde\chi^0}

\newcommand{\eeto}    {\ensuremath{ {\, e}^+ {e}^- \to}}
\begin{document}

\title{Non-Simplified SUSY: $\ttau$-Coannihilation at LHC and ILC}
\author{M.~Berggren$^1$, A.~Cakir$^1$, D. Kr{\"u}cker$^1$, J.~List$^1$, A.~Lobanov$^1$, I.-A.~Melzer-Pellmann$^1$\\[1ex]
$^1$DESY, Notkestra{\ss}e 85, 22607 Hamburg, Germany\\
}

\maketitle

\pubdate

\begin{abstract}
Simplified models have become a widely used and important tool to cover the more diverse phenomenology beyond 
constrained SUSY models. However, they come with a substantial number of caveats themselves, and great
care needs to be taken when drawing conclusions from limits based on the simplified approach. To illustrate this
issue with a concrete example, we examine the applicability of simplified model results to a series of full SUSY model points which all feature a 
small $\ttau$-LSP mass difference, and are compatible with electroweak and flavor precision
observables as well as current LHC results. Various channels have been studied using the Snowmass Combined LHC 
detector implementation in the Delphes simulation package, as well as the Letter of Intent or Technical Design Report
simulations of the ILD detector concept at the ILC. We investigated both the LHC and ILC capabilities for discovery, separation and identification of all parts of the spectrum. While parts of the spectrum would be discovered at the LHC, there is substantial room for further discoveries and property determination at the ILC.
\end{abstract}

%\clearpage

\section{Introduction} \label{sec:introduction}

In full SUSY models, the higher states of the spectrum can have many decay modes leading to
potentially long decay chains. This means that the simplified approach does in general not 
apply beyond the direct NLSP-production case, which
renders the interpretation of exclusion limits formulated in
the simplified approach non-trivial. Furthermore, also many production channels may be open, 
making SUSY the most serious background to itself. This becomes an issue especially for 
 interpreting a future discovery of a non-SM signal.

\begin{figure}[htb]
  \begin{center}
  \includegraphics[width=0.45\linewidth]{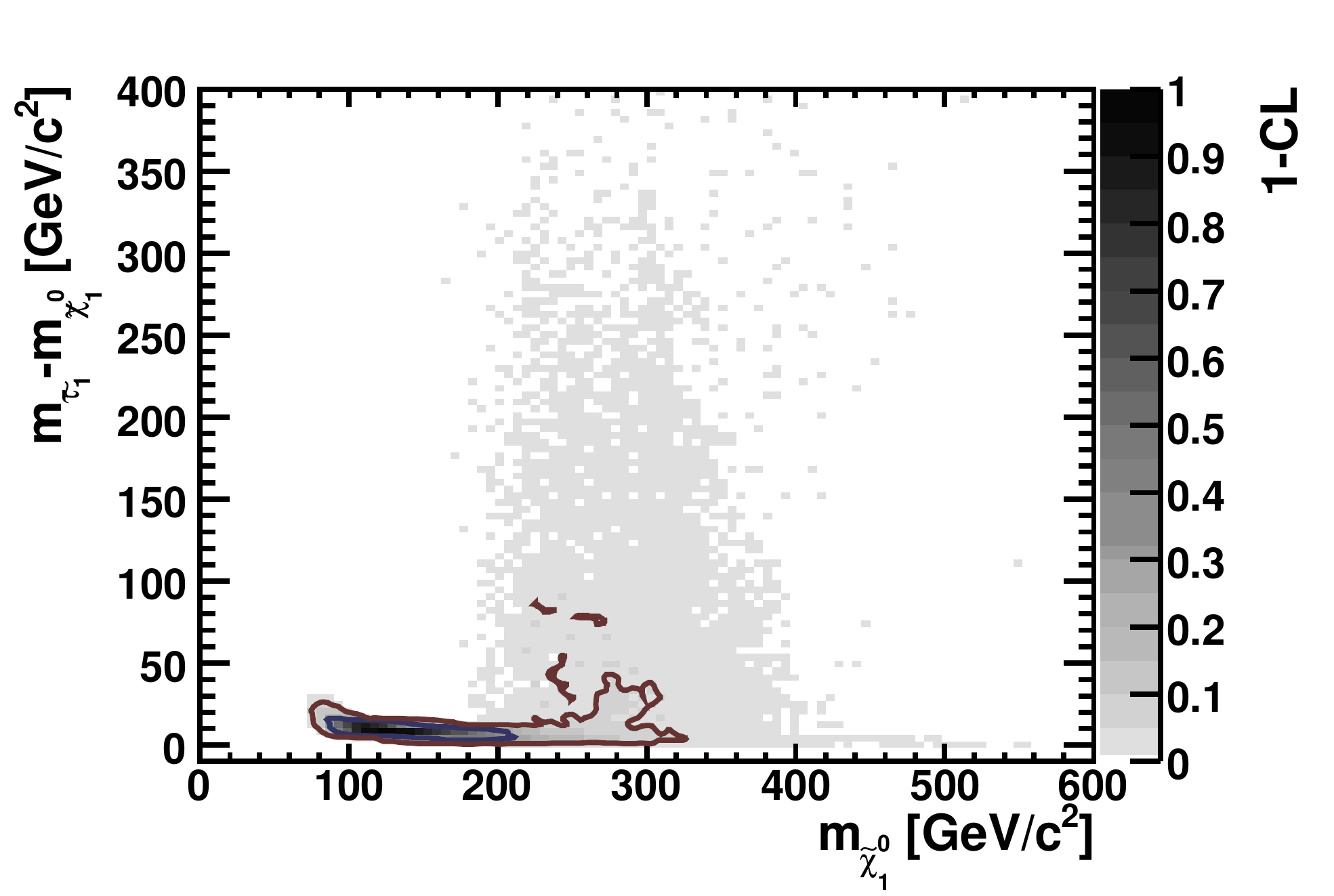}
  \vspace{-0.25cm}
  \includegraphics[width=0.45\linewidth]{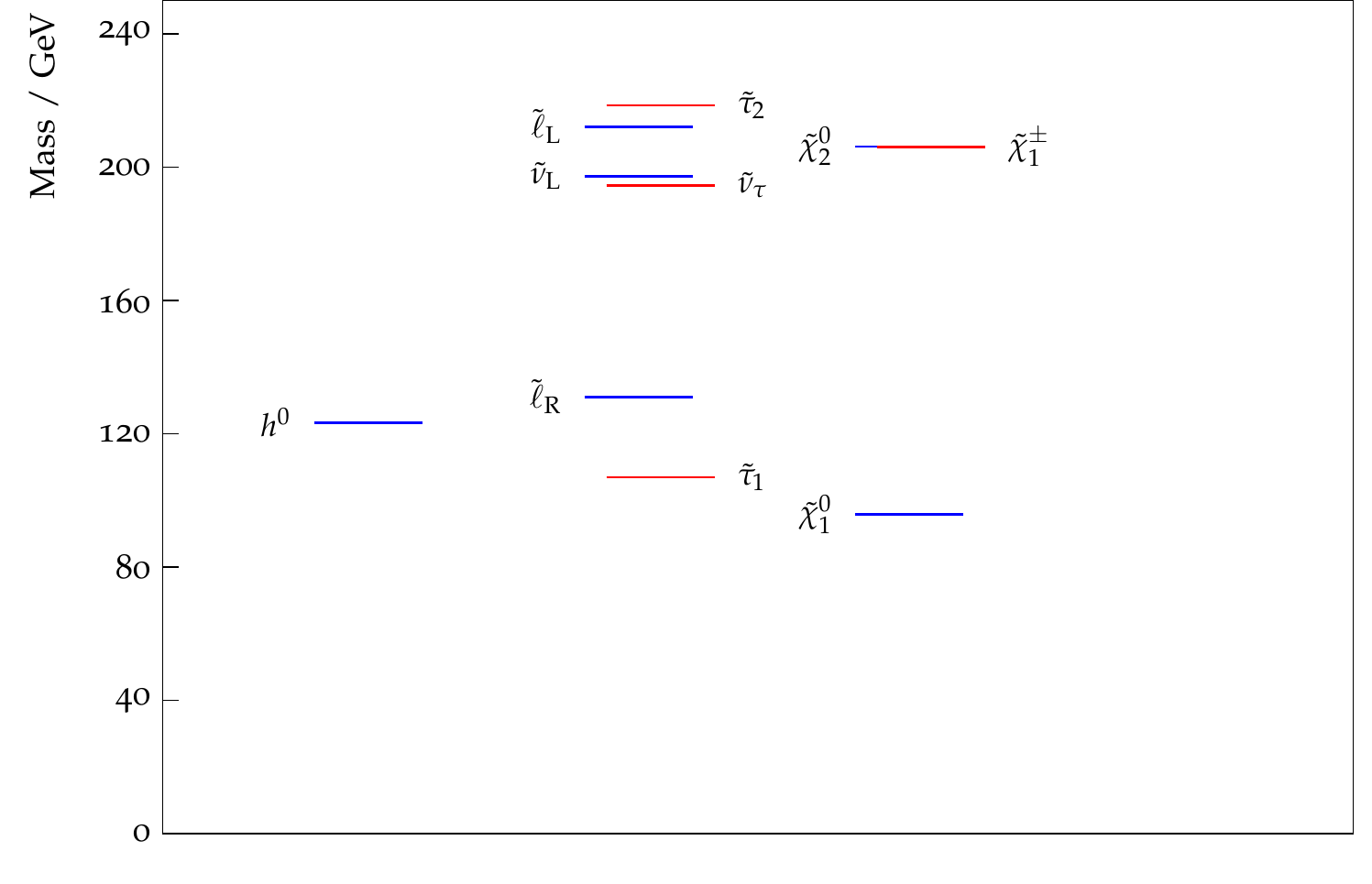}
 \end{center}
  \caption{\label{fig:deltaMstau} Left: Likelihood of a constrained MSSM fit to pre-LHC experimental data in
  the $\Delta M (\ttau, \tz_1) - M_{\tz_1}$ plane, showing a clear maximum around $\Delta M (\ttau, \tz_1) - M_{\tz_1}=10\,$GeV. From~\cite{Buchmueller:2009fn} Right: Lower part of the spectrum of the STC scenarios, which features $\Delta M (\ttau, \tz_1) - M_{\tz_1} \simeq 10\,$GeV.}
\end{figure}

We take as an example the regions in parameter space which gained the highest likelihood in fits to 
all pre-LHC experimental data within the constrained MSSM~\cite{Buchmueller:2009fn}. These fits preferred 
scenarios with a small mass difference of about $10$~GeV between the $\ttau$-NLSP and the 
$\tz_1$ as LSP, as illustrated by the likelihood distribution in the left panel of 
Fig.~\ref{fig:deltaMstau}.  Within the context of the cMSSM, this region is ruled out by LHC searches. However
this exclusion is based on the strongly interacting sector, which in constrained models is coupled to the 
electroweak 
sector by GUT-scale mass unification. Without the restriction of {\it mass} unification, the part of the spectrum which is of interest to electroweak and flavor precision observables and 
dark matter, 
ie. which is decisive for the fit outcome, is not at all in conflict with LHC results. This applies in particular
to the  $\ttau_1$ with a small mass difference to the LSP, which is essential to allow efficient (co-)annihilation of dark matter to lower the predicted relic density to its observed value:
Although first limits on direct $\ttau$ pair production from the LHC have been presented~\cite{ATLAS-CONF-2013-049, CMS-PAS-SUS-12-022},
they rapidly loose sensitivity if the $\ttau$ is not degenerate with the $\te$ and $\tmu$, and has a small mass
difference to the LSP. 

The right part of Fig.~\ref{fig:deltaMstau} shows the low mass part of an example spectrum which fulfills 
all
 constraints, including
a higgs boson with SM-like branching ratios at a mass in agreement with the LHC discovery within
the typical theoretical uncertainty of $\pm 3$~GeV on MSSM higgs mass calculations. The full definition
and further information can be found in~\cite{Baer:2013ula}.

When the 1st and 2nd generation squarks and the gluino are rather heavy, $\gtrsim 2\,$TeV, the size of the total SUSY
 cross-section at the LHC strongly depends on the mass of the lightest top squark. We therefore consider a series of
 points, called STC4 to STC8, whose physical spectra differ only by the $\tst_1$ mass. In this series, the mass parameter of the partner of the right-handed top quark, $M_{U3}$, is varied from $400\,$ to $800\,$GeV at a scale of $1\,$TeV, resulting in $m_{\tst_1} \simeq 300...700\,$GeV\footnote{The SLHA files are available at \url{http://www-flc.desy.de/ldcoptimization/physics.php}.}.

\begin{figure}[htb]
  \begin{center}
  \includegraphics[width=0.49\linewidth]{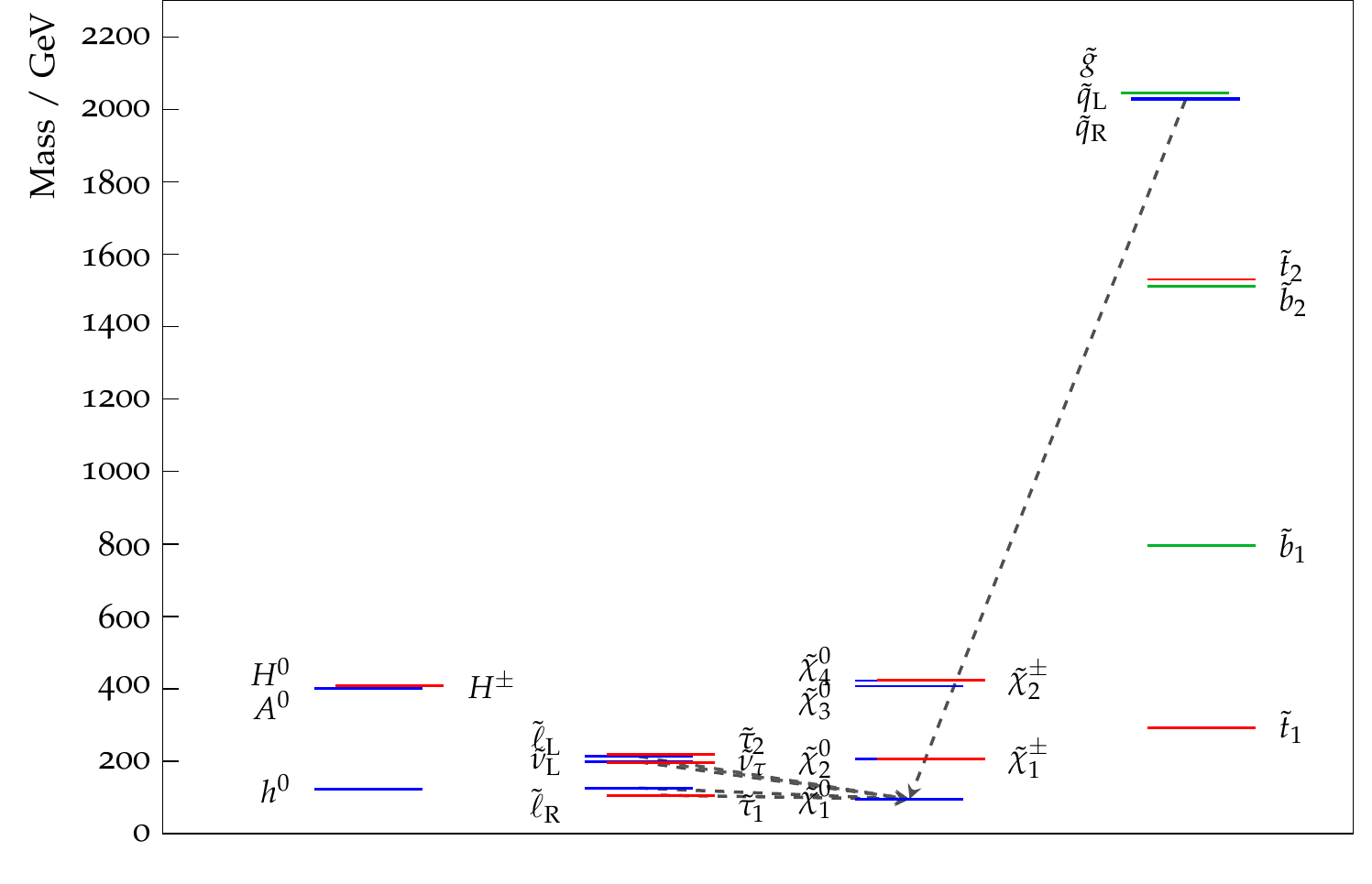}
  \includegraphics[width=0.49\linewidth]{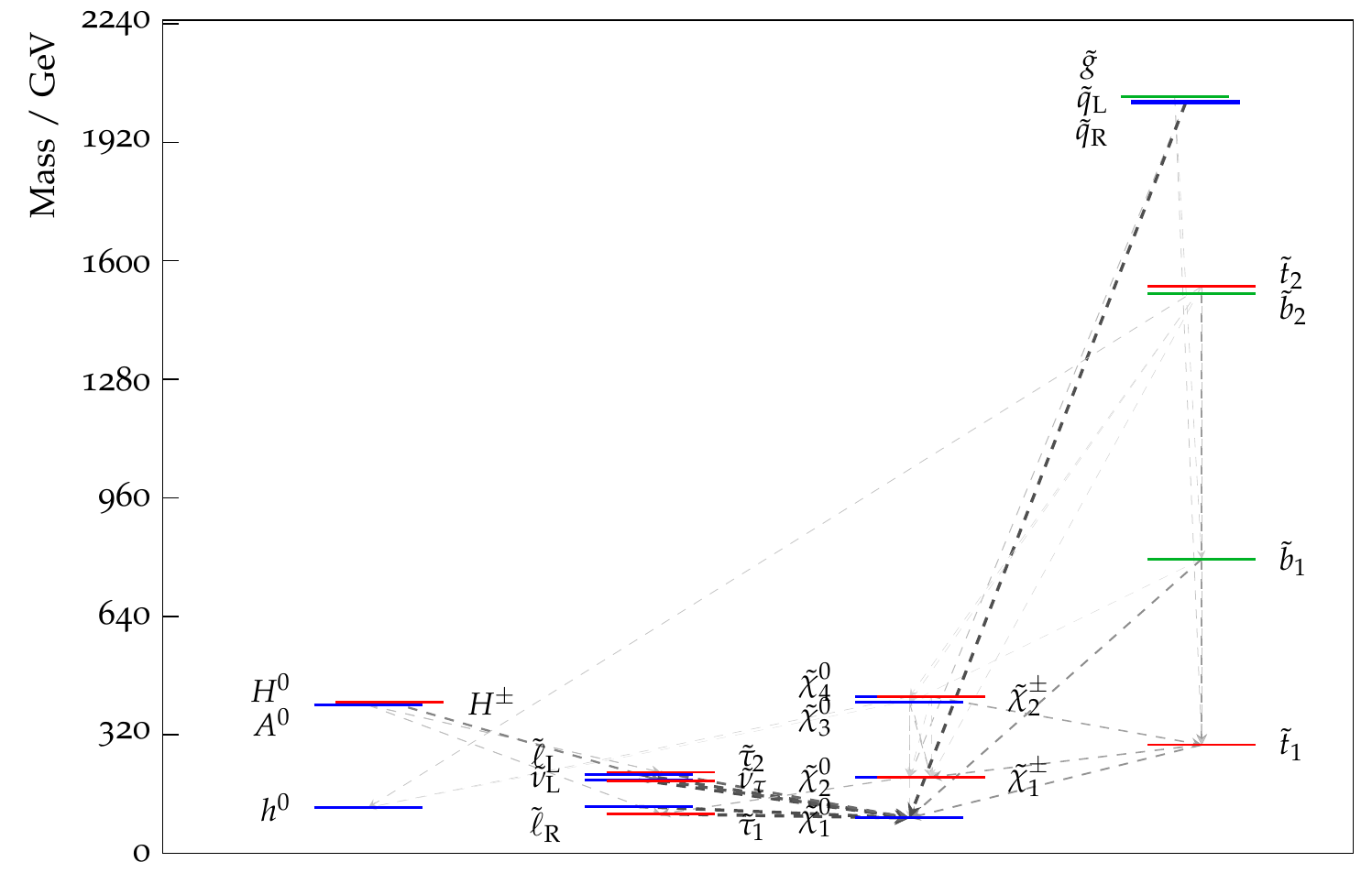}
 \end{center}
  \caption{\label{fig:stc4} Left: Full spectrum of STC4, with decay modes with a branching fraction of at 
  least $90\%$.
  Right: The same spectrum, but now indicating all decays with a branching fraction larger than $10\%$.}
\end{figure}

The full spectrum of STC4 is shown in Fig.~\ref{fig:stc4}. The dashed lines indicate the decay chains of 
the various sparticles. In the left part of the figure, only decays with a branching ratio (BR) larger 
than $90\%$ are shown, while the right part includes all decays with a branching fraction of at least 
$10\%$. The grey-scale of the lines indicates the size of the branching ratio. Only very few particles, 
namely the 1st/2nd generation squarks, the sneutrinos and the lighter set of charged sleptons have decay 
modes with $100\%$ BR. In particular the stop and sbottom, but also the electroweakinos have various decay 
modes, none of them with a BR larger than $50\%$. This plethora of decay modes makes it challenging to 
separate the various production modes and identify each sparticle.
  
Even if not all of them can be addressed on the timescale of the Snowmass process, the final goals of this study comprise the following questions for both LHC and the ILC in the example of the
STC scenarios:

\begin{itemize}
    \item Which signature will lead to the first discovery of a discrepancy from the SM? How much integrated
    luminosity and operation time would be needed for this?
    \item Which production modes of which sparticles contribute to this signal? Can we tell how many these are?
     And which masses and quantum numbers they have? 
    \item Which other signatures will show a signal? And again: Can one find out which production modes contribute?
    \item Which observables (masses, BRs, cross-sections) can be measured with which precision?
    \item Can we show that it is SUSY?
    \item Can we show that's the MSSM (and not eg. the NMSSM)?
    \item Can the $\tz_1$ be identified as Dark Matter particle?
\end{itemize}

In the next section, we will describe the phenomenology of our benchmark models at the LHC and summarize the
obtained simulation results. In section~\ref{sec:ILC}, we will do the same for the ILC case, before we
give our conclusions.

\section{Large Hadron Collider Studies} \label{sec:LHC}
The discovery potential of the LHC for the STC scenarios introduced above is described in this section. 
Here we study the LHC at 14 TeV with 300~fb$^{-1}$ accumulated luminosity and 50 pileup (PU) events.
Next steps would be to study these scenarios with 3000~fb$^{-1}$ and a pileup of 140 events, and a future proton-proton collider with 33 TeV. 

\begin{figure}[hbt]
\begin{center}
  \subfigure[]{ \includegraphics[width=0.45\textwidth]{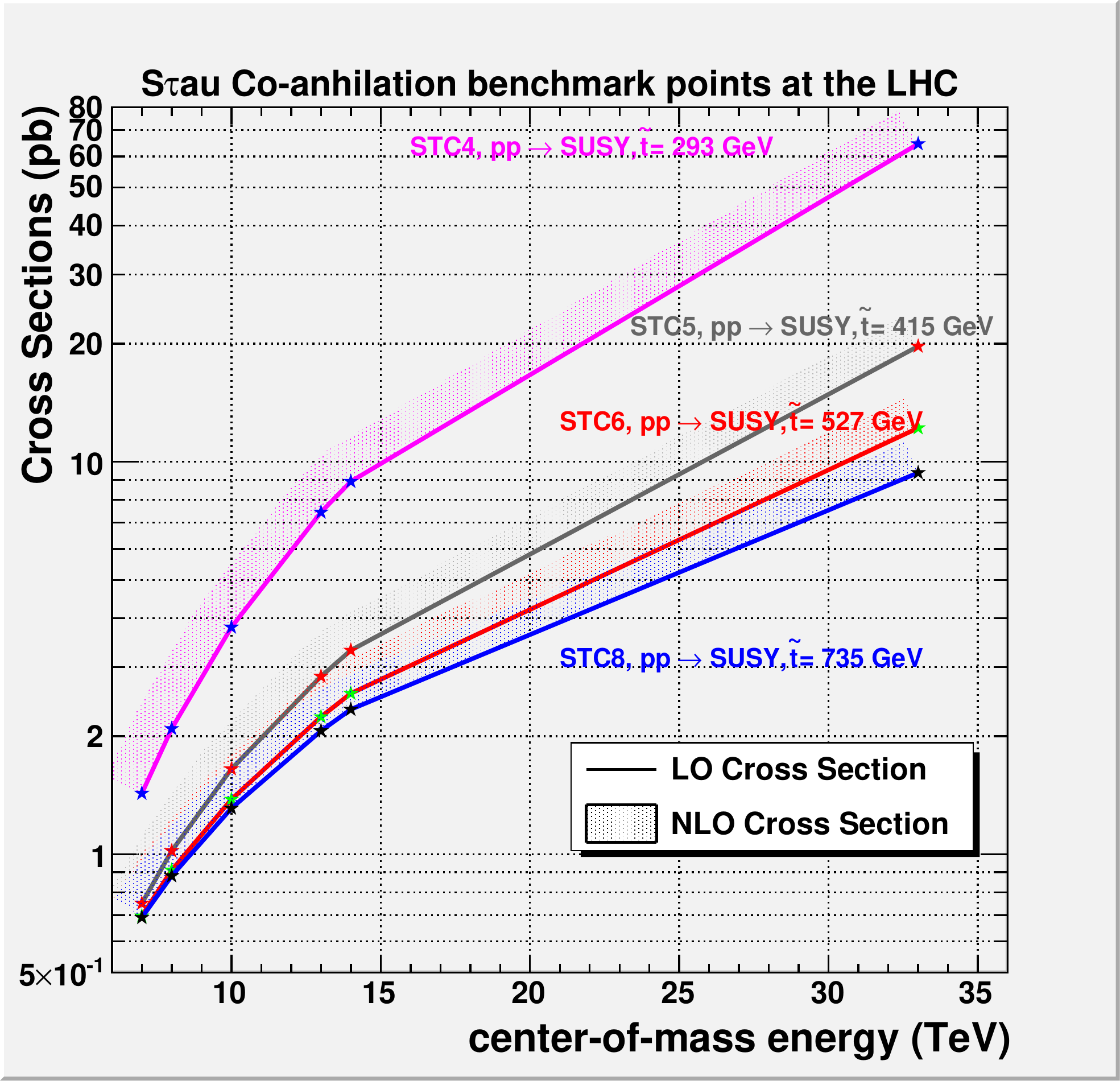} }
  \subfigure[]{ \includegraphics[width=0.45\textwidth]{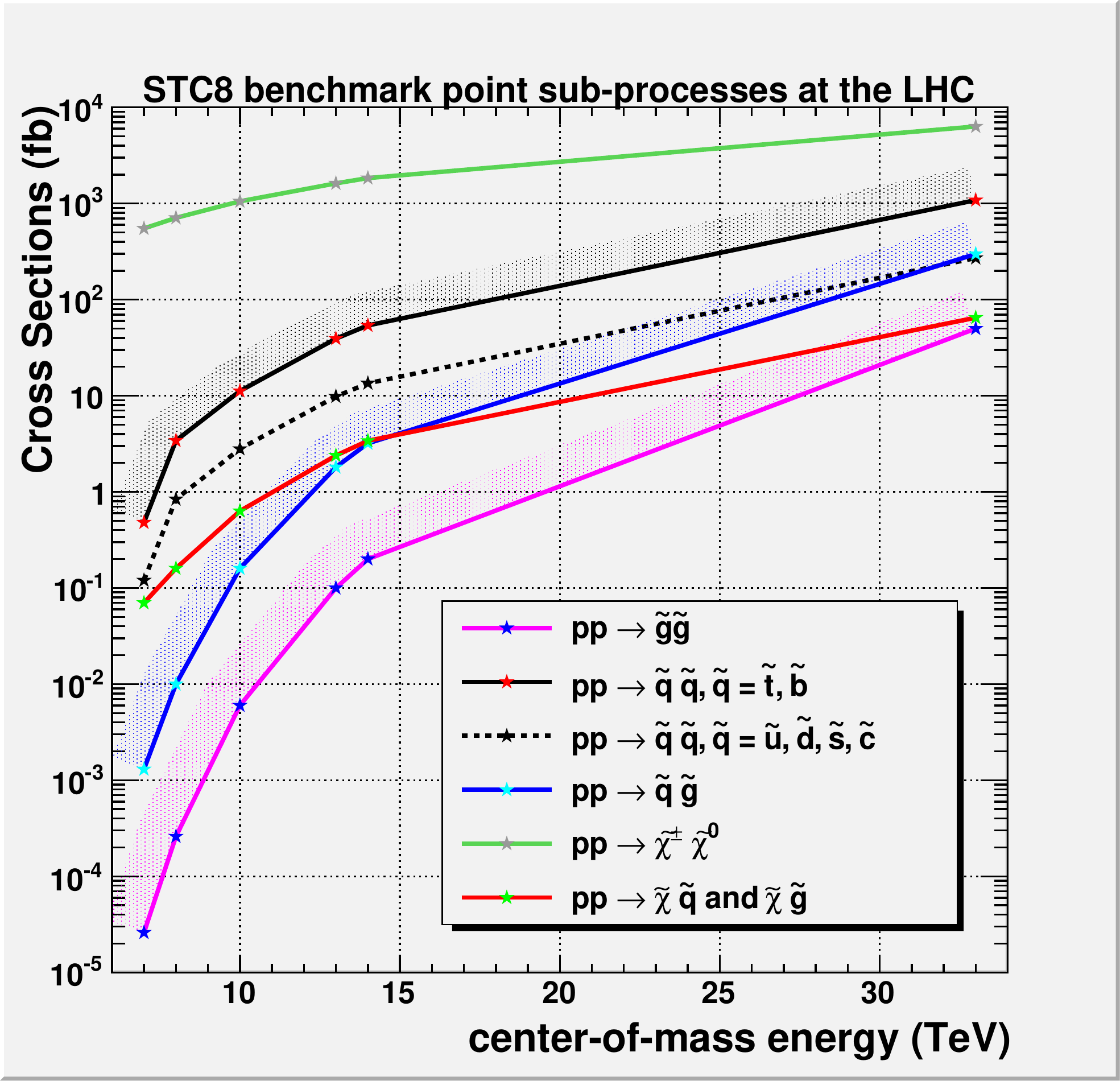} }
\caption{The inclusive cross sections of the four investigated models (a) and the cross sections of the subprocesses for the model STC8 (b). The lines correspond to the leading-order (LO) cross section and the upper end of the hatched area corresponds to the next-to-leading order (NLO) 
cross section.}
\label{fig:lhc_cs} 
\end{center}
\end{figure}

The cross sections for the signal models have been calculated at leading order with Pythia8~\cite{Sjostrand:2006za}, and for most
subprocesses at next-to-leading order with Prospino2.1~\cite{Beenakker:1996ch,Beenakker:1997ut}. As Prospino offers only cross section calculations up to 14~TeV, a private patch has been applied to calculate the cross sections at 33~TeV.
The inclusive cross sections for the four different models at different LHC energies are summarized in Fig.~\ref{fig:lhc_cs}(a).
The mass of the stop quarks rises from model STC4 to STC8 subsequently from 293~GeV to 735~GeV, reducing the cross section for stop production significantly, while the production cross section of the electroweak particles stays roughly the same. Already in STC5 the cross section for direct stop production is smaller than the one of chargino-neutralino production. 
The cross section of the different subprocesses for the model STC8 are shown in Fig.~\ref{fig:lhc_cs}(b). A table with the cross-sections of the dominant subprocesses for all four scenarios can be found in the appendix.

The stops predominantly decay to top quarks and the lightest neutralino (54\%), or to bottom quarks and the lightest chargino (46\%). Here, the chargino decays mainly to $\ttau_1$ and $\nu_{\tau}$ (70 \%) or $\tau$ and $\tnu_{\tau}$ (10\%), where the latter decays $93\%$ invisibly. This situation suggests different possible strategies to search for stop production:
\begin{itemize}
\item In the case that both stops decay via  $\tst_1 \to t \tz_1$, the $t \bar{t}$ plus missing transverse energy final state can be searched for either with no or one lepton. The sensitivity of these channels will be investigated in sections~\ref{subsec:Hadronic} and~\ref{subsec:LHCsemilep}, respectively.
\item In the case that both stops decay via $\tst_1 \to b \twpm_1$, searches for events with two $b$-jets and missing transverse energy could be 
sensitive since the decay products of the charginos are expected to be very soft. This case has been investigated recently by ATLAS~\cite{ATLAS-CONF-2013-001}, and therefore we studied the event yield expected from this analysis in the case of STC4, as described in section~\ref{subsec:LHC8TeV}. Due to time reasons we did not yet investigate the propects of this analysis at $14\,$TeV.
\item The largest branching fraction would be covered by the mixed case $\tst_1 \tst_1 \to t \tz_1 b \twpm_1$ 
\cite{Graesser:2012qy}. Currently to our knowledge such
      a signature is not targeted by any existing LHC search. Due to time reasons we could not develop a dedicated analysis in the context of this study either, but this decay mode is expected to be covered to some extent by the fully hadronic $t \bar{t} \tz_1 \tz_1$ analysis discussed in~\ref{subsec:Hadronic}.
\end{itemize}       

The sbottom mass is only slightly higher than the stop mass in scenario STC8, and will be produced with almost similar cross section in this scenario. Especially the analysis of the full-hadronic final state would also be sensitive to direct sbottom production, as the sbottom decays most of the time (58\%) to a bottom quark and the lightest neutralino. With high statistics (3000~fb$^{-1}$) it might even be possible to see slight differences in the energy 
spectrum of the $b$-tagged jets, a slightly harder spectrum is expected from the sbottom decay, but this has not yet been studied here.

The electroweak particles are very hard to identify at the LHC, as they mainly decay to final states containing rather soft tau leptons, but nevertheless we will present search prospects for electroweakino production in section~\ref{subsec:LHCewkino}.

For all analyses, signal and background are generated with Delphes 3.0.9~\cite{Ovyn:2009tx} as used by all Snowmass analyses~\cite{Snowmass_MC1, Snowmass_MC2}.
The efficiency of the reconstructed objects (muons, electrons, jets, etc.) is defined by Snowmass specific 
Delphes card files. In case of pileup, the fast jet correction is applied with active area 
correction~\cite{Cacciari:2011ma}. The jet energy resolution and the resolution for the different pileup 
scenarios is shown in the Appendix.

While the experiments usually try to estimate the background from data to the largest possible amount, we 
restrict ourselves here to a simple estimation based on the simulation of the above mentioned processes.

Systematic uncertainties have been considered in terms of a conservative and an optimistic scenario, which
assume global uncertainties on the background expectation of 25\% and 15\%, respectively, for the stop searches.
The search for the electroweak particles with the same-sign analysis is expected to suffer more from the
systematic uncertainties (due to the less well-known cross section of the di-boson production and analysis-specific problems like 
isolation in high pileup and identification of fake leptons), therefore we assume here uncertainties of 30\% and 20\%, respectively.
For the extrapolation to 3000~fb$^{-1}$ we expect that the systematic uncertainties will be further reduced, especially
as the backgrounds are determined from data, where higher statistics will decrease the uncertainty on the background estimation. Here
we assume an uncertainty of 10\% for all analyses. 

\subsection{Comparison to $8$~TeV search for final states with two $b$-jets}\label{subsec:LHC8TeV}

ATLAS published a search for final states with two $b$-jets from $\tb_1 \tb_1 \to b \bar{b} \tz_1 \tz_1$ 
based on 12.8~fb$^{-1}$ of $8\,$TeV data~\cite{ATLAS-CONF-2012-165}. They recently reinterpreted this 
analysis for $\tst_1 \to b \twpm_1$ with small $\twpm_1$-$\tz_1$ mass 
splittings~\cite{ATLAS-CONF-2013-001} based on a simplified model approach. Taking the obtained limits at 
face-value would lead to the conclusion that the STC4 point is excluded by this search. However, as 
discussed above, the concurring decay modes as well as the exact decay modes and mass splittings need to 
be taken into account correctly. Therefore, we reimplemented this analysis as closely as possible into 
the Snowmass analysis framework based on the Delphes detector
simulation program and evaluated the expected signal yield for STC4 for 12.8~fb$^{-1}$ at $8\,$TeV, following the 
selection requirements of the ATLAS analysis.
We show our results in Table~\ref{tab:STC4_Atlas8TeV} in comparison with the experimental results from 
ATLAS. The signal regions SR3a and SR3b with a b-tag veto on the leading jet are not included since they 
are even less sensitive for our model, and we only show STC4 which has the highest stop production 
cross-section of our series of points. It can be easily seen from Table~\ref{tab:STC4_Atlas8TeV} that this 
ATLAS analysis is not able to exclude STC4.

\begin{table}[htdp]

\caption{Number of signal events for the model STC4 after a selection as described for an ATLAS analysis 
performed on 12.8~fb$^{-1}$ of data recorded at a center-of-mass energy of 8~TeV. 
The detailed cut-flow is described in the ATLAS note~\cite{ATLAS-CONF-2012-165} and
$m_{\rm CT}$ is the boost-corrected contransverse mass~\cite{Polesello:2009rn}.
  }
\begin{center}
\begin{tabular}{|r|c|c|c|c|c|}
\hline 
Description    & \multicolumn{5}{c|}{Signal Region} \\ 
\hline
\hline
               & \multicolumn{4}{c|}{SR1} & SR2  \\ 
\hline
               & $m_{\rm CT} > 150$ & $m_{\rm CT} > 200$ &$m_{\rm CT}
               > 250$ &$m_{\rm CT} > 300$ &\\
\hline 
ATLAS \hfill observed & 172 &  66 & 16  & 8   & 104 \\
expected
      SM bgrd. & 176 &  71 & 25  & 7.4 &  95 \\ 
95\% CL UL on exp. bgrd. 
               & 55  &  25 & 12.5& 5.5 &  32\\ 
\hline
\hline 
STC4           &  18 & 13 & 9.0 & 6.6 & 18 \\ 
\hline 									
\end{tabular} 
\end{center}
\label{tab:STC4_Atlas8TeV}
\end{table}%

\subsection{Stop search with full-hadronic final states}\label{subsec:Hadronic}

%Though the masses of stop and $\tz_1$ as given in the STC4 model are in the excluded area for the direct 
%stop production discussed in the ATLAS analysis, where the stop decays to a 
%top quark and the $\tz_1$, the discovery of the direct stop production with fully hadronic decay with the 
%current data is not possible with the ATLAS cut flow, showing that the simplified models sometimes give 
%an 
%over-optimistic picture due to the assumed branching ratio of 100\%, which is not expected to happen in a 
%full model.

In the followig we define a simple hadronic cut-and-count search without leptons at 14~TeV sensitive for 
our model points. The cut flow is summarized in Table~\ref{tab:event_selection_fullhadronic}.
The main backgrounds in this search are:
\begin{itemize}
\item t$\bar{\rm t}$ + jets
\item W+jets
\item Single top production
\item Z+jets with Z$ \rightarrow \nu \bar{\nu}$
\item QCD multijet production
\end{itemize}

Several kinematic variables are exploited to separate the signal from background. We calculate the missing 
transverse energy $E_T^{\rm miss}$ as the vectorial sum of pileup subtracted jets with
%dk $p_T > 20$~GeV and with $|\eta| < 2.8$ 
$p_T > 20$~GeV and $|\eta| < 2.5$ 
and leptons with 
%dk $p_T > 10$~GeV and with $|\eta| < 2.4$. 
$p_T > 5$~GeV and $|\eta| < 2.5$. Another variable to separate signal from background is the minimum 
angle $\Delta \Phi_{\rm min}$ between the leading jets and $E_T^{\rm miss}$, which is small for QCD multi 
jet background, while signal leads preferably to larger values.
The scalar sum of jets with $p_T > 20$~GeV and 
%DK with $|\eta| < 2.8$,
$|\eta| < 2.5$ added to the missing energy, which is called $m_{\rm eff}=H_T+E_T^{\rm miss}$, can separate 
events with higher mass SUSY particles from Standard Model processes. In this analysis we use the ratio of 
$E_T^{\rm miss}$ and $m_{\rm eff}$ calculated with the three leading jets. After the large $E_T^{\rm miss}$  
requirement the QCD multijet background is expected to be negligible.

\begin{table}[htdp]
\caption{Overview of the event selection requirements.}
\begin{center}
\begin{tabular}{|c|c|}
\hline
Description & \multicolumn{1}{c|}{Selection} \\ \hline \hline
Lepton veto & No $e$ or $\mu$ with $p_T >10$~GeV  \\ \hline
Leading jet $p_T$ & $>120$~GeV\\ \hline 
2nd leading jet $p_T$ & $>70$~GeV \\ \hline 
3rd leading jet $p_T$ & $>60$~GeV \\ \hline 
No. of b-tagged jets & $\ge 2$ \\ \hline
$H_{\rm T}$ & $>1000$~GeV \\ \hline
$\Delta \Phi (E_T^{\rm miss}, p_T^{\rm jet 1,2})$ &  $>0.5$ \\ \hline 
$E_T^{\rm miss}/m_{\rm eff}$ & $>0.2$ \\ \hline
$E_T^{\rm miss}$  & $>750$~GeV  \\ \hline
\end{tabular}
\end{center}
\label{tab:event_selection_fullhadronic}
\end{table}%

The cutflow for the inclusive signal and background events for 14 TeV center-of-mass energy and 
300~fb$^{-1}$ is summarized in Table~\ref{tab:fullhadronic_result_noPU} for no pileup and in 
Table~\ref{tab:fullhadronic_result_50PU} for 50 pileup events.

\begin{figure}[htbp]
\begin{center}
  \subfigure[]{ \includegraphics[width=0.45\textwidth]{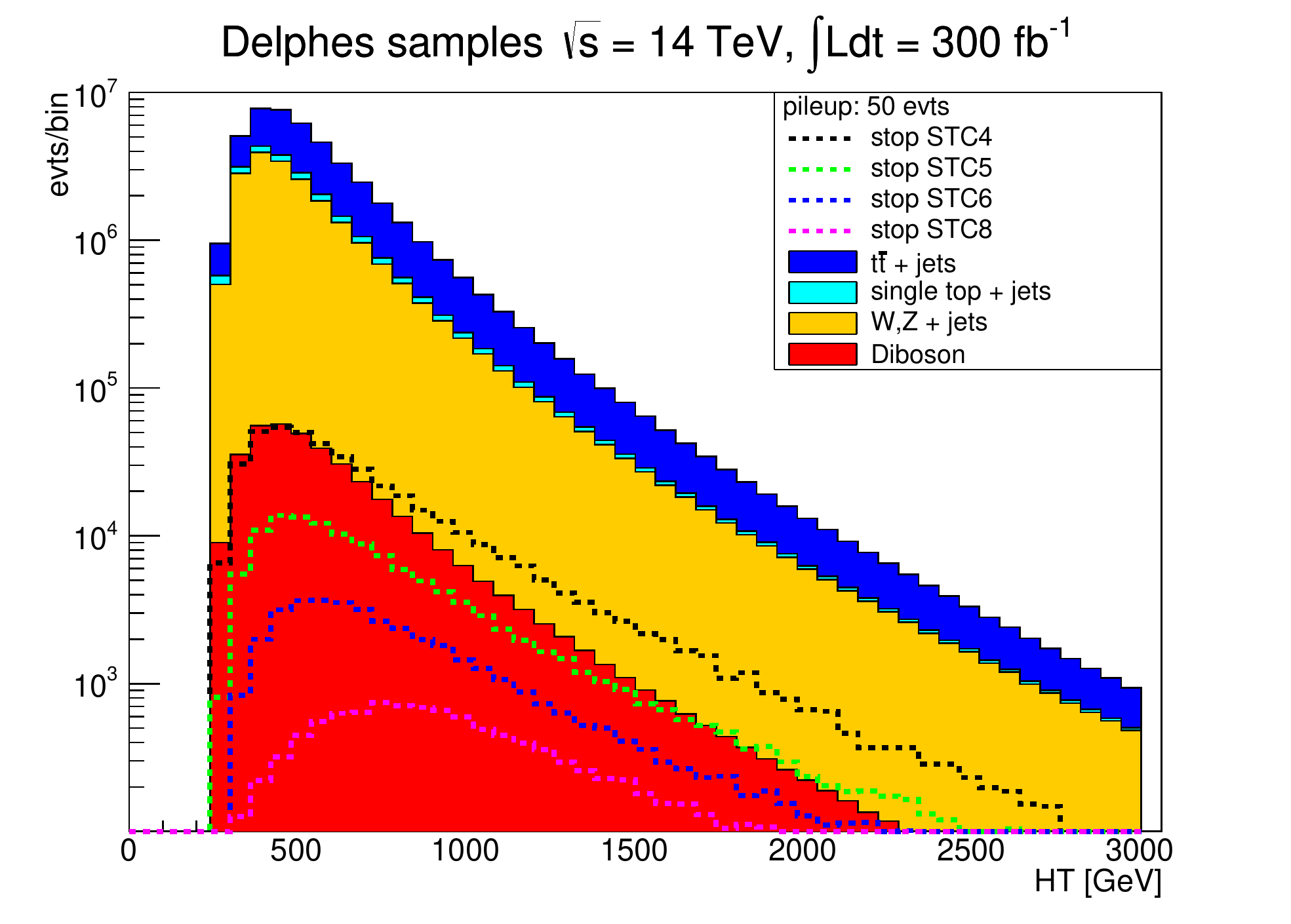} }
  \subfigure[]{ \includegraphics[width=0.45\textwidth]{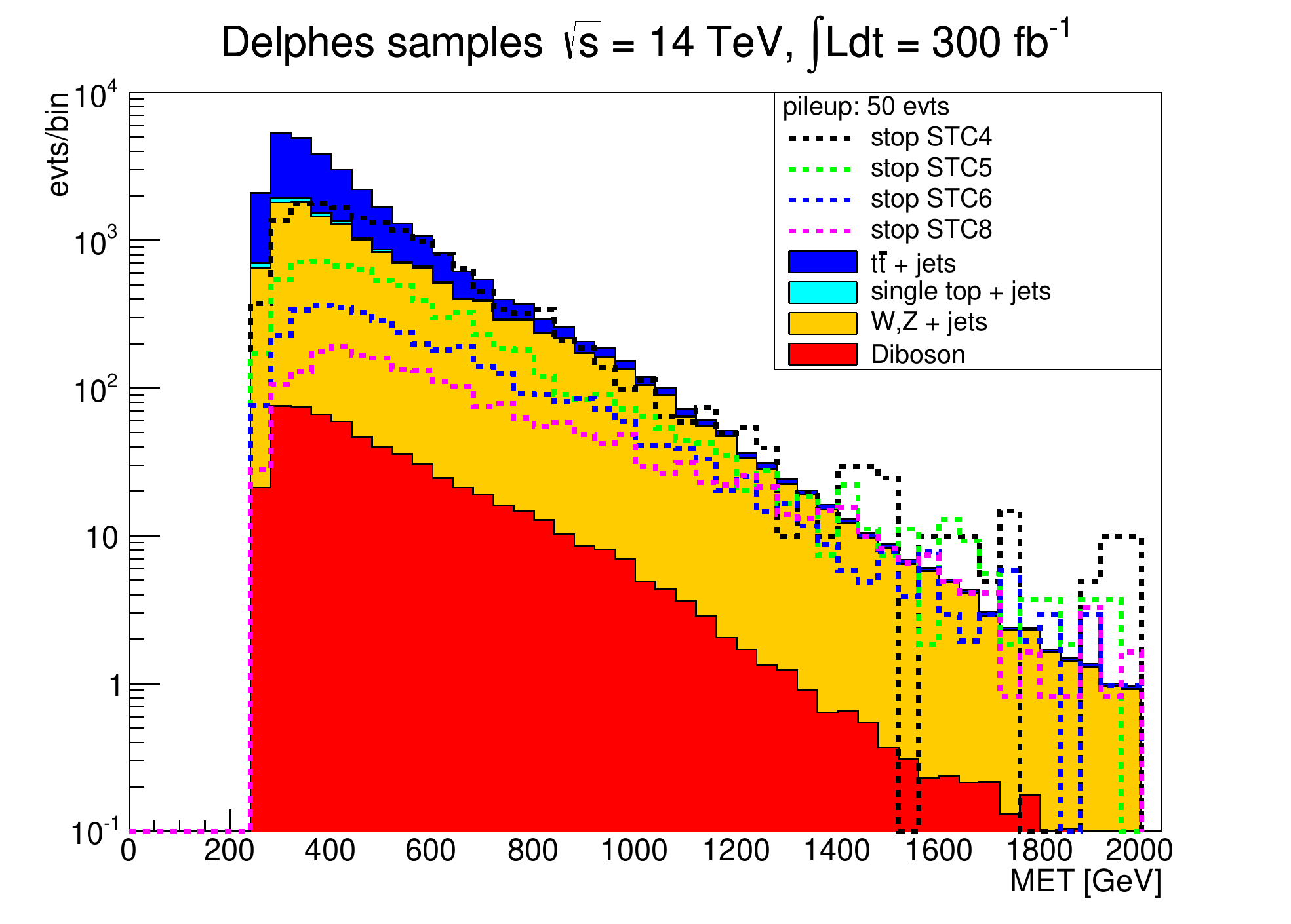} }
\caption{The scalar sum of the jets $H_T$ after jet and b-jet requirements and lepton veto (a) and 
the missing transverse energy (b) after the full-hadronic event selection for 50 pileup events. The full 
histograms describing the backgrounds are stacked, and the four inclusive signal models are shown as 
dotted lines (not stacked). }
\label{fig:fullhadronic_MET_HT}
\end{center}
\end{figure}

Figure~\ref{fig:fullhadronic_MET_HT} shows two control plots after the application of a part of the 
selection requirements as described in Table~\ref{tab:event_selection_fullhadronic}. The variable $H_T$ is 
shown after the lepton veto and the jet and b-jet requirements, and $E_T^{\rm miss}$ after the full 
selection except for the $E_T^{\rm miss}$ requirement itself.

\begin{sidewaystable}[htdp]
\caption{
Cutflow: number of events for the inclusive signal samples and several important backgrounds for the 
full-hadronic stop analyses with 300~fb$^{-1}$ at 14~TeV without pileup. The last two lines show the 
significances %without and 
with an additional systematic background uncertainty of 25\% and, as an 
optimistic scenario, of 15\%.
}
\begin{center}
\begin{tabular}{|c||c|c|c|c||c||c|c|c|c|}
\hline
Description   &diboson  &  ttbar+jets   &boson+jets& single top&sum bgrds  &    STC4  &     STC5  &    STC6  &     STC8 \\ \hline \hline
  
preselection                       &  110817000 &  215894000 & 16840400000 &   62062700 & 17229173700 &    3840000   &    1146000   &     759000   &     657000   \\ \hline
lepton veto                        &   93950100 &  150178000 & 15967400000 &   50836600 & 16262364700 &    2992790   &     867266   &     573276   &     505996   \\ \hline
n jets $\ge 3$                    &   14108800 &  123304000 & 1612230000 &   22900100 & 1772542900 &    1918060   &     327129   &     106267   &      49223   \\ \hline
jet1 $> 120$~GeV                   &    6665570 &   57008900 &  614067000 &    9833780 &  687575250 &    1266390   &     282534   &      89244   &      35846   \\ \hline
jet2 $> 70$~GeV                    &    6102030 &   53676000 &  555486000 &    9053350 &  624317380 &    1177530   &     267671   &      83758   &      31973   \\ \hline
jet3 $> 60$~GeV                    &    4306200 &   43971700 &  351147000 &    6102790 &  405527690 &     962921   &     222022   &      69959   &      24684   \\ \hline
bjets $ge 2$                       &     389723 &   24659200 &   21764800 &    2211580 &   49025303 &     551566   &     135976   &      41827   &      11298   \\ \hline
$H_T > 1000$~GeV                   &      29056 &    1347070 &     884348 &      72649 &    2333123 &      67195   &      22797   &      10468   &       5264   \\ \hline 
$\Delta \Phi > 0.5$                &      19028 &     850594 &     582261 &      44547 &    1496430 &      40860   &      16951   &       8262   &       4281   \\ \hline 
$E_T^{
m miss}/m_{
m eff}>0.2$   &        639 &      16317 &      14545 &        576 &      32079 &      16538   &       7486   &       4162   &       2297   \\ \hline \hline
$E_T^{
m miss}>750$~GeV           &         97 &        331 &       1908 &         12 &       2350 &       1843   &       1048   &        775   &        636   \\ \hline
$s/\sqrt{b+(0.25*b)^2}$           &      &      &      &      &      &        3.1 &        1.8 &        1.3 &        1.1 \\ \hline
$s/\sqrt{b+(0.15*b)^2}$           &      &      &      &      &      &        5.2 &        2.9 &        2.2 &        1.8 \\ \hline
\end{tabular}
\end{center}
\label{tab:fullhadronic_result_noPU}

\vspace{1cm}
\caption{
Cutflow: number of events for the inclusive signal samples and several important backgrounds for the 
full-hadronic stop analyses with 300~fb$^{-1}$ at 14~TeV and with 50 pileup events. The last two lines show 
the significances %without and 
with an additional systematic background uncertainty of 25\% and, as an 
optimistic scenario, of 15\%.
  }
\begin{center}
\begin{tabular}{|c||c|c|c|c||c||c|c|c|c|}
\hline
Description   &diboson  &  ttbar+jets   &boson+jets& single top&sum bgrds  &    STC4  &     STC5  &    STC6  &     STC8 \\ \hline \hline
  
preselection                       &  110822000 &  216124000 & 16842600000 &   62086200 & 17231632200 &    3840000   &    1146000   &     759000   &     657000   \\ \hline
lepton veto                        &   94040900 &  148709000 & 15970700000 &   50821500 & 16264271400 &    2939780   &     858682   &     569011   &     502994   \\ \hline
n jets $\ge 3$                    &   15503200 &  118324000 & 1987680000 &   22550600 & 2144057800 &    1749960   &     317637   &     103979   &      48157   \\ \hline
jet1 $> 120$~GeV                   &    6945990 &   54696900 &  668990000 &    9576410 &  740209300 &    1052210   &     259433   &      84595   &      34240   \\ \hline
jet2 $> 70$~GeV                    &    6300220 &   51083400 &  597564000 &    8686940 &  663634560 &     960943   &     242314   &      78652   &      30258   \\ \hline
jet3 $> 60$~GeV                    &    4355270 &   41560400 &  359923000 &    5790680 &  411629350 &     774987   &     199271   &      65688   &      23520   \\ \hline
bjets $ge 2$                       &     382265 &   23020600 &   21429600 &    2091020 &   46923485 &     433180   &     121225   &      39601   &      11163   \\ \hline
$H_T > 1000$~GeV                   &      28432 &    1241760 &     870796 &      69562 &    2210551 &      59848   &      21130   &       9763   &       4912   \\ \hline 
$\Delta \Phi > 0.5$                &      18856 &     800289 &     578069 &      43795 &    1441010 &      36608   &      15623   &       7720   &       4020   \\ \hline 
$E_T^{
m miss}/m_{
m eff}>0.2$   &        619 &      15769 &      12902 &        577 &      29869 &      15716   &       7067   &       3854   &       2192   \\ \hline \hline
$E_T^{
m miss}>750$~GeV           &         92 &        334 &       1721 &         13 &       2161 &       1920   &       1109   &        815   &        633   \\ \hline
$s/\sqrt{b+(0.25*b)^2}$           &      &      &      &      &      &        3.5 &        2.0 &        1.5 &        1.2 \\ \hline
$s/\sqrt{b+(0.15*b)^2}$           &      &      &      &      &      &        5.9 &        3.4 &        2.5 &        1.9 \\ \hline

\end{tabular}
\end{center}

\label{tab:fullhadronic_result_50PU}
\end{sidewaystable}%

We assume 
two cases for the systematic uncertainty: 25\%  and a more optimistic scenario of 15\%. It is possible to see an excess due to the signal for the direct stop and sbottom production subprocess for all four scenarios, but only 
for STC4 and STC5  a significance of more than 3$\sigma$ can be observed if the background uncertainty is around 15\% or lower. Due to the smaller stop production 
cross section in the models STC6 and STC8 they are more challenging, as the background is higher, but it might still be possilbe to raise the significance 
if the selection and background determination is further developed. The pileup does not have a large influence here.

\subsection{Stop search with final states including one electron or muon}\label{subsec:LHCsemilep}

In addition to the full-hadronic decay channel, we also investigate the discovery reach for stop decays 
including one electron or muon in the final state with an analysis similar to the currently performed 
analysis by CMS~\cite{CMS-PAS-SUS-13-011} on the 2012 data. The main backgrounds in this search are:
\vspace{1em}
\begin{itemize}
\item t$\bar{\rm t}$ + jets
\item W+jets
\item Single top production
\item Z+jets (with one lepton not identified)
\end{itemize}

The dominant $W$ and t$\bar{\rm t}$ backgrounds can be controlled efficiently by analyzing the event 
kinematics. For this purpose two additional kinematical variables are introduced: $M_T$ and 
$M_{T2}^W$~\cite{Bai:2012gs}. The transverse mass, defined as
\begin{equation} 
	M_T = \sqrt{2 p_T^{\rm lep} E_T^{\rm miss} - 2 \vec{p_T^{\rm lep}} \vec{p_T^{\rm miss}}} 
\end{equation}
allows to reject events with leptonically decaying $W$ bosons, while
the $M_{T2}^W$ variable, defined as the minimum 'mother' particle mass compatible with all
the transverse momenta and mass-shell constraints, 
\begin{equation} \label{eq:MT2W_def}
  M_{T2}^W = \rm{minimum} \left\{ \text{$m_y$ consistent with:} \left[ \begin{array}{r}
  \vec{p}_1^T + \vec{p}_2^T =  \vec{E}_T^{miss}, p_1^2 = 0, \left(p_1 + p_l \right)^2 = p_2^2 = M_W^2, \\
  \left(p_1 + p_l + p_{b_1} \right)^2 = \left(p_2 + {p_{b_2}} \right)^2 =m_y^2
  \end{array}
  \right] \right\}.
\end{equation}
exploits the event topology to reject semileptonic t$\bar{\rm t}$ events.
By construction, $M_{T2}^W$ has an endpoint at the top mass
for the dilepton t$\bar{\rm t}$ background.

For events with two identified $b$ jets the calculation relies on the correct pairing of the lepton and 
the $b$ jets. For events with only one identified $b$ jet one of the none b-tagged jets has to be 
included. The definition \ref{eq:MT2W_def} is therefore extended by minimizing over all possible lepton, 
jet, and $b$-jet combinations within an event.

The missing transverse energy $E_T^{\rm miss}$ is calculated as for the fully hadronic case 
\ref{subsec:Hadronic}, and similar the minimum angle $\Delta \Phi_{\rm min}$ which is now calculated for 
only the two highest $p_T$ jets.

\begin{figure}[htbp]
\begin{center}
  \subfigure[]{ \includegraphics[width=0.45\textwidth]{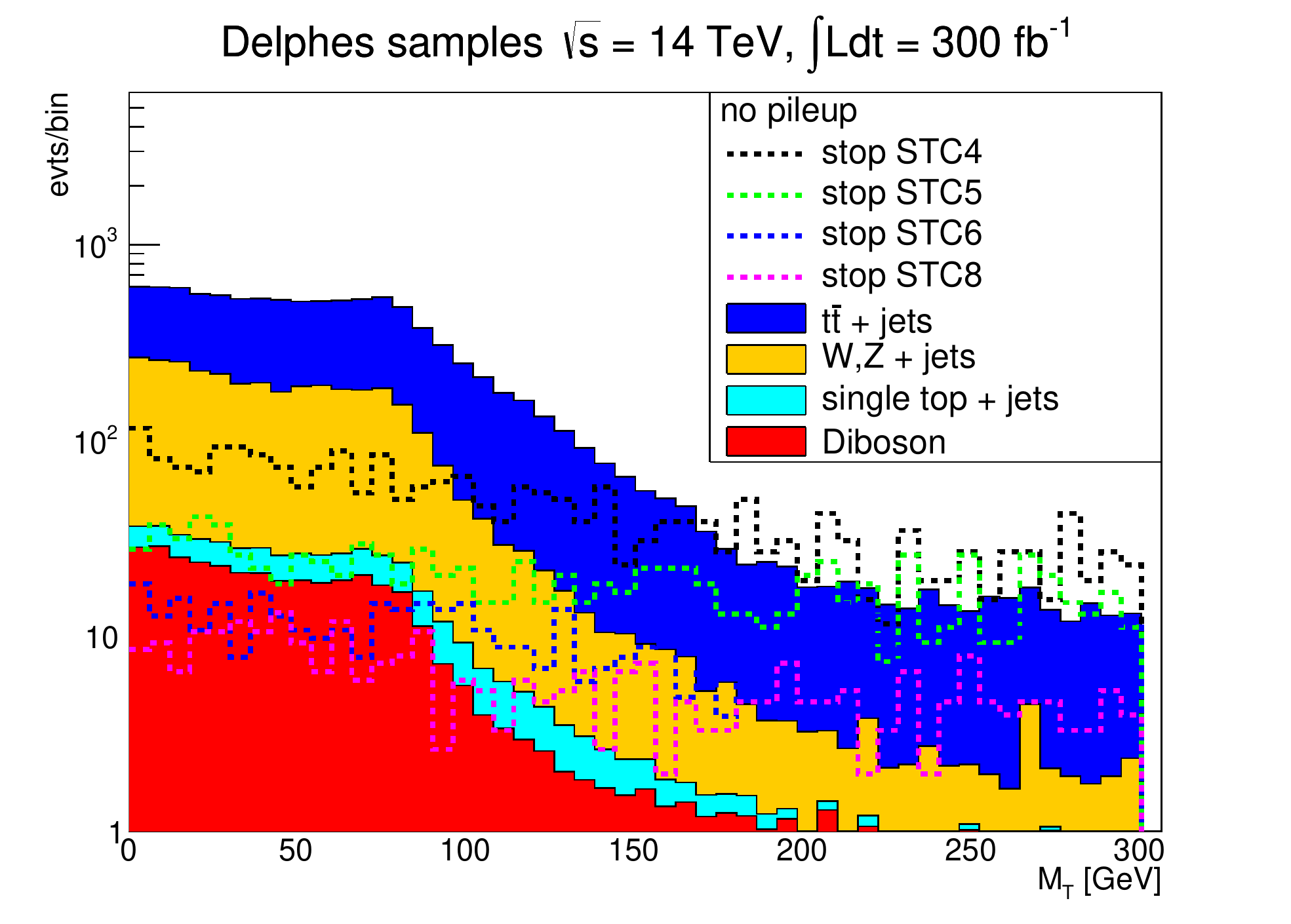} }
  \subfigure[]{ \includegraphics[width=0.45\textwidth]{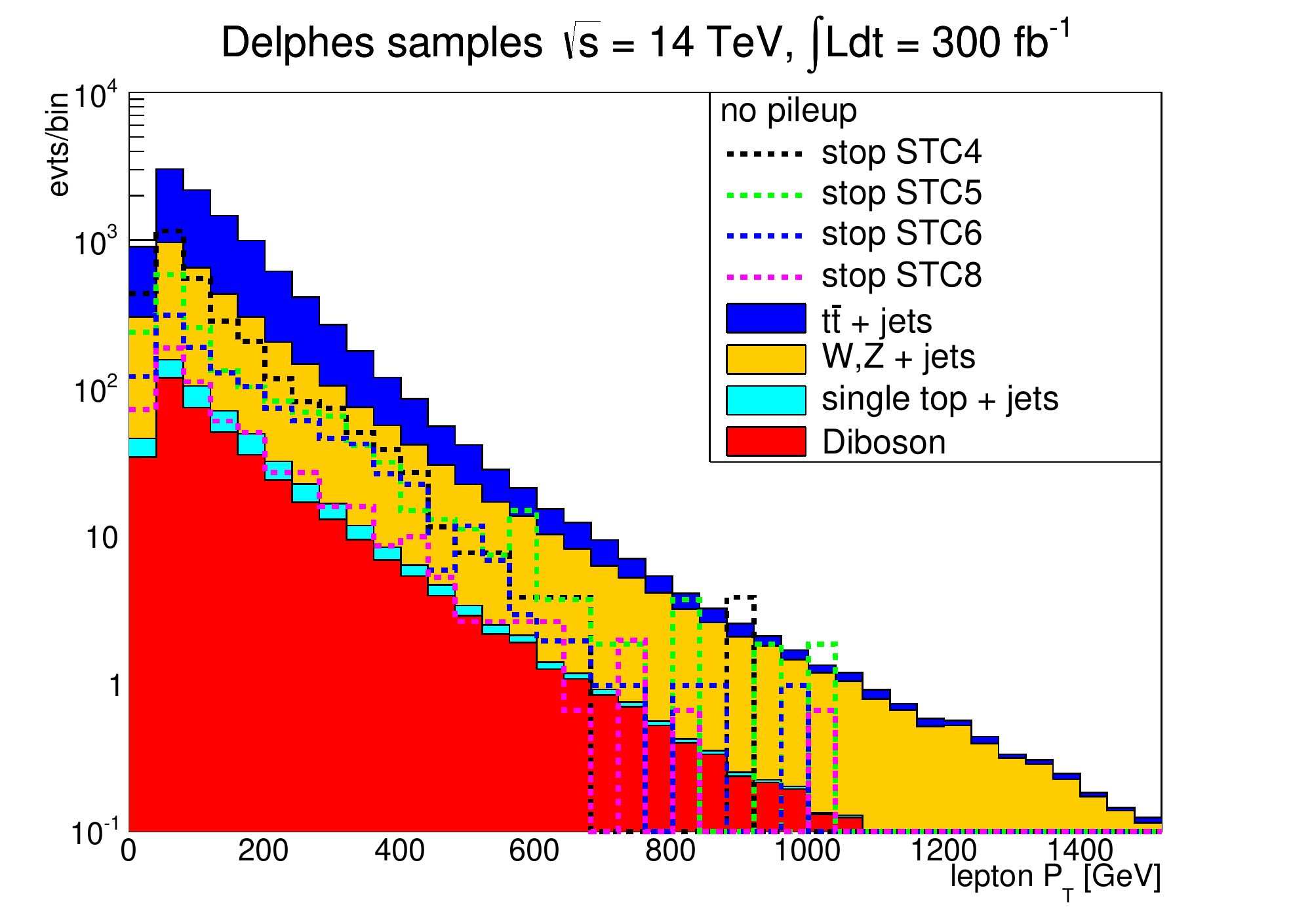} }
\caption{ $M_T$ distribution after all lepton and jet requirements (a) and the lepton transverse momentum (b) after the 
 $M_T$ requirement. The full histograms describing the backgrounds are stacked, and the inclusive signals for all four models are shown as dotted lines (not stacked).  }
\label{fig:single_MT_LPT}
\end{center}
\end{figure}

An overview of the event selection is given in Table~\ref{tab:event_selection_stop_1l}. A cutflow with 
these requirements is given in Table~\ref{tab:single_lepton_result_noPU} for no pileup and in Table~\ref{tab:single_lepton_result_50PU}
for 50 pileup events.
Figure~\ref{fig:single_MT_LPT} contains the $M_T$ distribution after all lepton and jet requirements (a) and the lepton transverse momentum (b) after the $M_T$ requirement.

\begin{table}[htdp]
\caption{Overview of the event selection requirements for the leptonic direct stop search.}
\begin{center}
\begin{tabular}{|c|c|}
\hline
Description & {Selection} \\ 
\hline \hline
Exactly 1 lepton  &  {$e$ or $\mu$ with $p_T >30$~GeV} \\ 
                  &  {no other $e$/$\mu$ with $p_T >10$~GeV} \\ 
\hline
Number of jets    &  {n $\ge$ 4 with  $p_T$  $>40$~GeV}  \\ 
\hline 
b-tagged jets     &  { 1 or 2 }  \\ 
\hline
$\Delta \Phi (E_T^{\rm miss}, p_T^{\rm jet 1,2})$ \rule{0em}{1.1em} &   {$>0.5$} \\  
\hline
$E_T^{\rm miss}$  &  {$>500$~GeV}  \\ 
\hline
$H_{\rm T}$ &  {$>500$~GeV} \\ 
\hline
$M_{\rm T}$ &  {$>120$~GeV} \\ 
\hline\hline
\multirow{2}{*}{$M_{\rm T2}^{\rm W}$  $>250$~GeV }&topness $>9.5$\\&$p_{T}$ asymmetry  $>-0.2$\\
\hline

\end{tabular}
\end{center}
\label{tab:event_selection_stop_1l}
\end{table}%

As an alternative to $M_{\rm T2}^{\rm W}$ for suppressing the dileptonic top background we consider the {\it topness} 
variable as defined in \cite{Graesser:2012qy}. The unknown lepton momenta, i.e.\ the lost electron or muon and the 
two 
neutrinos, are reconstructed by numerically minimizing a $\chi^2$-expression that also takes into account the 
invariant mass of the $t\tilde{t}$ system. We have applied the different parameters, i.e.\ resolution terms, as given 
in \cite{Graesser:2012qy} and 
did not optimized for the 14 TeV and high pileup case. We also cut on 
the $p_T$ asymmetry between the lepton and the leading $b$ jet
\begin{equation}
p_T\,\,{\rm asymmetry} =\frac{P^{\rm b-jet}_T-p_T^{\rm lep} }{ p^{\rm b-jet}_T+p_T^{\rm lep}}.
\end{equation}
In Table 
\ref{tab:single_lepton_result_50PU} both methods are compared for the case of 50 pileup events. Topness in general 
performances well. Typically, it 
has a higher selection efficiency compared to $M_{\rm T2}^{\rm W}$ and improves the discovery potential (section 
\ref{sec:project}) in most cases.

\begin{sidewaystable}[htdp]
\caption{
Cutflow: number of events for the inclusive signal samples and several important backgrounds for the 
single-lepton stop analyses with 300~fb$^{-1}$ at 14~TeV without pileup. The last two lines show the 
significances %without and 
with an additional systematic background uncertainty of 25\% and, as an 
optimistic scenario, of 15\%.\label{tab:single_lepton_result_noPU}
}
\begin{center}
\begin{tabular}{|c||c|c|c|c||c||c|c|c|c|}
\hline

Description   &diboson  &  ttbar+jets   &boson+jets& single top&sum bgrds  &    STC4  &     STC5  &    STC6  &     STC8 \\ \hline \hline
  
preselection                       &  110817000 &  215894000 & 16840400000 &   62062700 & 17229173700 &    3840000   &    1146000   &     759000   &     657000   \\ \hline
singl. lep. and $\tau$ veto       &    9910050 &   40839900 &  552664000 &    7768710 &  611182660 &     431923   &     112517   &      63185   &      46360   \\ \hline
n jets $\ge 4$                    &     253994 &   10916500 &    4208940 &     160321 &   15539755 &     166663   &      33967   &      12546   &       4136   \\ \hline
b-tagged jets = 1 or 2             &      91026 &    8727330 &    1225930 &     131791 &   10176077 &     134556   &      26899   &       9141   &       2411   \\ \hline
$E_T^{\rm miss}>500$~GeV          &        383 &       6552 &       2784 &        130 &       9852 &       2246   &       1282   &        903   &        423   \\ \hline
$\Delta \Phi >0.5$               &        358 &       5779 &       2612 &        111 &       8861 &       1682   &       1054   &        783   &        375   \\ \hline
$H_T>1500$~GeV                     &         49 &        622 &        366 &         12 &       1051 &        460   &        282   &        214   &        147   \\ \hline
$M_T>120$~GeV                      &          6 &        108 &         21 &          1 &        137 &        273   &        176   &        141   &        101   \\ \hline \hline
%$M_{\rm T2}^{\rm W}>250$~GeV     &          4 &         52 &         17 &          0 &         76 &        120   &        103   &        102   &         83   \\ \hline
%$s/\sqrt{b+(0.25*b)^2}$           &      &      &      &      &      &        5.7 &        5.0 &        4.9 &        4.0 \\ \hline
%$s/\sqrt{b+(0.15*b)^2}$           &      &      &      &      &      &        8.4 &        7.2 &        7.1 &        5.8 \\ \hline \hline
%topness $> 9.5$ and b-asym.$<-0.2$  
\multicolumn{1}{|p{3cm}||}{\centering topness $> 9.5$ and $p_T$ asym.$<-0.2$  }
&          4 &         61 &         12 &          1 &         79 &        168   &        124   &        111   &         85   \\ \hline 
$s/\sqrt{b+(0.25*b)^2}$           &      &      &      &      &      &        7.7 &        5.7 &        5.1 &        3.9 \\ \hline
$s/\sqrt{b+(0.15*b)^2}$           &      &      &      &      &      &       11.3 &        8.3 &        7.5 &        5.7 \\ \hline

\end{tabular}
\end{center}

\vspace{1.0cm}

\caption{
Cutflow: number of events for the inclusive signal samples and several important backgrounds for the 
single-lepton stop analyses with 300~fb$^{-1}$ at 14~TeV and with 50 pileup events. The last two lines show 
the significances %without and 
with an additional systematic background uncertainty of 25\% and, as an 
optimistic scenario, of 15\%. 
Also shown is an example for the differences between the {\it topness} based selection
and the $M_{\rm T2}^{\rm W}$ approach.
}
\begin{center}
\begin{tabular}{|c||c|c|c|c||c||c|c|c|c|}
\hline
Description   &diboson  &  ttbar+jets   &boson+jets& single top&sum bgrds  &    STC4  &     STC5  &    STC6  &     STC8 \\ \hline \hline
  
preselection                       &  110822000 &  216124000 & 16842600000 &   62086200 & 17231632200 &    3840000   &    1146000   &     759000   &     657000   \\ \hline
singl. lep. and $\tau$ veto       &    9652070 &   40031600 &  536493000 &    7465180 &  593641850 &     427551   &     110574   &      61849   &      45070   \\ \hline
n jets $\ge 4$                    &     277933 &   10210600 &    5009660 &     181852 &   15680045 &     155328   &      34542   &      13211   &       4300   \\ \hline
b-tagged jets = 1 or 2             &      96490 &    8161060 &    1388700 &     148269 &    9794519 &     122321   &      26681   &       9285   &       2416   \\ \hline
$E_T^{\rm miss}>500$~GeV          &        384 &       6776 &       2920 &        145 &      10227 &       2434   &       1286   &        855   &        414   \\ \hline
$\Delta \Phi >0.5$               &        356 &       5968 &       2733 &        123 &       9181 &       1725   &       1057   &        751   &        370   \\ \hline
$H_T>1500$~GeV                     &         49 &        640 &        374 &         12 &       1076 &        492   &        277   &        209   &        129   \\ \hline
$M_T>120$~GeV                      &          6 &        110 &         25 &          1 &        143 &        269   &        162   &        140   &         85   \\ \hline \hline
$M_{\rm T2}^{\rm W}>250$~GeV     &          4 &         52 &         19 &          0 &         77 &        116   &         79   &        102   &         62   \\ \hline
$s/\sqrt{b+(0.25*b)^2}$           &      &      &      &      &      &        5.5 &        3.7 &        4.8 &        2.9 \\ \hline
$s/\sqrt{b+(0.15*b)^2}$           &      &      &      &      &      &        8.0 &        5.5 &        7.0 &        4.3 \\ \hline \hline
%topness $> 9.5$ and b-asym.$<-0.2$  
\multicolumn{1}{|p{3cm}||}{\centering topness $> 9.5$ and $p_T$ asym.$<-0.2$  }
&          4 &         62 &         15 &          1 &         83 &        155   &         98   &        107   &         67   \\ \hline 
$s/\sqrt{b+(0.25*b)^2}$           &      &      &      &      &      &        6.9 &        4.3 &        4.7 &        3.0 \\ \hline
$s/\sqrt{b+(0.15*b)^2}$           &      &      &      &      &      &       10.1 &        6.4 &        6.9 &        4.3 \\ \hline

\end{tabular}
\end{center}
\label{tab:single_lepton_result_50PU}
\end{sidewaystable}%

All the above kinematic variables are sensitive to pileup but the sensitivity can be mitigated by 
adjusting the jet and lepton $p_T$ cuts. The high momentum lepton from the stop decay provides an 
additional handle to select stop decays in a high pileup environment. After all selection requirements, 
the analysis is sensitive to all four models with 300~fb$^{-1}$. While this analysis, 
containing one lepton which is expected to originate from a top decay, is especially sensitive to decays 
containing top quarks, the full-hadronic analysis is also sensitive to decays containing only bottom 
decays. By comparison of these two analyses it will be possible to draw some conclusion on whether a stop 
or sbottom decay is observed in data.

%%%%%%%%%%%%%%%%%%%%%%%%%%%%%%%%%%%%%%%%%%

\subsection{Search for decays of the electroweak subprocess pp $\rightarrow \twpm_1 \tz_2$}\label{subsec:LHCewkino}
 
Once stop production is discovered, it needs to be clarified whether the stops are accompanied by sleptons 
and/or electroweak bosinos, which could very well hide at lower masses. Therefore, we study here the 
possibility to explore the electroweak spectrum of the investigated models. Current 
analyses~\cite{CMS-SUS-12-022, ATLAS-CONF-2013-028} do not have the power to see any of the four studied 
models, as the processes of interest cannot be excluded by simplified models that assume a 100\% branching 
ratio of $\twpm_1 \rightarrow \tilde{\tau} \nu_{\tau}$ and of $\tz_2 \rightarrow \tilde{\tau} \tau$, where 
the $\tilde{\tau}$ mass is defined as $m_{\tilde{\tau}}=0.5m_{\twpm_1} +0.5m_{\tz_1}$. Models where the 
$\tilde{\tau}$ mass is closer to the $\tz_1$ are in general more difficult, as the final objects are 
softer.

\begin{figure}[htbp]
\begin{center}
  \subfigure[]{ \includegraphics[width=0.45\textwidth]{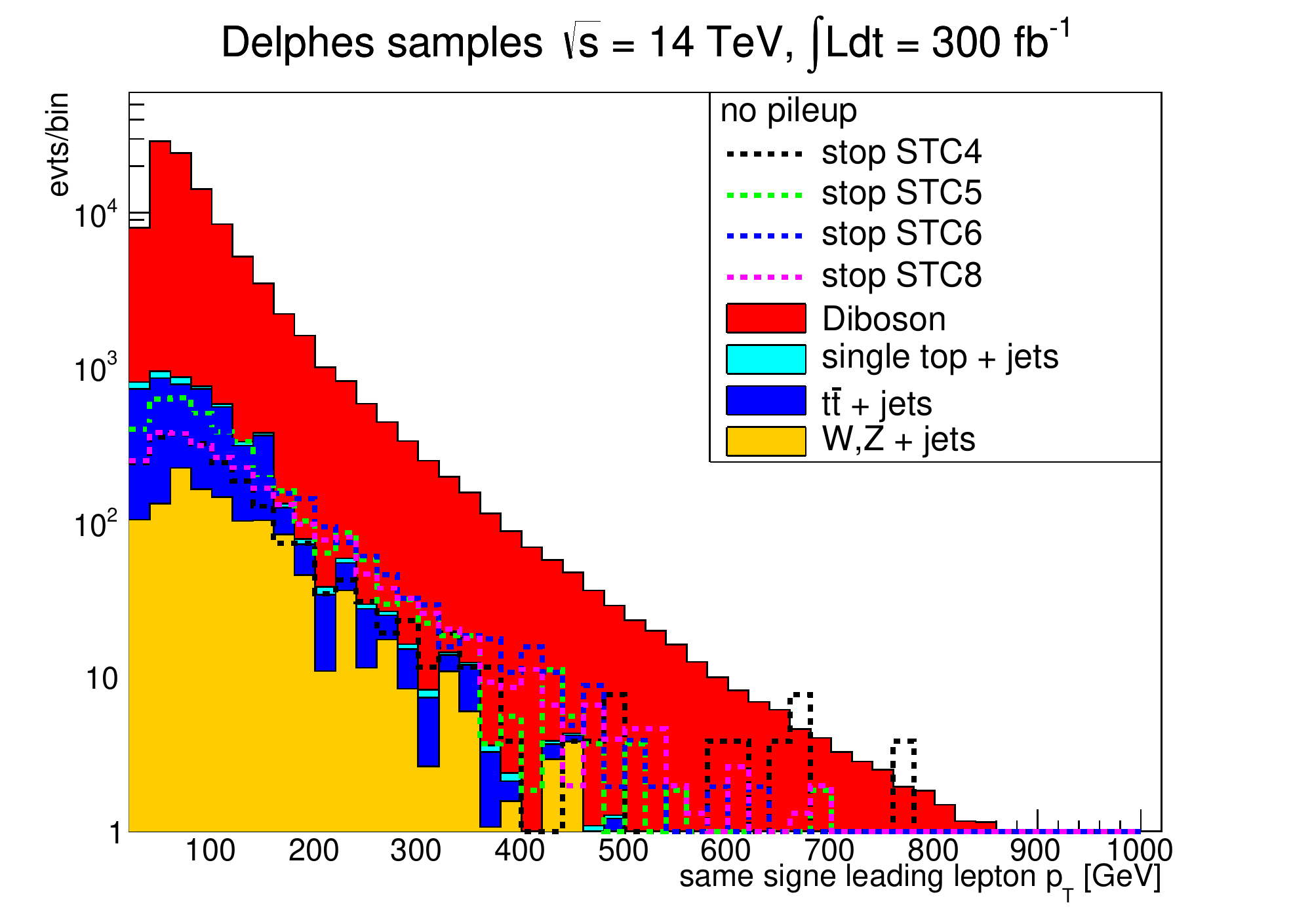} }
\caption{
The lepton transverse momentum of the leading lepton of the same-sign pair in the electroweakino analysis. 
The full histograms describing the backgrounds are stacked, and the four inclusive signal models are shown 
as dotted lines (not stacked).
}
\label{fig:ewkino}
\end{center}
\end{figure}

We are here investigating a final state containing two same-sign leptons, where only electrons and muons 
are taken into account. These are expected in case of leptonic $\tau$ decays of the above particles, where 
one lepton of the $\tz_2$ decay is lost, and the other one has the same charge as the lepton from the 
$\twpm_1$ decay. The lepton is softer than in the previous analysis, as can be seen from 
Fig.~\ref{fig:ewkino}, which displays the transverse momentum of the leading lepton of the same-sign pair.

\begin{sidewaystable}[htdp]
\caption{
Cutflow: number of events for the inclusive signal samples and several important backgrounds for the 
same-sign di-lepton analysis targeting electroweak particles with $300\,$fb$^{-1}$ at $14\,$TeV and without
pileup. The last two lines show the significances 
%without and 
with an additional systematic 
background uncertainty of 30\% and, as an optimistic scenario, of 20\%.
  }
\begin{center}
\begin{tabular}{|c||c|c|c|c||c||c|c|c|c|}
\hline
Description   &diboson  &  ttbar+jets   &boson+jets& single top&sum bgrds  &    STC4  &     STC5  &    STC6  &     STC8 \\ \hline \hline
  
preselection              &  110817000 &  215894000 & 16840400000 &   62062700 & 17229173700 &    3840000   &    1146000   &     759000   &     657000   \\ \hline
2 lepton req.             &    2001300 &    6362250 &   53165400 &       7180 &   61536130 &      72024   &      38810   &      30136   &      23439   \\ \hline
$E_T^{miss} >$ 120~GeV    &     111344 &    1005900 &     686304 &        872 &    1804420 &      35841   &      21353   &      16059   &      11472   \\ \hline
same-sign req.            &       7126 &        741 &        536 &        105 &       8510 &       1264   &       2431   &       2254   &       1625   \\ \hline
Z veto                    &       3021 &        731 &        390 &        104 &       4247 &       1015   &       1975   &       1801   &       1323   \\ \hline
b-jet veto                &       2640 &         98 &        304 &         46 &       3091 &        729   &        957   &        851   &        730   \\ \hline 
$E_T^{miss} >$ 200~GeV    &        738 &         43 &         89 &         12 &        883 &        408   &        512   &        474   &        411   \\ \hline\hline
$E_T^{miss} >$ 400~GeV    &         88 &          4 &         17 &          0 &        110 &         86   &        116   &        131   &        115   \\ \hline
$s/\sqrt{b+(0.3*b)^2}$    &      &      &      &      &      &        2.5 &        3.3 &        3.8 &        3.3 \\ \hline
$s/\sqrt{b+(0.2*b)^2}$    &      &      &      &      &      &        3.5 &        4.7 &        5.3 &        4.7 \\ \hline

\end{tabular}
\end{center}

\label{tab:ewkino_noPU}
\vspace{1.0cm}
\caption{
Cutflow: number of events for the inclusive signal samples and several important backgrounds for the 
same-sign di-lepton analysis targeting electroweak particles with $300\,$fb$^{-1}$ at $14\,$TeV and with 50
pileup events. The last two lines show the significances 
%without and 
with an additional systematic 
background uncertainty of 30\% and, as an optimistic scenario, of 20\%.
  }
\begin{center}
\begin{tabular}{|c||c|c|c|c||c||c|c|c|c|}
\hline
Description   &diboson  &  ttbar+jets   &boson+jets& single top&sum bgrds  &    STC4  &     STC5  &    STC6  &     STC8 \\ \hline \hline
  
preselection              &  110822000 &  216124000 & 16842600000 &   62086200 & 17231632200 &    3840000   &    1146000   &     759000   &     657000   \\ \hline
2 lepton req.             &    1914290 &    6745420 &   50864800 &      66503 &   59591013 &      80459   &      40117   &      30936   &      23903   \\ \hline
$E_T^{miss} >$ 120~GeV    &     125330 &    1203360 &     883404 &       8527 &    2220621 &      39799   &      22116   &      16700   &      11604   \\ \hline
same-sign req.            &       7920 &       7424 &       2511 &        548 &      18405 &       1385   &       2431   &       2284   &       1659   \\ \hline
Z veto                    &       3546 &       7356 &       2115 &        546 &      13565 &       1121   &       1988   &       1829   &       1335   \\ \hline
b-jet veto                &       3071 &       2183 &       1172 &        201 &       6629 &        701   &        930   &        838   &        740   \\ \hline 
$E_T^{miss} >$ 200~GeV    &        783 &        413 &        414 &         46 &       1657 &        378   &        526   &        460   &        424   \\ \hline\hline
$E_T^{miss} >$ 400~GeV    &         90 &         21 &         66 &          3 &        182 &         91   &        109   &        114   &        101   \\ \hline
$s/\sqrt{b+(0.3*b)^2}$    &      &      &      &      &      &        1.6 &        1.9 &        2.0 &        1.8 \\ \hline
$s/\sqrt{b+(0.2*b)^2}$    &      &      &      &      &      &        2.4 &        2.8 &        2.9 &        2.6 \\ \hline

\end{tabular}
\end{center}

\label{tab:ewkino_50PU}
\end{sidewaystable}

The main backgrounds originate from (di-)vectorboson+jets events. Background from $t\bar{t}$ can rarely 
occur from dileptonic decays, where the charge of an electron is reconstructed wrongly (we did not study whether such effects are simulated properly in Delphes), or from semileptonic top decays, where a second 
lepton originates from the $b$ decay and is (in rare cases) isolated.
Typically, such a background can be reduced by tightening the isolation criterium, which is not possible 
for the Snowmass Delphes samples. Also $t\bar{t}$ production in association with a vectorboson can lead 
to a same-sign signature. All backgrounds containing $Z$ bosons are reduced by rejecting events that, 
after applying looser electron and muon selection criteria, contain an OSSF pair within 15 GeV of the $Z$ 
boson mass ('$Z$-veto').

Tables~\ref{tab:ewkino_noPU} and \ref{tab:ewkino_50PU} summarize the expected number of events for this 
analysis at the LHC with a center-of-mass energy of 14~TeV and an integrated luminosity of 300~fb$^{-1}$. 
After the requirement of two same-sign leptons, the SM background is strongly reduced, though this effect 
is reduced in the case with 50 pileup events. The number of events in the signal regions differ for the 
different models. The difference is reduced if one requires zero b-tags, but then the significance of the 
signal is reduced as well. 
%This leads to the conclusion that part of the observed signal is not coming 
%from the electroweak processes in this selection, but from some corners of the phase space of the direct 
%stop production. %DK: stop? then stc4 should be larger not smaller - some unexplored phase space of MC generation
%DK better keep the tag for consisitent results
The b-tag requirement renders a clean elektroweakino signal, as can be seen by the fact that final event yields for 
all model points are similar. 
The expected systematic uncertainty is worse for this analysis compared to the stop analyses in current 
data, therefore we assume a larger uncertainty also for 300 fb$^{-1}$.

%In summary, a signal is expected to be statistically visible, but a $5\sigma$ discovery requires an 
In summary, the signal is at the edge of visibility, and a $5\sigma$ discovery requires an 
understanding of the background to better than $5\%$. 
A deviation from the analysis design from \cite{CMS-SUS-12-022},
can improve the sensitivity for our signal points but probably only at the cost of a higher stop contamination.
Once a signal is observed, the selection 
requirements would have to be developed further, and also 3-lepton final states should be taken into 
account to enhance the significance. If more than the expected number of events are observed with this 
analysis, it would be a hint for additional electroweak production, which then could be determined further at the 
ILC. But based on the current studies, it could also be possible that the electroweakinos remain burried 
below the background and its systematics. This is an example where the ILC would serve as discovery 
machine, and with its precise measurements could help to narrow the LHC analyses such that a signal could 
also be extracted from the LHC data.

\subsection{LHC projections in view of luminosity and systematic uncertainties}\label{sec:project}

In this section we summarize the results of the three performed analyses (in the case of 50 pileup events) and compare the 
sensitivity (in terms of standard deviations $\sigma$) for exclusion or discovery of the STC models.
We use the observed number of events as test statistics, corresponding to a pure counting experiment.
A discovery with a certain significance can be claimed if the background-only hypothesis can be excluded
at this significance: 
\begin{equation}
\sigma_{{\boldmath{disc}}} = S / \sqrt{B + (\delta B_{{\boldmath{sys}}})^2 }
\end{equation}
Here, $S$ and $B$ are the respective numbers of events expected for signal and (SM) background at a certain integrated luminosity.
The significance for exclusion of the signal-plus-background hypothesis is defined analoguosly, but replacing $B$ by $S+B$ in the above
formula.

\begin{figure}[htbp]
\vspace*{-2cm}
\begin{center}
  \subfigure[0-lepton analysis, discovery sensitivity]{ \includegraphics[width=0.44\textwidth]{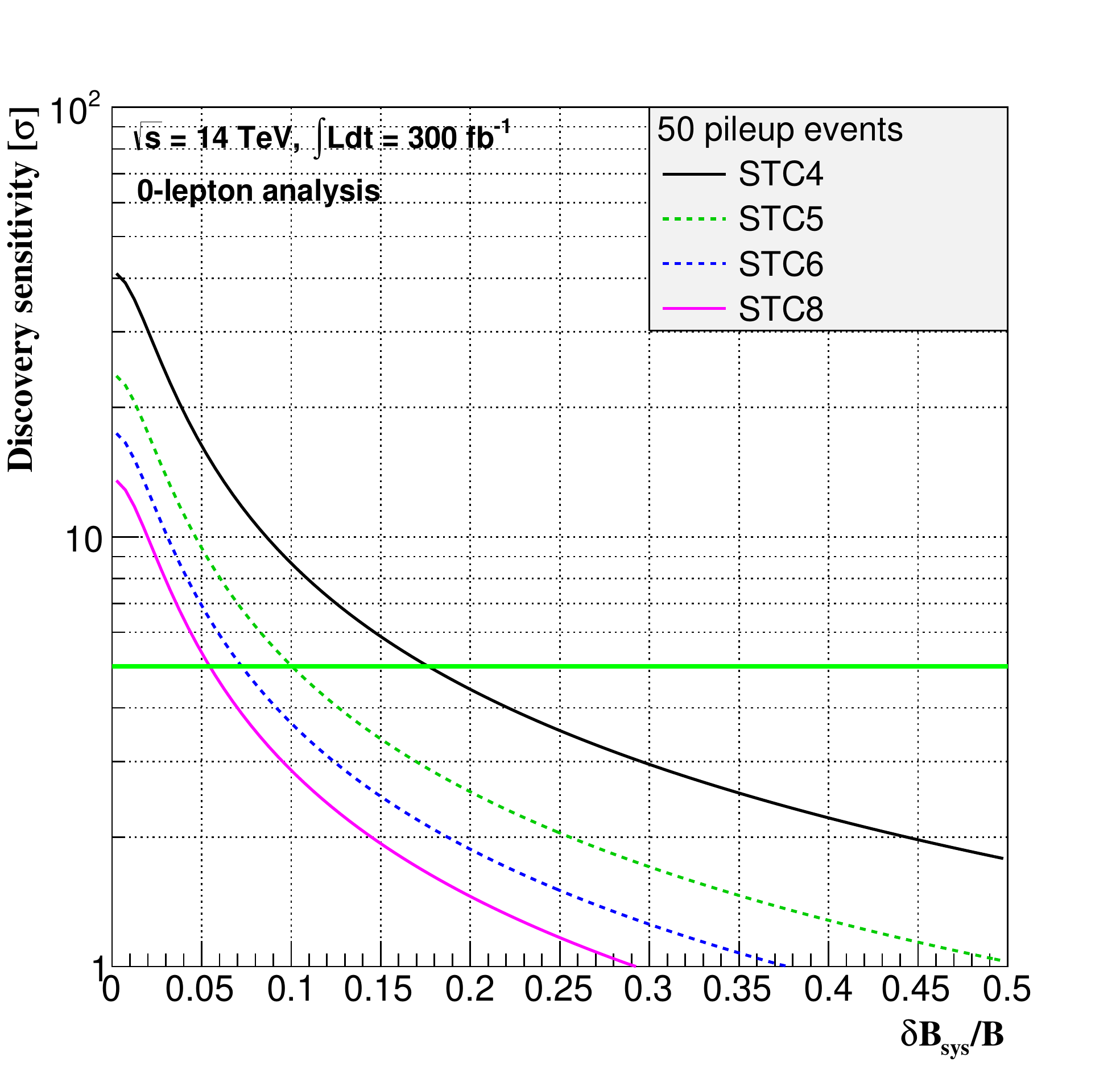} }
  \subfigure[0-lepton analysis, exclusion sensitivity]{ \includegraphics[width=0.44\textwidth]{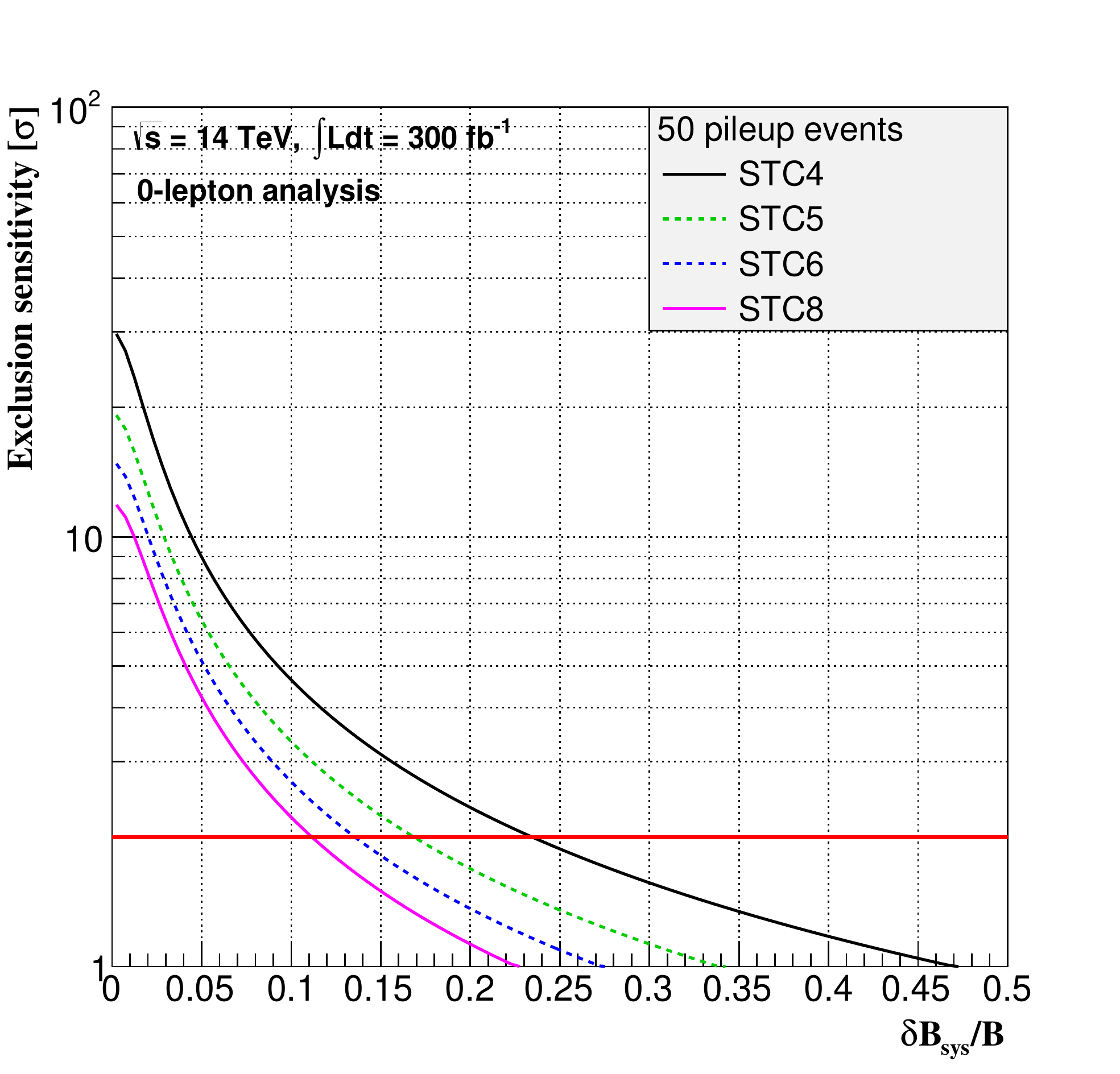} }
  \subfigure[1-lepton analysis, discovery sensitivity]{ \includegraphics[width=0.44\textwidth]{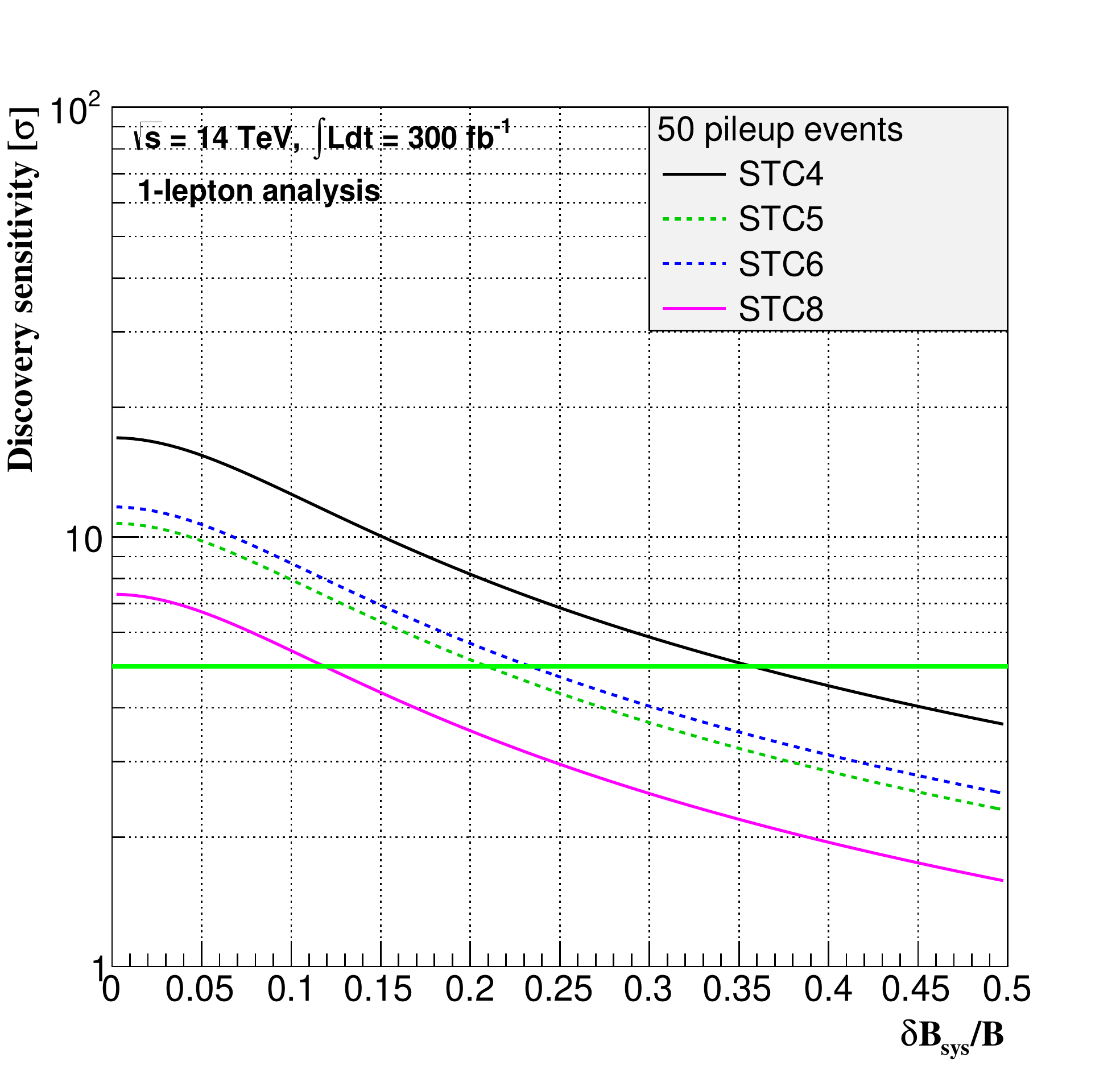} }
  \subfigure[1-lepton analysis, exclusion sensitivity]{ \includegraphics[width=0.44\textwidth]{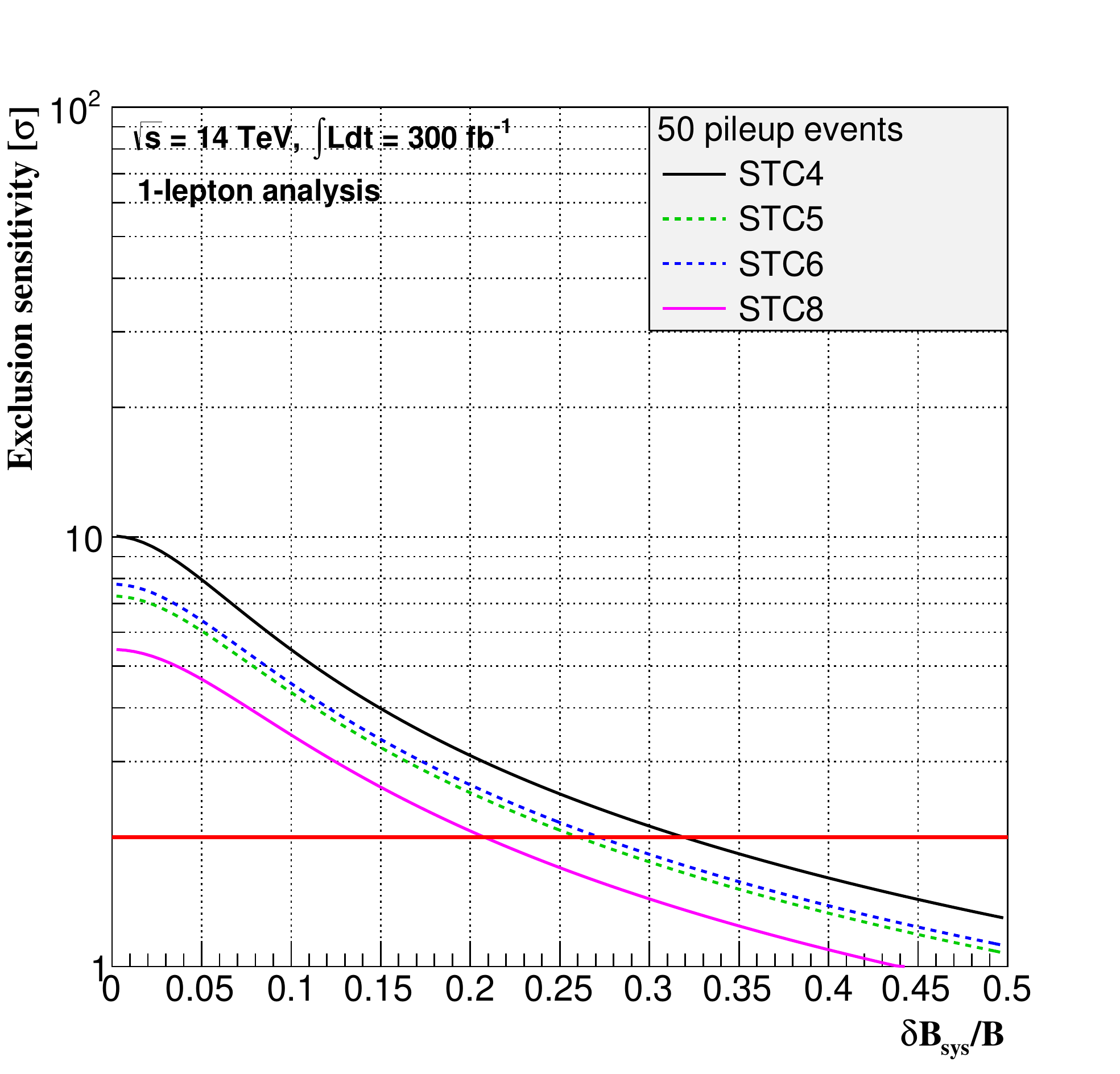} }
  \subfigure[2-lepton analysis, discovery sensitivity]{ \includegraphics[width=0.44\textwidth]{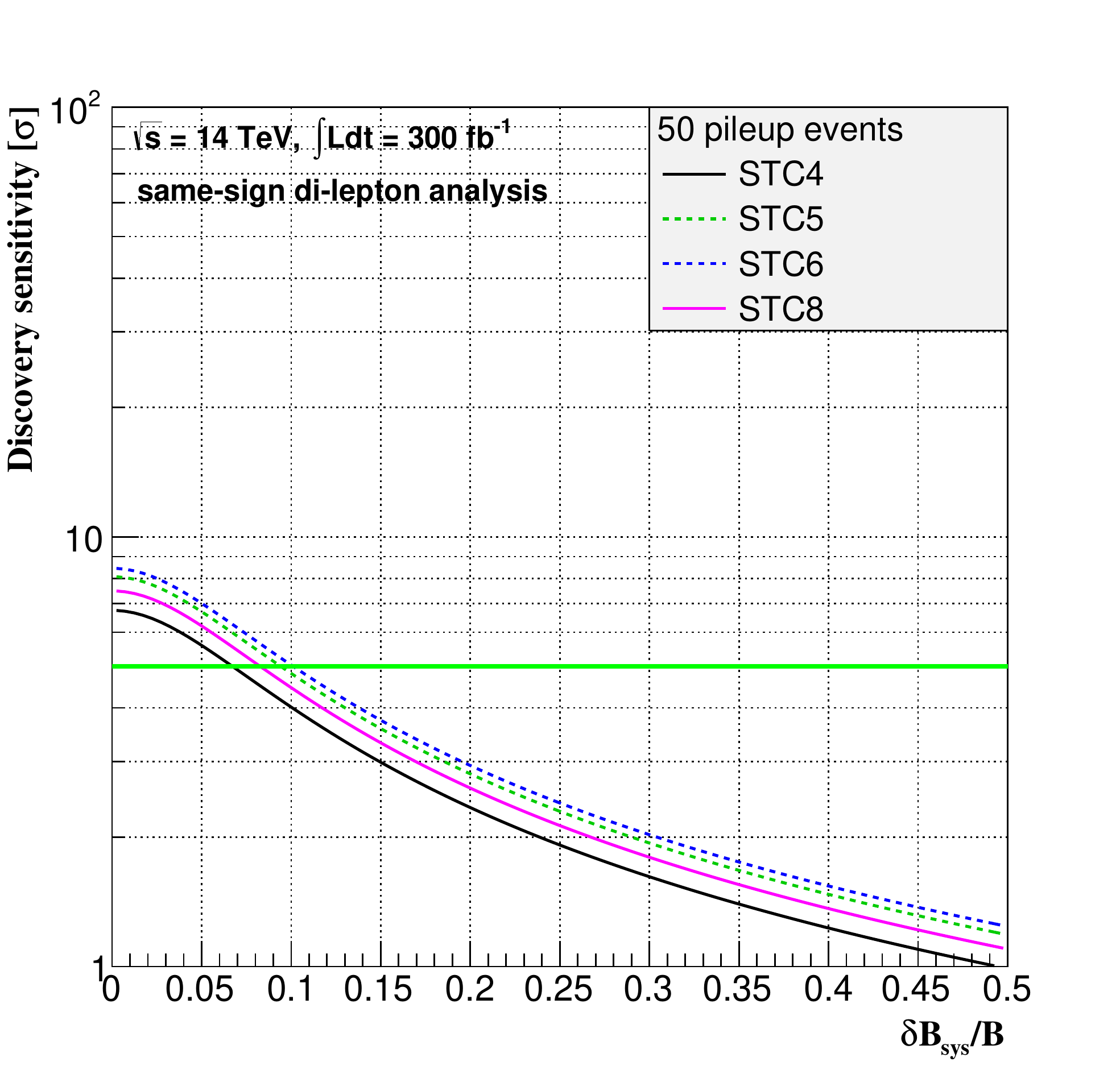} }
  \subfigure[2-lepton analysis, exclusion sensitivity]{ \includegraphics[width=0.44\textwidth]{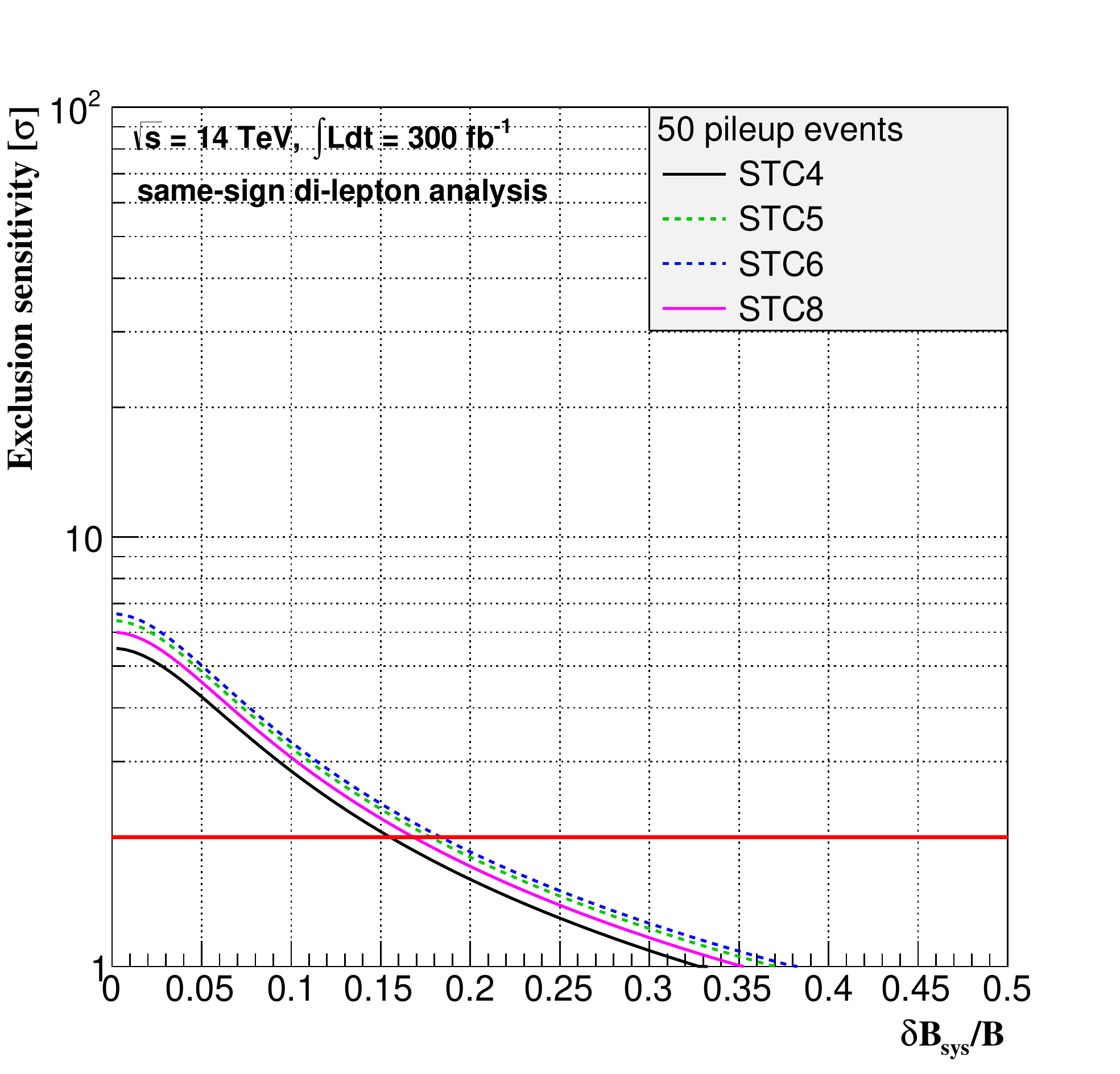} }
\caption{(a),(c),(e): Discovery significances for the 4 STC models by the 3 considered analyses as a function of the relative systematic uncertainty on the background.
The green horizontal line indicates the $5$-$\sigma$-level.
(b),(d),(f): Same for the exclusion sensitivity - the red horizontal line indicates the $2$-$\sigma$-level (ie. exclusion at $95\%$ CL).}
\label{fig:sens_lumi300}
\end{center}
\end{figure}

As already indicated in the cutflow tables, the assumptions on how well systematic uncertainties can be controled will be
decisive. Figure~\ref{fig:sens_lumi300} shows the discovery and exclusion sensitivities as a function of the relative systematic
uncertainty, based on an integrated luminosity fixed to $300\,$fb$^{-1}$. It leads to the following 
observations\footnote{The statistical uncertainty of these reults are dominated by the statistical uncertainty of our 
signal samples which is typically in the order of 10\%.}:
\begin{itemize}
\item Already at $300\,$fb$^{-1}$, all three searches are limited by systematics in all scenarios.
\item The 1-lepton analysis as the cleanest selection is the most robust one against systematic uncertainties, but also this
      search is systematically limited for uncertainties larger than $10\%$.
\item The 1-lepton analysis can discover the STC4 , STC5 and STC6 scenarios with at least $5\sigma$ if the 
systematics can be controlled better
      than $20\%$ to $35\%$. Discovery of STC8 needs precision better than $12\%$.
\item In case of the 0-lepton analysis, systematic uncertainties smaller than $18\%$ (STC4) to $5\%$ (STC8) are needed for a $5\sigma$ 
discovery. However in case of small systematics, potentially much larger sensitivities can be reached than with the 1-lepton analysis.      
\item The same-sign di-lepton analysis is most fragile with respect to systematic uncertainties. Exclusion or discovery requires a control
of the systematics to at least $10\%$ or $7\%$, respectively.
\item In case of the same-sign di-lepton, the sensitivity is very similar for all four scenarios, remaining differences reflect the
available MC statistics. This demonstrates that indeed the same-sign di-lepton analysis selects electroweakino production, which
is the same in all four scenarios, with rather little contamination by the strongly varying stop production.
\end{itemize}

\begin{figure}[htbp]
\vspace*{-2cm}
\begin{center}
%  \subfigure[0-lepton analysis, $20\%$ uncertainty]{ \includegraphics[width=0.48\textwidth]{fig/hadron_disc_sys20} }
%  \subfigure[0-lepton analysis, $10\%$ uncertainty]{ \includegraphics[width=0.48\textwidth]{fig/hadron_disc_sys10} }
%  \subfigure[1-lepton analysis, $20\%$ uncertainty]{ \includegraphics[width=0.48\textwidth]{fig/lepton_disc_sys20} }
%  \subfigure[1-lepton analysis, $10\%$ uncertainty]{ \includegraphics[width=0.48\textwidth]{fig/lepton_disc_sys10} }
%  \subfigure[2-lepton analysis, $20\%$ uncertainty]{ \includegraphics[width=0.48\textwidth]{fig/ewkino_disc_sys20} }
%  \subfigure[2-lepton analysis, $10\%$ uncertainty]{ \includegraphics[width=0.48\textwidth]{fig/ewkino_disc_sys10} }
  \subfigure[0-lepton analysis, $25\%$ uncertainty]{ \includegraphics[width=0.44\textwidth]{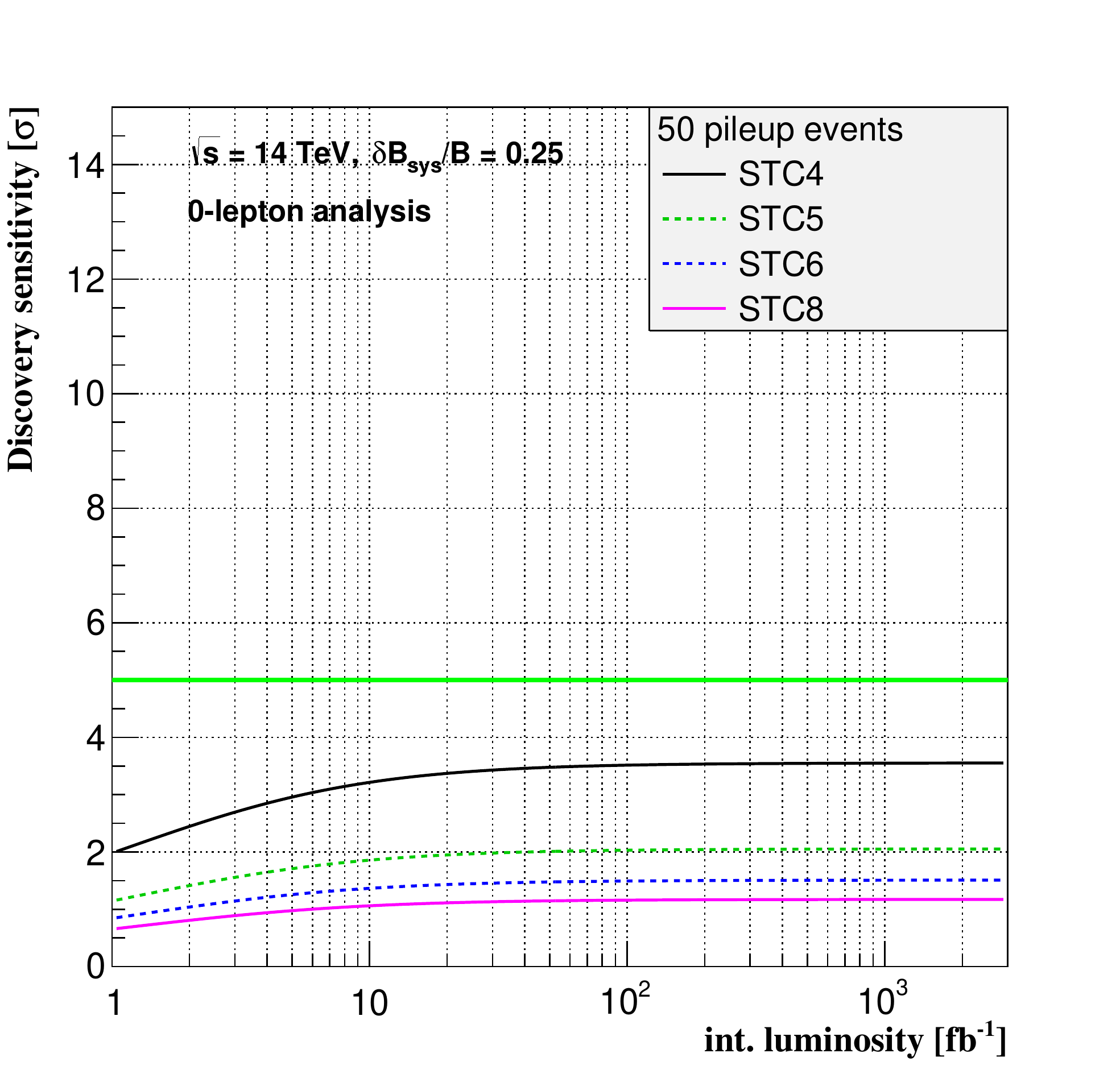} }
  \subfigure[0-lepton analysis, $15\%$ uncertainty]{ \includegraphics[width=0.44\textwidth]{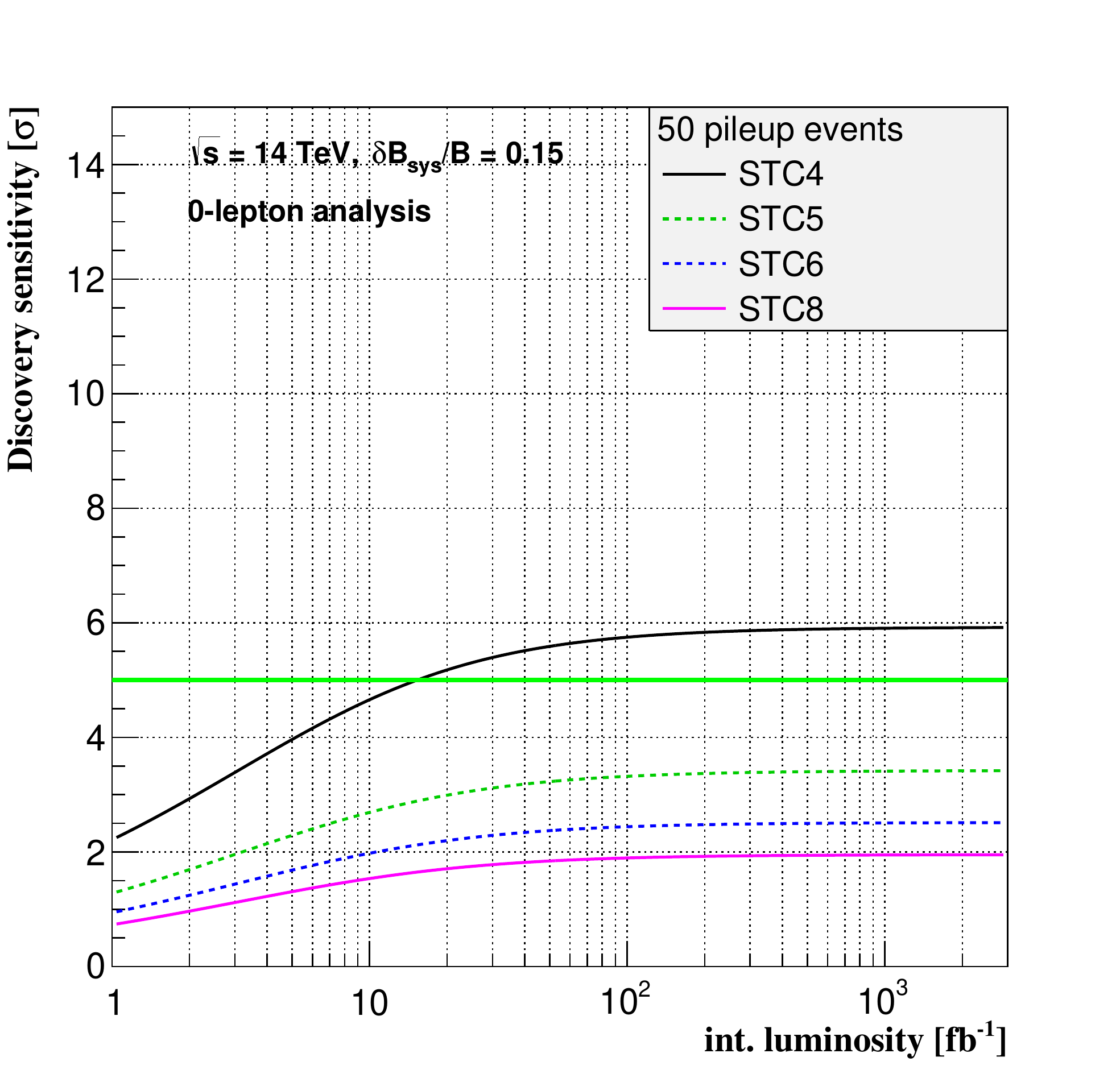} }
  \subfigure[1-lepton analysis, $25\%$ uncertainty]{ \includegraphics[width=0.44\textwidth]{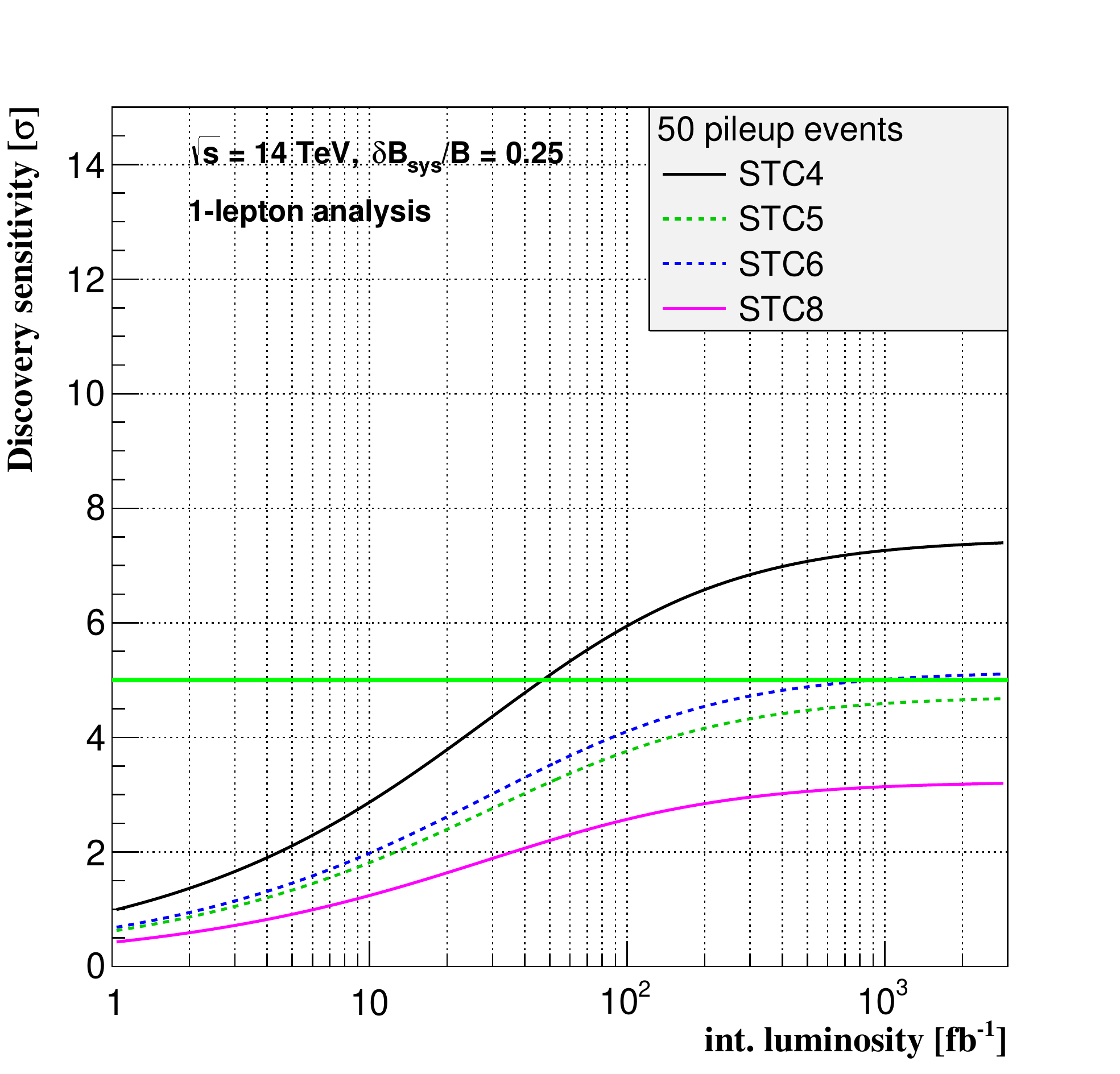} }
  \subfigure[1-lepton analysis, $15\%$ uncertainty]{ \includegraphics[width=0.44\textwidth]{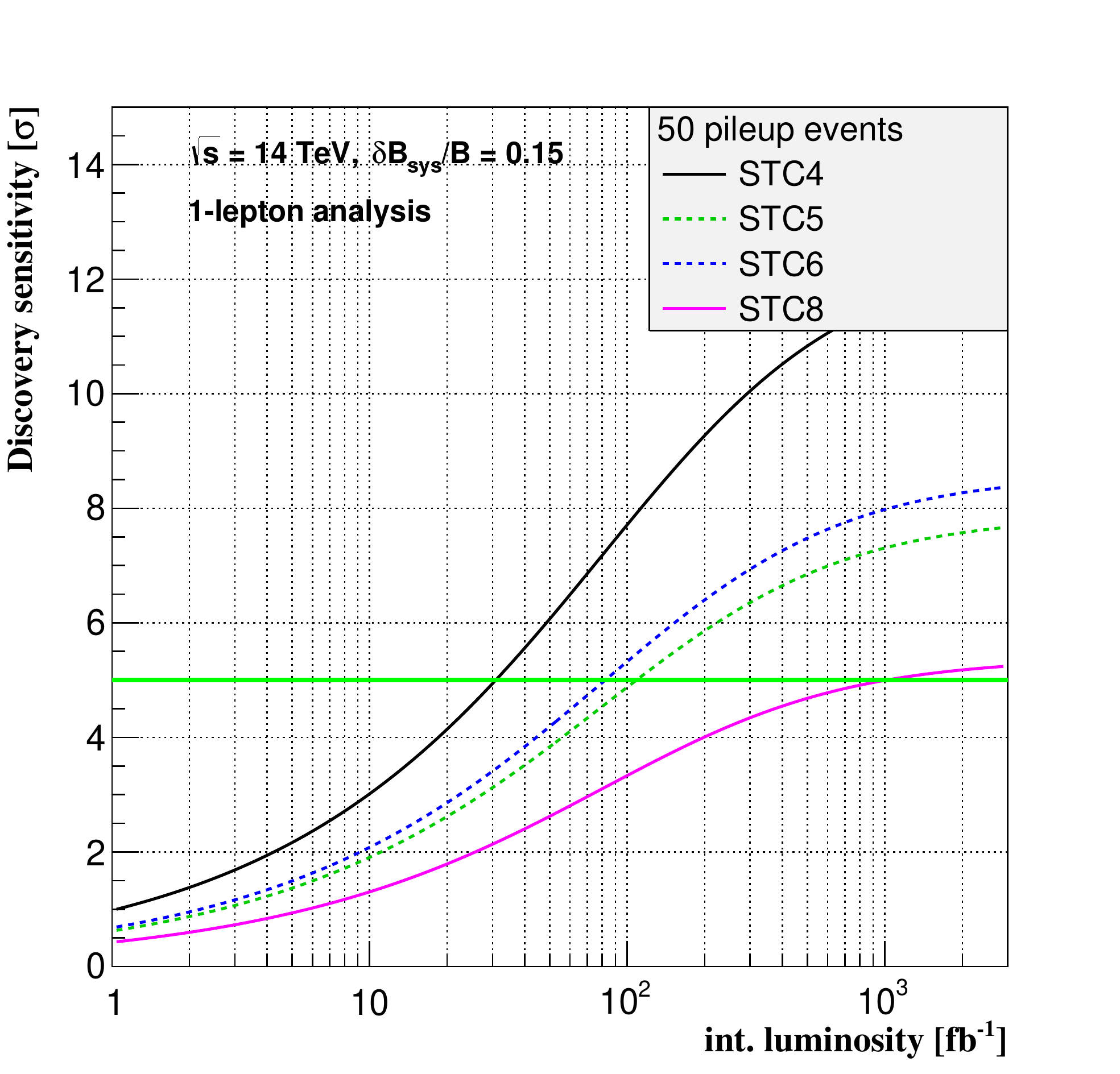} }
  \subfigure[2-lepton analysis, $30\%$ uncertainty]{ \includegraphics[width=0.44\textwidth]{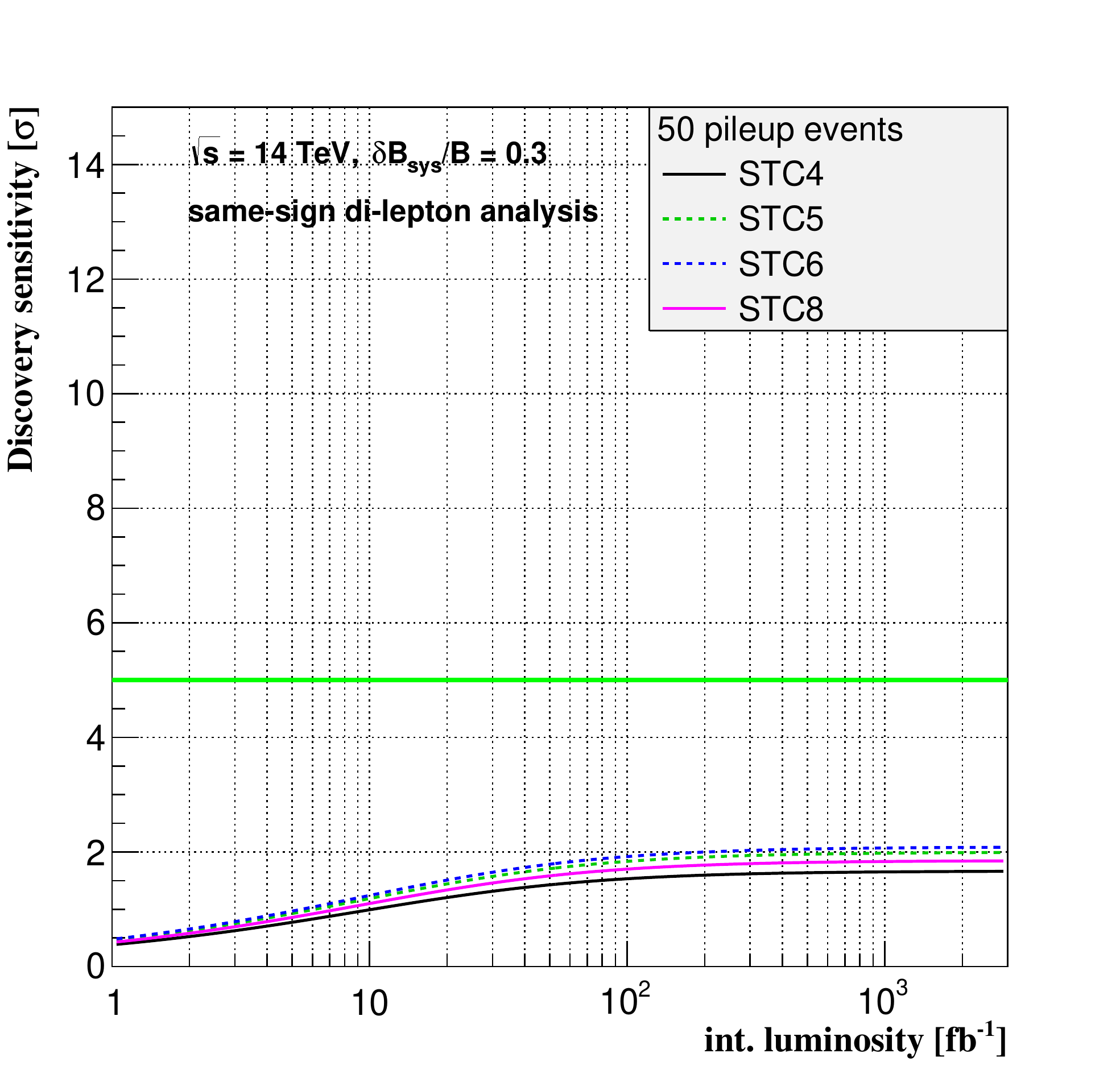} }
  \subfigure[2-lepton analysis, $20\%$ uncertainty]{ \includegraphics[width=0.44\textwidth]{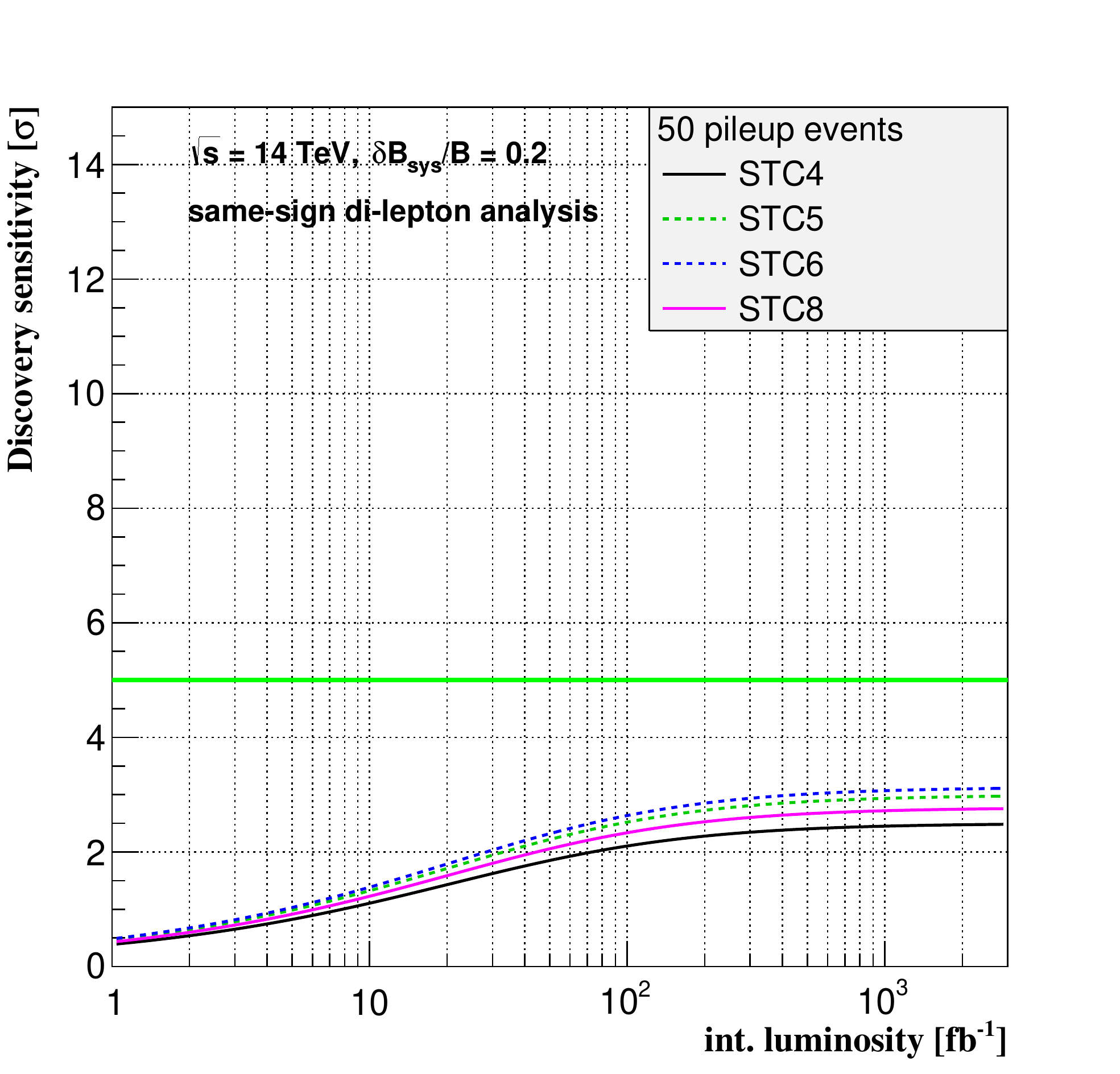} }
\caption{(a),(c),(e): Discovery significances for the 4 STC models by the 3 considered analyses as a function of the integrated luminosity 
assuming a relative systematic uncertainty of $25\%(30\%)$ on the background.
(b),(d),(f): Same but assuming a reduced  systematic uncertainty of $15\%(20\%)$. The green horizontal line indicates the $5$-$\sigma$-level.}
\label{fig:sens_disc}
\end{center}
\end{figure}

\begin{figure}[htbp]
\vspace*{-2cm}
\begin{center}
%  \subfigure[0-lepton analysis, $20\%$ uncertainty]{ \includegraphics[width=0.48\textwidth]{fig/hadron_excl_sys20} }
%  \subfigure[0-lepton analysis, $10\%$ uncertainty]{ \includegraphics[width=0.48\textwidth]{fig/hadron_excl_sys10} }
%  \subfigure[1-lepton analysis, $20\%$ uncertainty]{ \includegraphics[width=0.48\textwidth]{fig/lepton_excl_sys20} }
%  \subfigure[1-lepton analysis, $10\%$ uncertainty]{ \includegraphics[width=0.48\textwidth]{fig/lepton_excl_sys10} }
%  \subfigure[2-lepton analysis, $20\%$ uncertainty]{ \includegraphics[width=0.48\textwidth]{fig/ewkino_excl_sys20} }
%  \subfigure[2-lepton analysis, $10\%$ uncertainty]{ \includegraphics[width=0.48\textwidth]{fig/ewkino_excl_sys10} }
  \subfigure[0-lepton analysis, $25\%$ uncertainty]{ \includegraphics[width=0.44\textwidth]{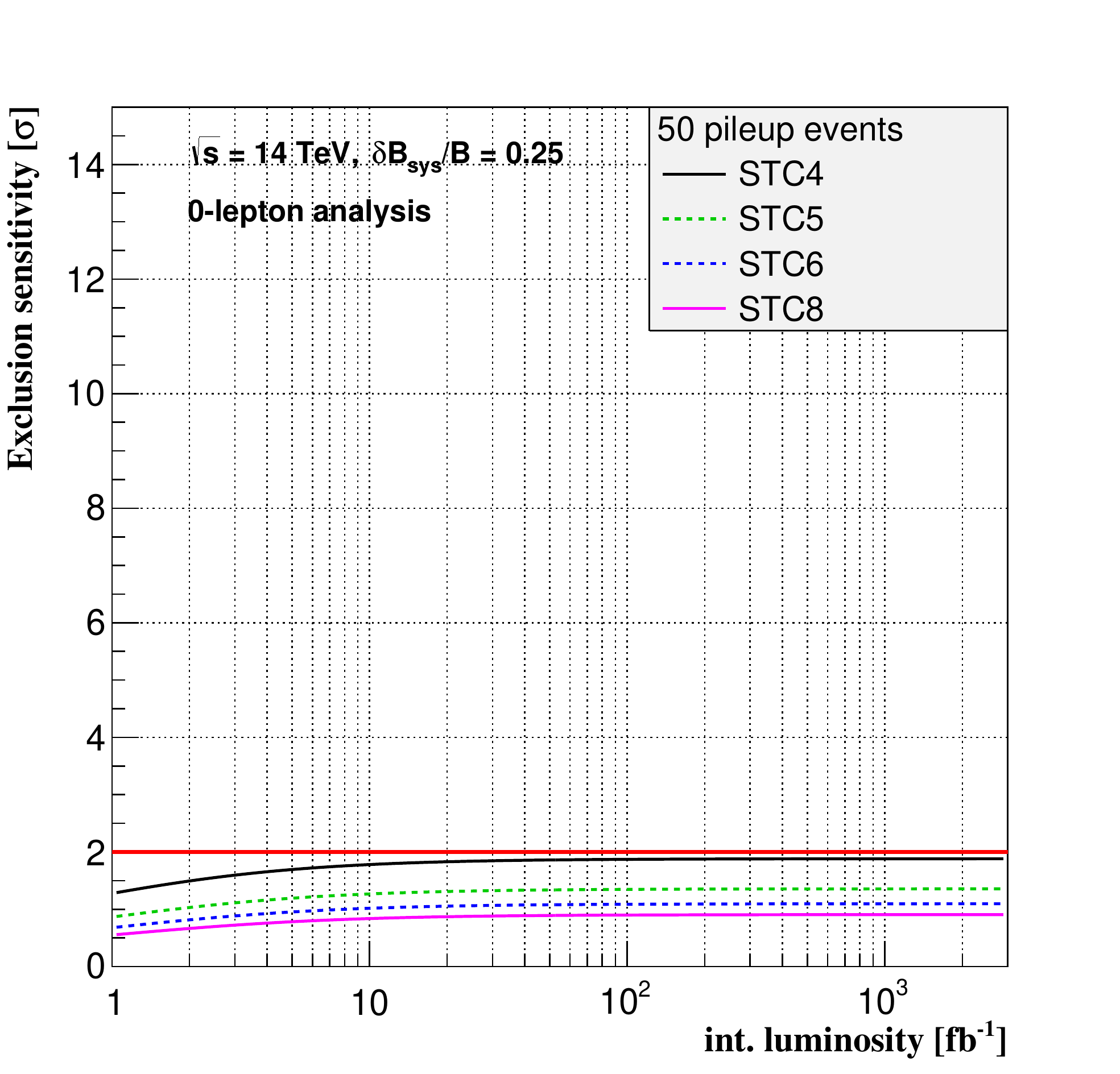} }
  \subfigure[0-lepton analysis, $15\%$ uncertainty]{ \includegraphics[width=0.44\textwidth]{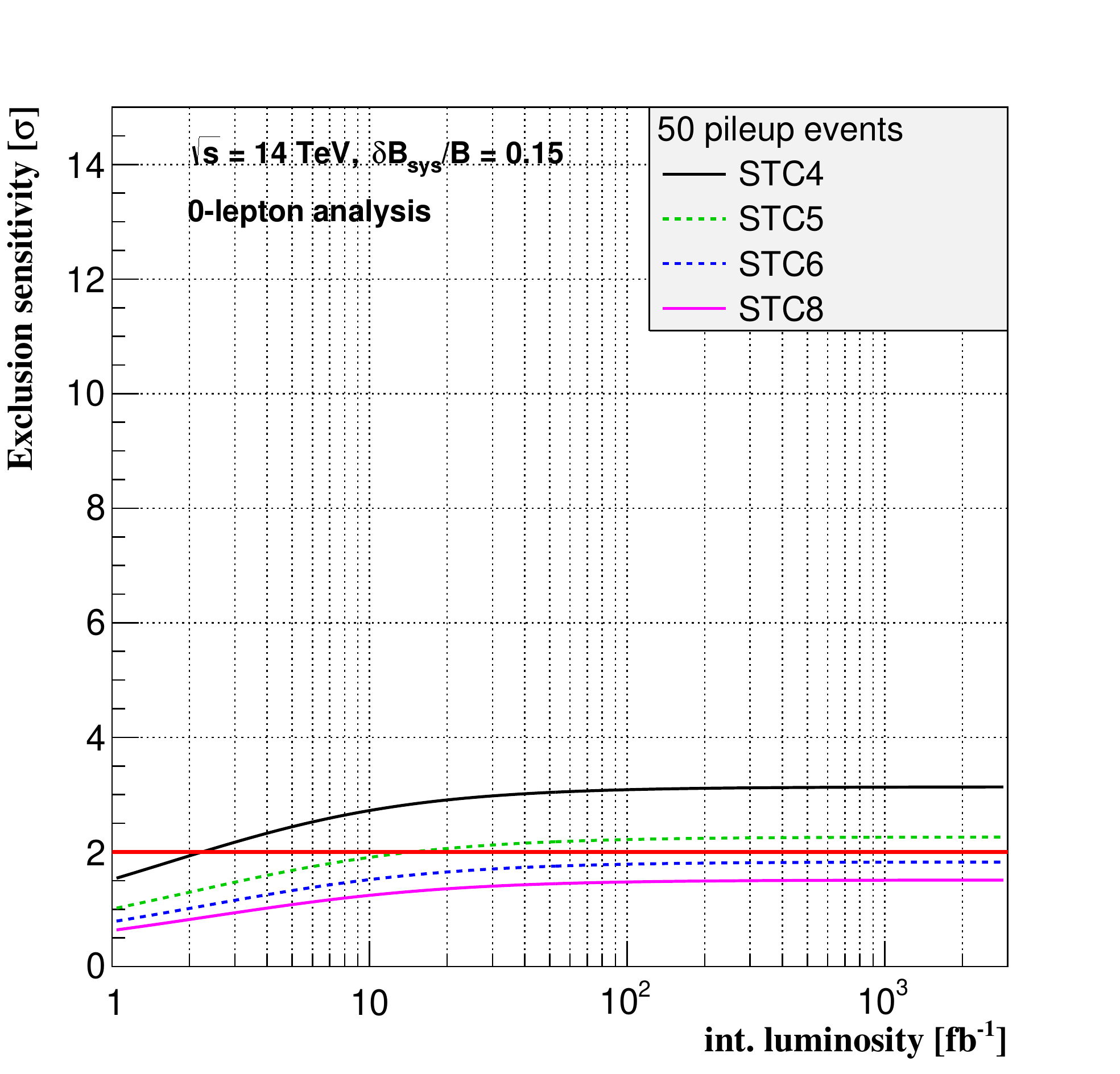} }
  \subfigure[1-lepton analysis, $25\%$ uncertainty]{ \includegraphics[width=0.44\textwidth]{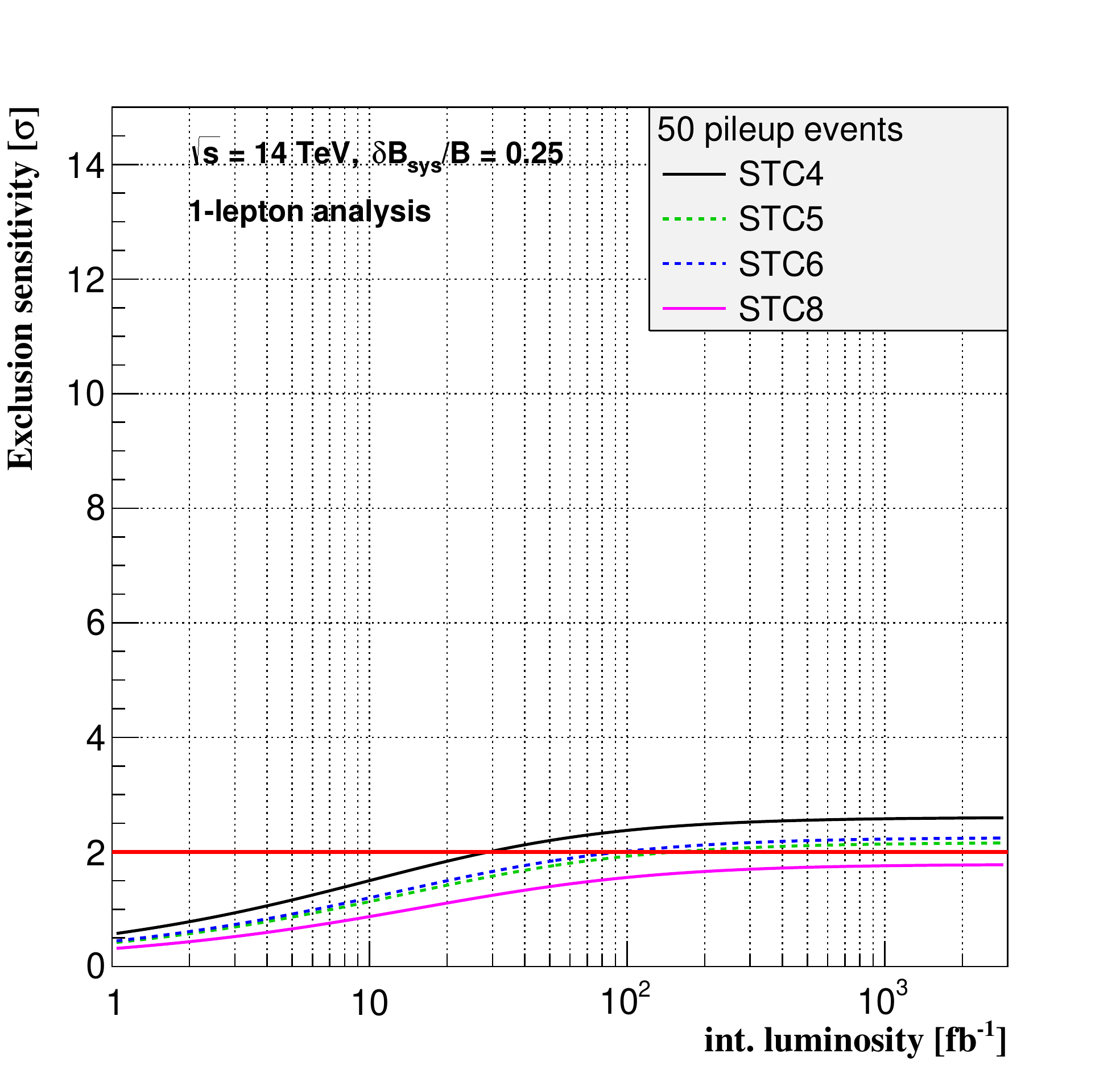} }
  \subfigure[1-lepton analysis, $15\%$ uncertainty]{ \includegraphics[width=0.44\textwidth]{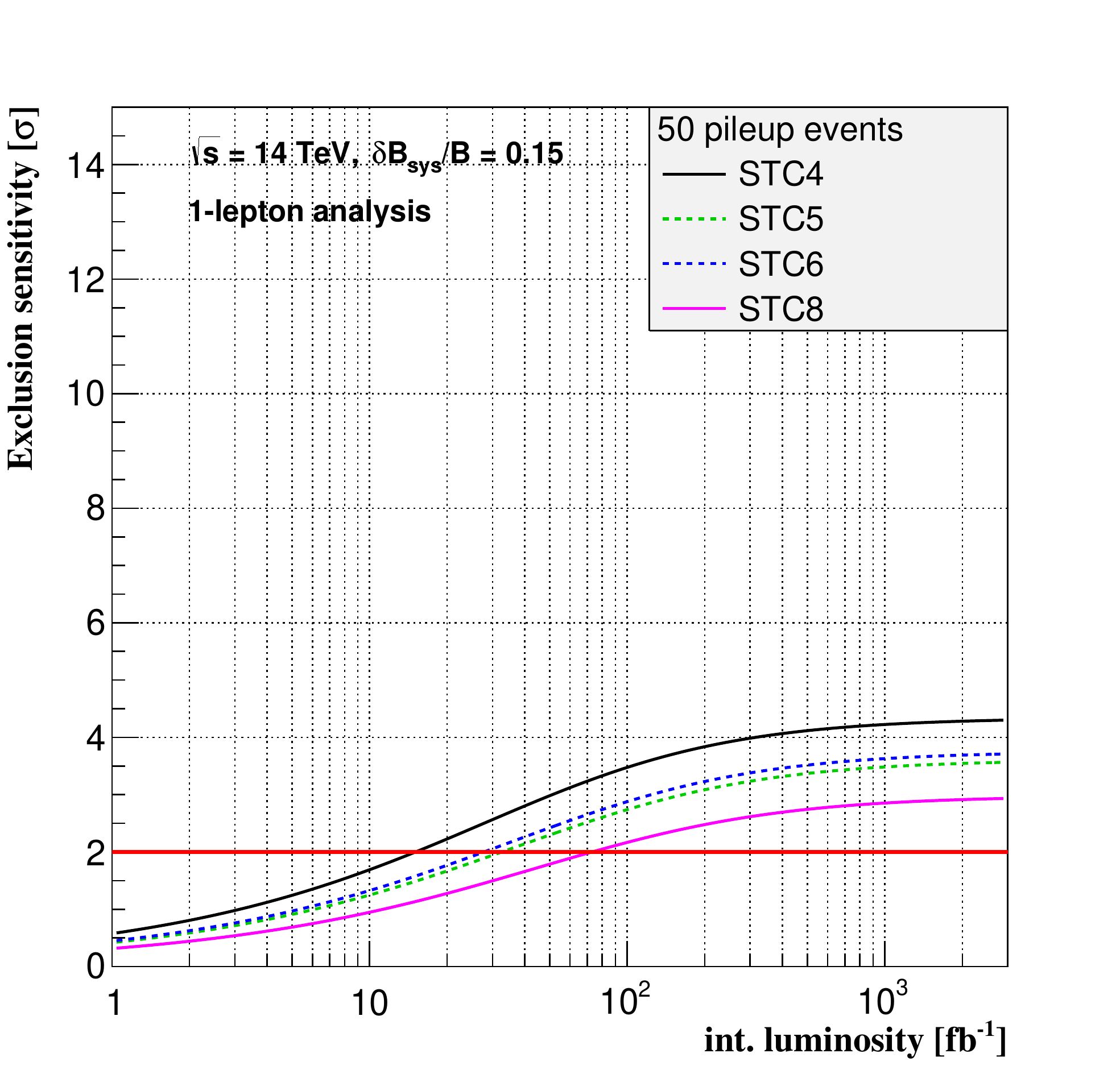} }
  \subfigure[2-lepton analysis, $30\%$ uncertainty]{ \includegraphics[width=0.44\textwidth]{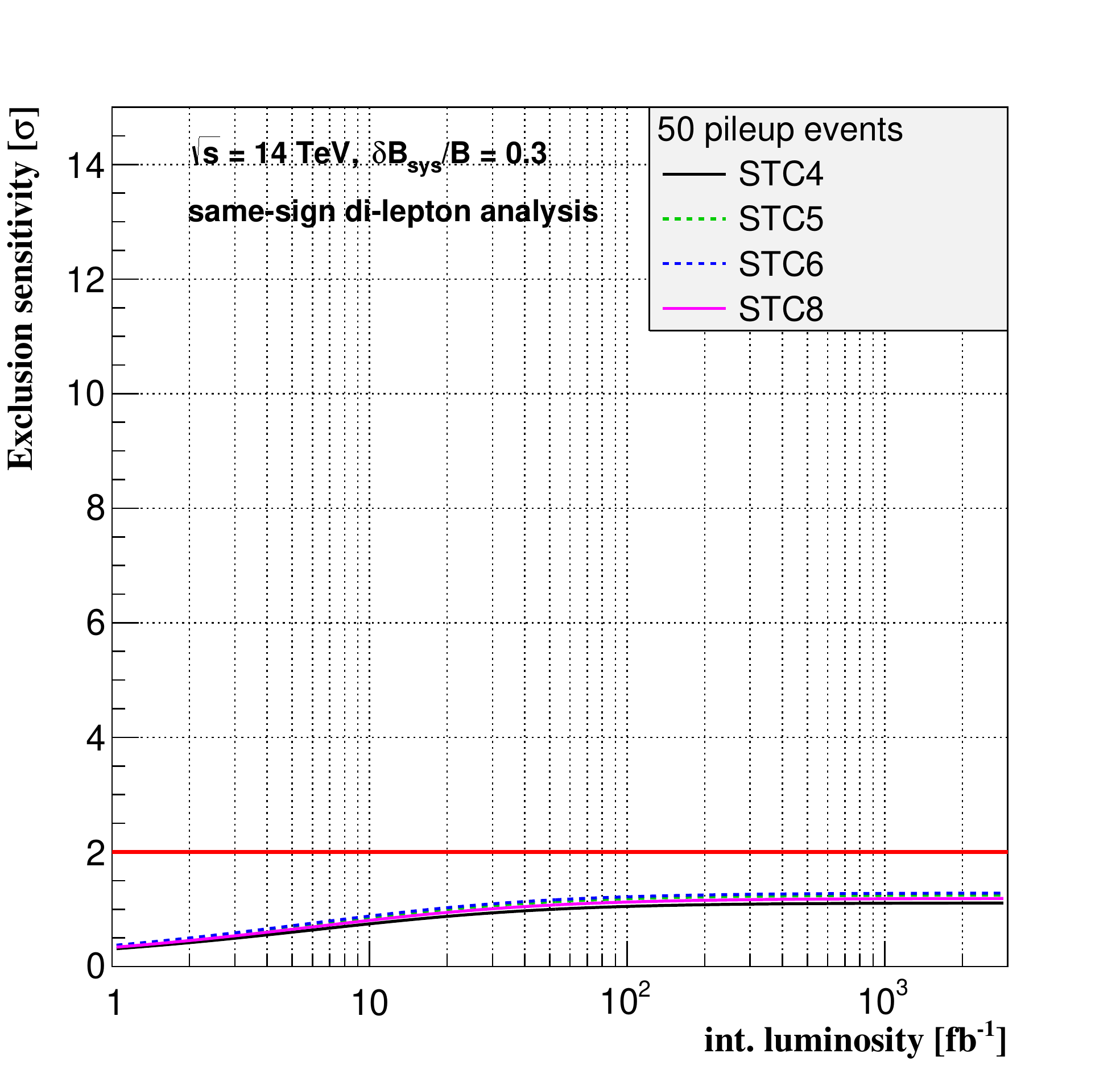} }
  \subfigure[2-lepton analysis, $20\%$ uncertainty]{ \includegraphics[width=0.44\textwidth]{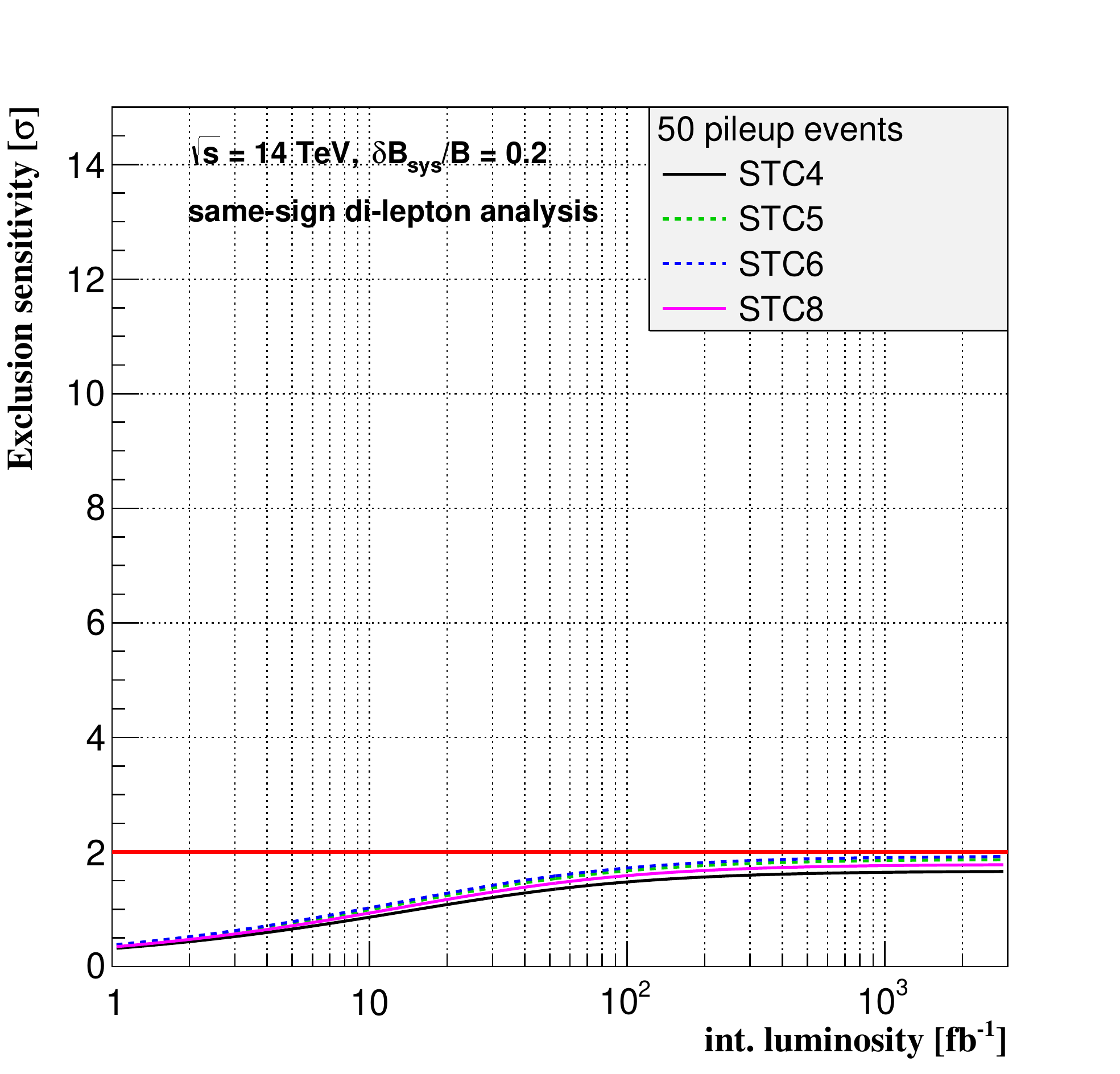} }
\caption{(a),(c),(e): Exclusion significances for the 4 STC models by the 3 considered analyses as a function of the integrated luminosity 
assuming a relative systematic uncertainty of $25\%(30\%)$ on the background.
(b),(d),(f): Same but assuming a reduced  systematic uncertainty of $15\%(20\%)$. The red horizontal line indicates the $2$-$\sigma$-level 
(ie. exclusion at $95\%$ CL). }
\label{fig:sens_excl}
\end{center}
\end{figure}

Figure~\ref{fig:sens_disc} shows the discovery sensitivity of all three analyses as a function of the integrated luminosity
for two different assumptions on the systematic uncertainties: $25\%$ or $15\%$ ($30\%$ or $20\%$ for the 2-lepton analysis). 
Sensitivity numbers for additional values of the systematic
uncertainty can be found in the cutflow tables. A thorough estimate of the achievable precision for each analysis is beyond the
scope of this study, so we leave it to the judgement of the reader which scenario to consider the most realistic. We draw the following
conclusions:
\begin{itemize}
\item In case of $25\%$ systematic uncertainty, the 1-lepton analysis will be the first -- and only -- analysis which allows 
a stand-alone $5\sigma$ discovery of STC4 after accumulating $45\,$fb$^{-1}$. STC5 and STC6 are at the edge of 
discovery.
\item In all other scenarios and analyses no $5\sigma$ discovery is possible assuming an uncertainty of $25\%$, even with $3000\,$fb$^{-1}$.
\item Assuming a modelling of all backgrounds at the level of $15\%$ is possible, the first discovery will move to the 0-lepton analysis in case
      of STC4, requiring not even $15\,$fb$^{-1}$ at $14\,$TeV!
\item Staying with $15\%$ systematic uncertainty, STC5 and STC6 would be first discovered in the 1-lepton search after accumulating $100\,$ fb$^{-1}$.
\item STC8  would first -- and only -- be discovered in the 1-lepton analysis with $1000\,$fb$^{-1}$, again assuming
a systematic uncertainty of $15\%$. 
\item No analysis would gain significantly from a luminosity increase to $3000\,$fb$^{-1}$. Only the 1-lepton analysis may improve, if the
systematic uncertainties can be controled beyond the $15\%$ level.
\end{itemize}

Figure~\ref{fig:sens_excl} finally gives the same set of plots for the case of exclusion sensitivity.
\begin{itemize}
\item In case of a systematic uncertainty of $15\%$, all models can be excluded by the 1-lepton analysis, and with the exception of STC6 and STC8 in
the 0-lepton analysis. 
The required integrated luminosities for exclusion at $95\%\,$CL range from less than $2\,$fb$^{-1}$ (STC4, 
0-lepton) to a few $100\,$fb$^{-1}$.
\item The electroweakino sector of the STC models cannot be excluded with any amount of
integrated luminosity, including $3000\,$fb$^{-1}$.
\item The stop sector of STC4 would be excluded by the 0-lepton analysis, requiring only little more than 
$2\,$fb$^{-1}$  ($15\%$ systematics). 
\item Stop production in STC5 and STC6 would be first excluded by the 1-lepton analysis with $30\,$fb$^{-1}$ (again $15\%$ systematics).
 STC8 could be excluded at $80\,$fb$^{-1}$ in this case. 
\item  In case of $25\%$ systematics, STC8 would not be excluded by any of the analyses at any luminosity.
\end{itemize}

Of course these results are based on only three cut based analyses which could be implemented and roughly optimised during the time available
with respect to the Snowmass study. Further optimisation is most probably possible. With increasing luminosity, the optimal working point of
the analyses is likely to be found for harder cuts, since a purer selection is less vulnerable to systematic 
uncertainties. On the other hand
this means cutting further out in the tails of distributions, where the relative systematic uncertainties might be larger. It should also be noted
that the extrapolation to $3000\,$fb$^{-1}$ is done here under the assumption of 50 pileup events\footnote{Results for simulations with 140 pileup events are 
given in the appendix.}.

Nevertheless the potentially large impact of systematic uncertainties on the discovery of difficult signatures should be taken note of. At some point, better 
control of backgrounds might be more important than increase in luminosity. Furthermore precise knowledge of the lower lying states (e.g.~the EWKinos, but also the 
sleptons) from a Linear Collider could predict the decay chains of the heavy states, including their kinematics, and thus give important input to the study of the 
heavier states (e.g.~stop /sbottom) at the LHC.

\def\XIPM#1{\ensuremath{ \tilde{\chi}^{\pm}_#1}}
\def\XI0#1{\ensuremath{ \tilde{\chi}^0_#1}}
\def\stau#1{\ensuremath{ \tilde{\tau}_#1}}
\def\smu#1{\ensuremath{ \tilde{\mu}_#1}}
\def\sel#1{\ensuremath{ \tilde{e}_#1}}

\section{International Linear Collider Studies} \label{sec:ILC}

\begin{figure}[htb]
  \begin{center}
\includegraphics[width=0.49\linewidth]{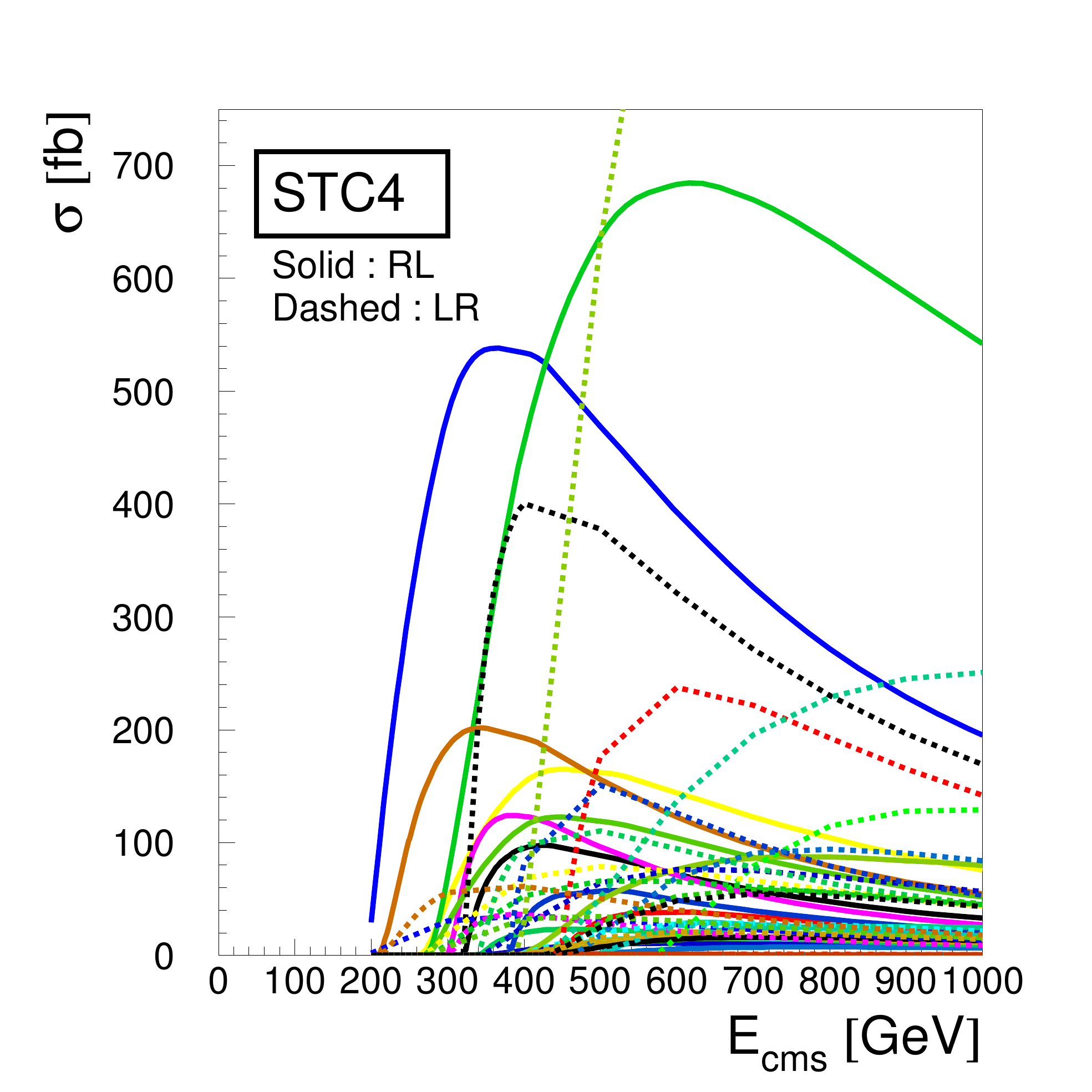}
\includegraphics[width=0.49\linewidth]{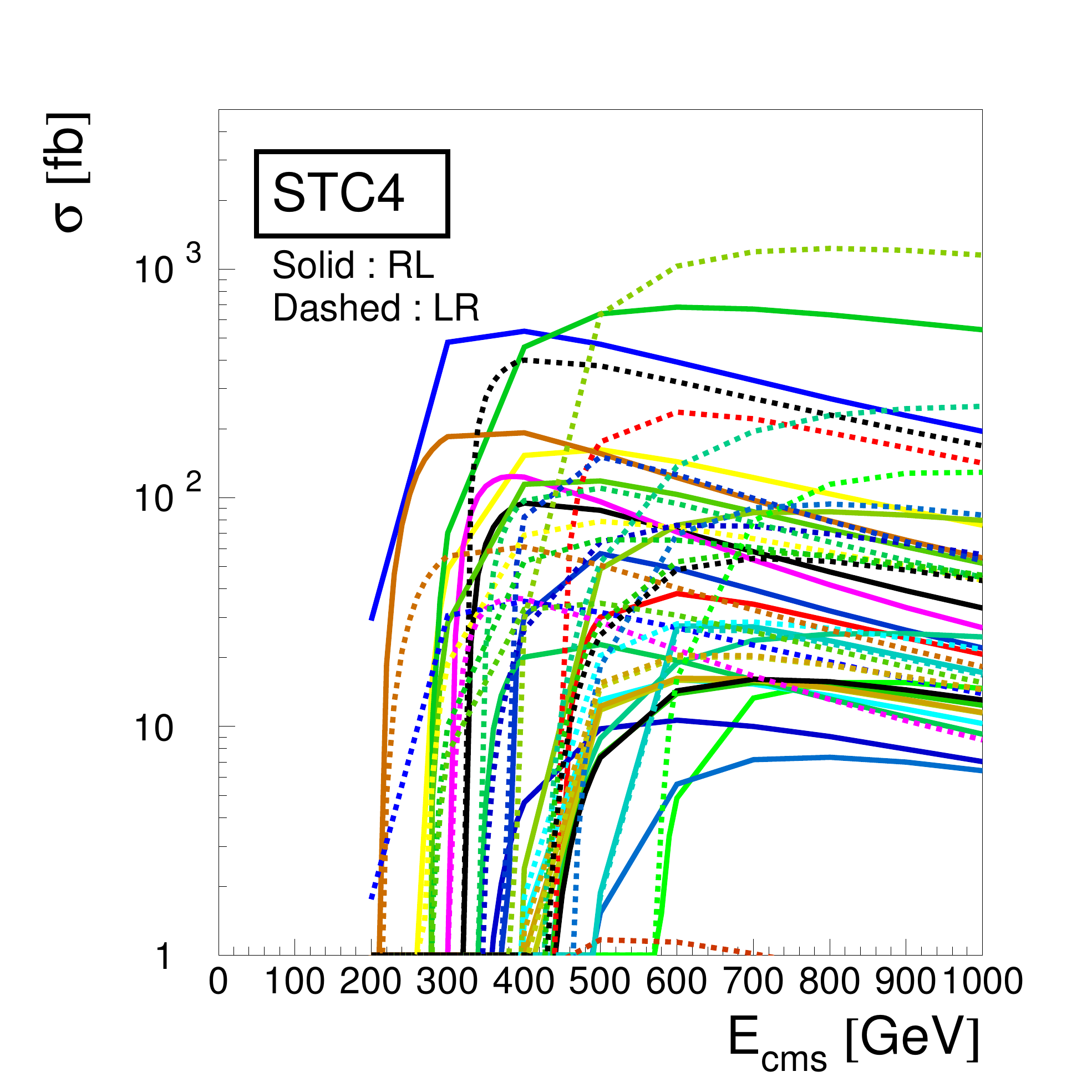}
 \end{center}
  \caption{\label{fig:xsect-ilc} STC4 cross sections for sparticle production as a function of $E_{{\mathrm{cms}}}$ at the ILC.
  Full lines correspond to $P(e^+e^-)=(-0.3,+0.8)$, dashed lines to $P(e^+e^-)=(-0.3,+0.8)$.
  Left: linear scale; Right: logarithmic scale.
  The most prominent channels at $P(e^+e^-)=(-0.3,+0.8)$ are
  $\tw^0_1\tw^0_1$ (blue), $\te_R\te_R$ (green), $\ttau_1\ttau_1$ (brown),
  and $\tw^0_1\tw^0_2$ (yellow).
  At $P(e^+e^-)=(-0.3,+0.8)$ they are $\tnu_e\tnu_e$ (olive green), 
  $\twp_1\twm_1$ (black), $\twpm_1\twpm_2$ (red), 
  and  $\te_L\te_L$ (blue-green). 
  %Also note the marine-blue dashed curve
  %(seventh from the top at the right border of the linear plot):
  %it is the  process $\tw^0_2\tw^0_4$ and is the 
  %one with highest 
  %cross-section for producing $\tw^0_4$.
  }
\end{figure}

At the ILC running at $E_{{\mathrm{cms}}} = 500\,$GeV, 
all sleptons and the lighter set of electroweakinos of the STC scenarios can be produced. \XI0{3} and \XI0{4} become accessible in associated production around $E_{{\mathrm{cms}}} = 600\,$ GeV and in pair production at around $E_{{\mathrm{cms}}} = 850\,$GeV, along with \XIPM{2} pair production. 
The cross sections are sizable -- only one of the kinematically allowed processes would have
a production cross section below $1\,$fb for both beam polarization configurations.
The total SUSY cross section is over 3 pb in both cases.
Figure~\ref{fig:xsect-ilc} shows the polarized cross sections for various processes as
 a function of the  center-of-mass energy in linear and logarithmic scale.

Although the \stau{1} is the NLSP,
almost all electroweakinos have sizable branching fractions to other final states
than the notoriously difficult $\tau$-lepton. This also means that signatures with electrons or muons in the final state can originate either from slepton or electroweakino production. The ability to operate at any desired center-of-mass
energy between $200$ and $500\,$GeV (or even $1\,$TeV) and to switch the sign of the beam polarizations are unique tools to identify each of these processes. The low SM background levels
allow in many cases a full and unique kinematic reconstruction of cascade decays.

\subsection{First observation channels}
The first channel to manifest itself at the ILC depends on the assumed running scenario. If the ILC
starts out as a Higgs factory at $E_{{\mathrm{cms}}}=250\,$GeV, then $\eeto \ttau_1 \ttau_1$ and $\tz_1 \tz_1 \gamma$ would
be the first observable channels, while $\te_R$ and $\tmu_R$ pair production is just beyond reach. 
The measurement of the $\ttau$ mass however would be challenging close to threshold, since both upper and lower
edge of the $\tau$-lepton energy spectrum would be in the region affected by background from multi-peripheral 
two-photon processes.

\begin{figure}[htb]
  \begin{center}
\includegraphics[width=0.40\linewidth]{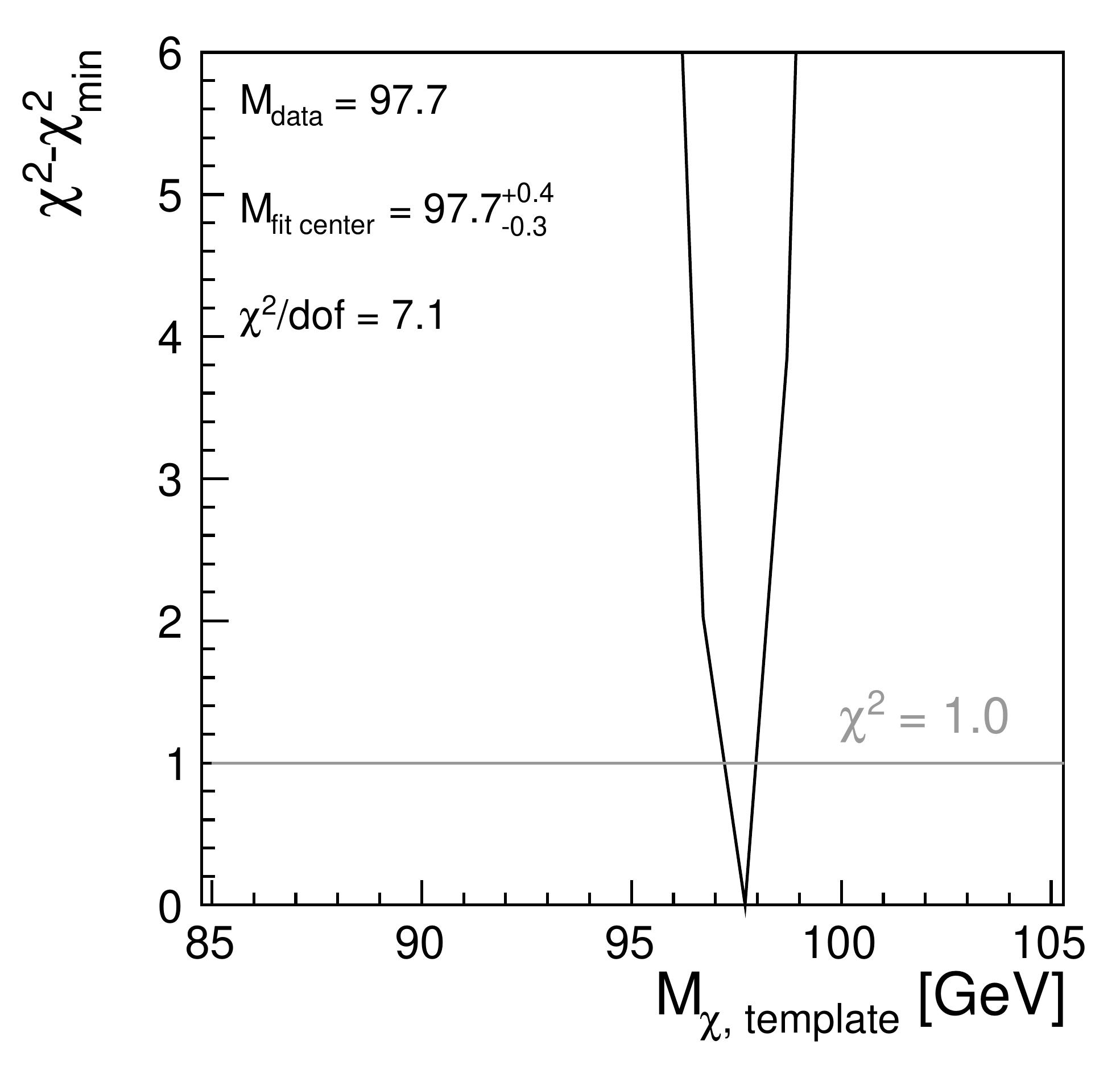}
\hspace{0.1cm}
\includegraphics[width=0.49\linewidth]{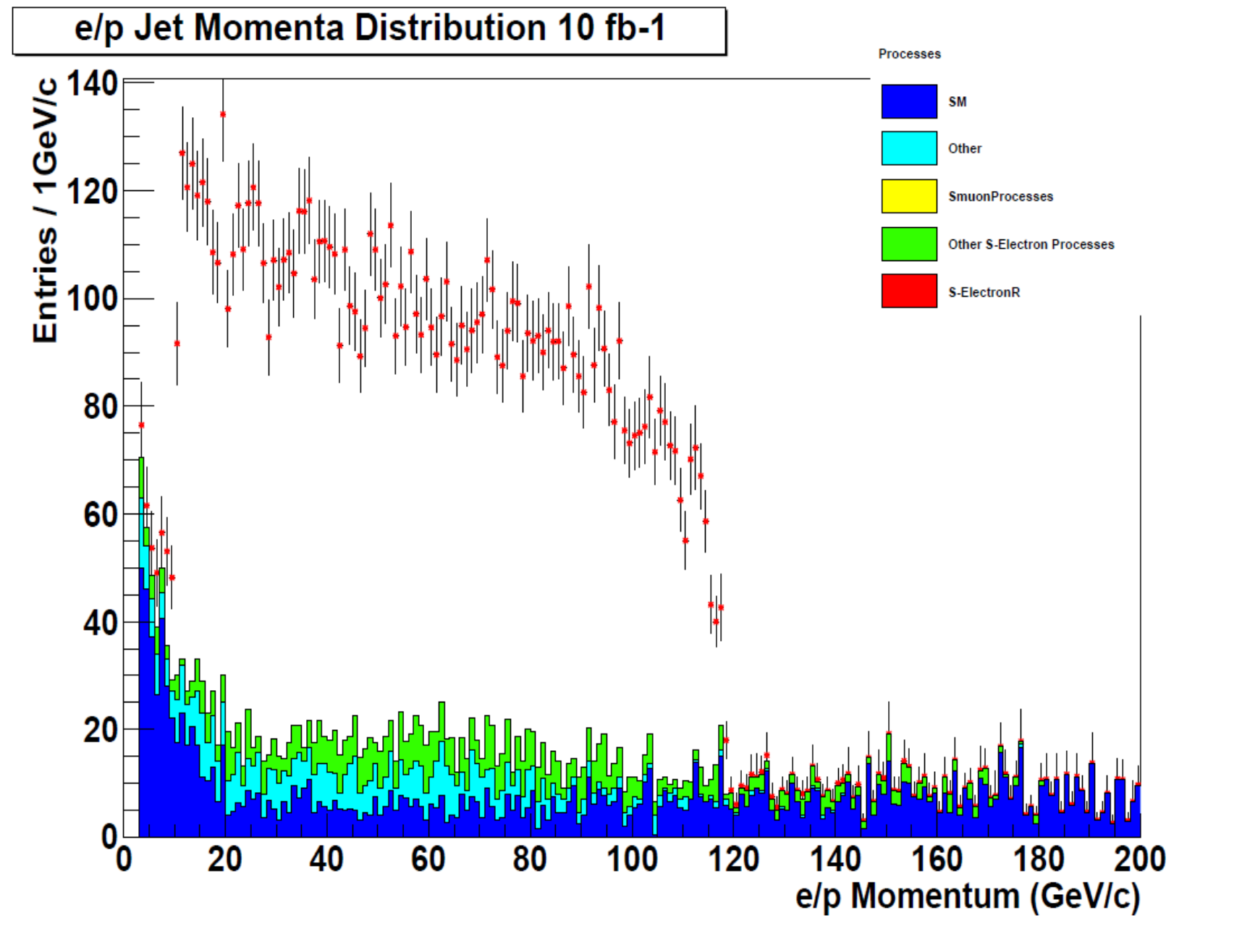}
 \end{center}
  \caption{\label{fig:selectrons} Left: Determination of $m_{\tz_1}$ from a template fit to the photon energy spectrum in $\eeto \tz_1 \tz_1 \gamma$. From~\cite{bib:thesis_bartels}. Right: Momentum spectrum of events with $e^+e-$ and missing $4$-momentum. The assumed luminosity of
  $10\,$fb$^{-1}$ corresponds to one week data taking at design luminosity. From~\cite{bib:stefano_ecfa}.}
\end{figure}

On the other hand, the LSP mass and pair production cross-section could be measured at least
with a few percent precision from the energy (or recoil mass) spectrum of the accompanying ISR photons~\cite{Bartels:2012rg}. The left part of 
Fig.~\ref{fig:selectrons} illustrates the
precision achievable on the neutralino mass from a template fit\footnote{This study has been performed at a higher center-of-mass energy. At $E_{{\mathrm{cms}}}=250\,$GeV, the cross-section is similar, and the photon energy spectrum
less spread out, so that the quality of the mass determination is expected to be comparable.}. Since the neutralino pair production is dominated by $t$-channel selectron exchange,
the mass of the lighter selectron and its helicity can be determined from the measurement of the polarized cross-sections. 

As soon as the center-of-mass energy is raised past the pair production threshold for right-handed sleptons, in our case when $E_{{\mathrm{cms}}} \gtrsim 270\,$GeV, the $e^+e^- +$ missing $4$-momentum signature would see a striking signal
within a few days. Figure~\ref{fig:selectrons} shows the SM and all SUSY contributions to this signature after a simple event selection
on just $10\,$fb$^{-1}$ of data at $E_{{\mathrm{cms}}}=500\,$GeV, which corresponds to one week of data taking at design luminosity.

\begin{figure}[htb]
  \begin{center}
\includegraphics[width=0.375\linewidth]{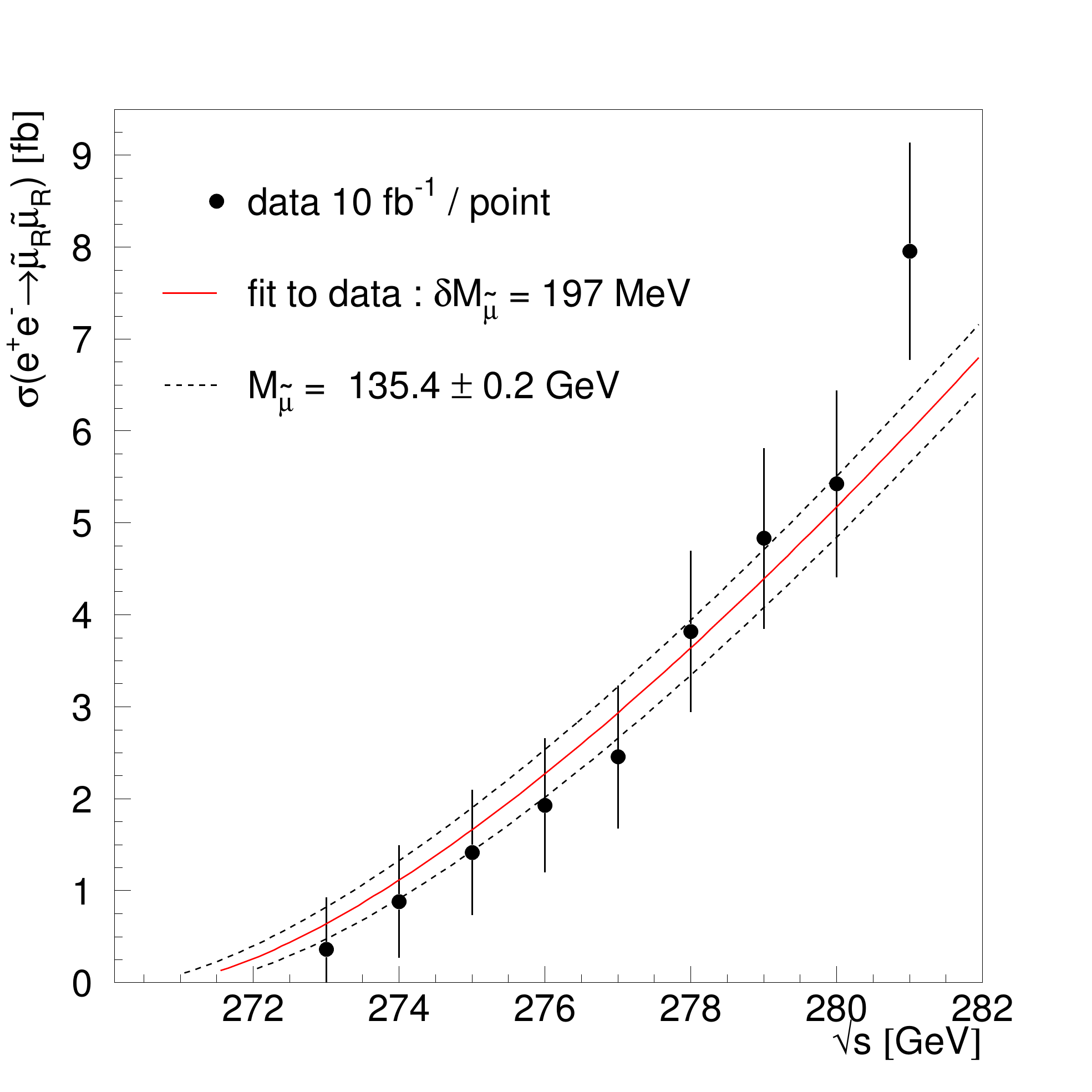}
 \end{center}
  \caption{\label{fig:smu_thresh} Threshold scan at the $\eeto \smu{R} \smu{R}$ threshold.
   From Ref.~\cite{Brau:2012hv}}.
  %% OR if I can find the plot:
  %% Right: Kinematicly constrained \smu{R} mass-spectrum from \XI0{2} $\to$ \smu{R}$\mu \to \mu \mu \XI0{1}.
  %% From~\cite{...}}.
\end{figure}

The cross-section for \smu{R} pair production is much lower due to the absence of a $t$-channel. Still the \smu{R} mass can be determined to $\sim 200\,$MeV by scanning the production threshold near $270\,$GeV~\cite{Brau:2012hv}, as illustrated by Fig.~\ref{fig:smu_thresh}. Note that 
this can be improved to $\sim 10\,$MeV in the continuum if $\XI0{2}\XI0{2}$ is accessible and
$\XI0{2}$ has a non-vanishing branching fraction to $\smu{R} \mu$, cf. below.

\subsection{Sleptons and Electroweakinos in the Continuum}

Several of the channels in the slepton and electroweakino sector are being studied, or have been in the past in very similar models, assuming only a moderate amount of integrated luminosity of
 $500\,$fb$^{-1}$ at $E_{{\mathrm{cms}}}=500\,$GeV unless stated otherwise. This corresponds to two years of ILC operation at design parameters.

\subsubsection{The $\tilde{\tau}$-Sector}

\begin{figure}[htb]
  \begin{center}
\includegraphics[width=0.45\linewidth]{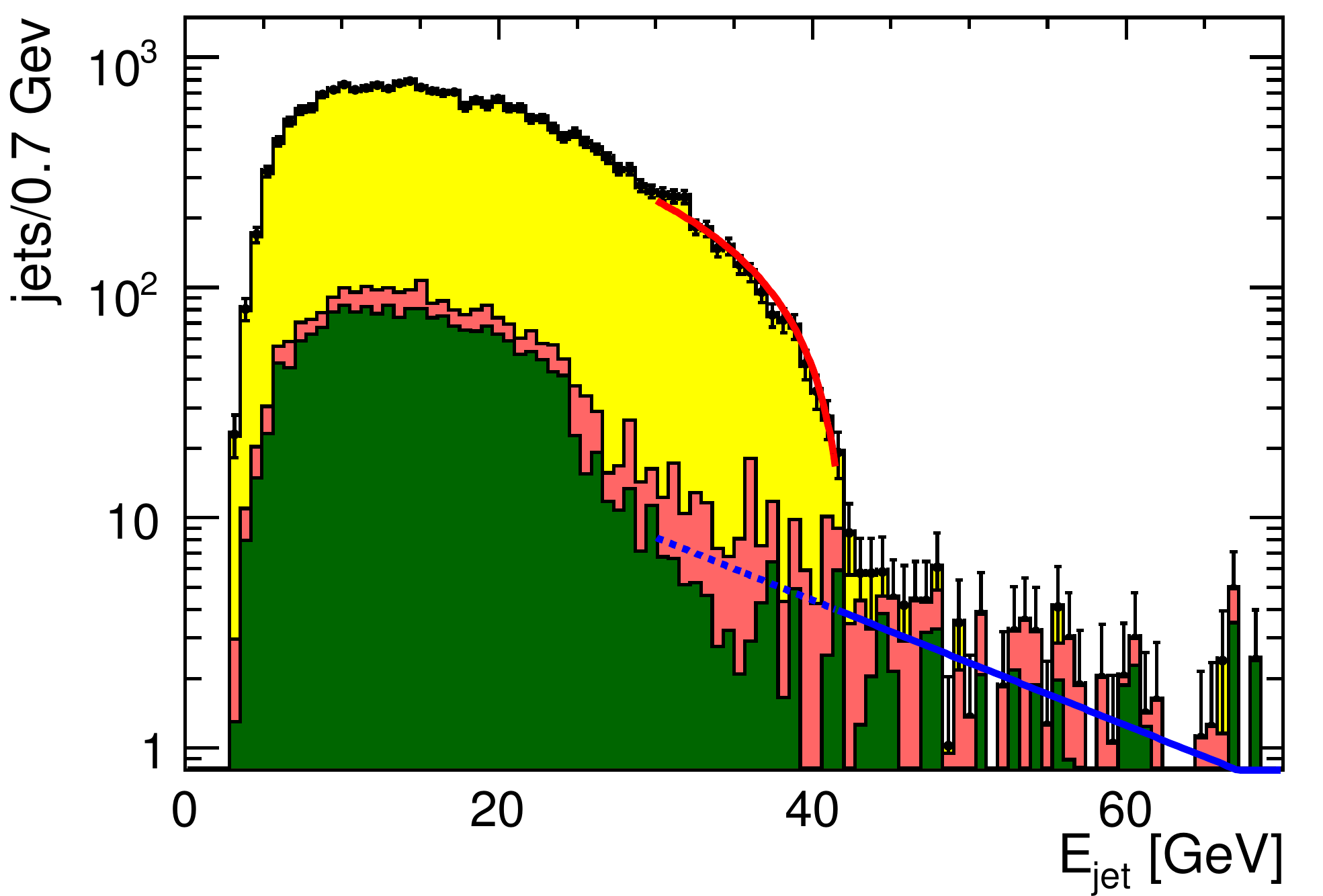}
\hspace{0.1cm}
\includegraphics[width=0.45\linewidth]{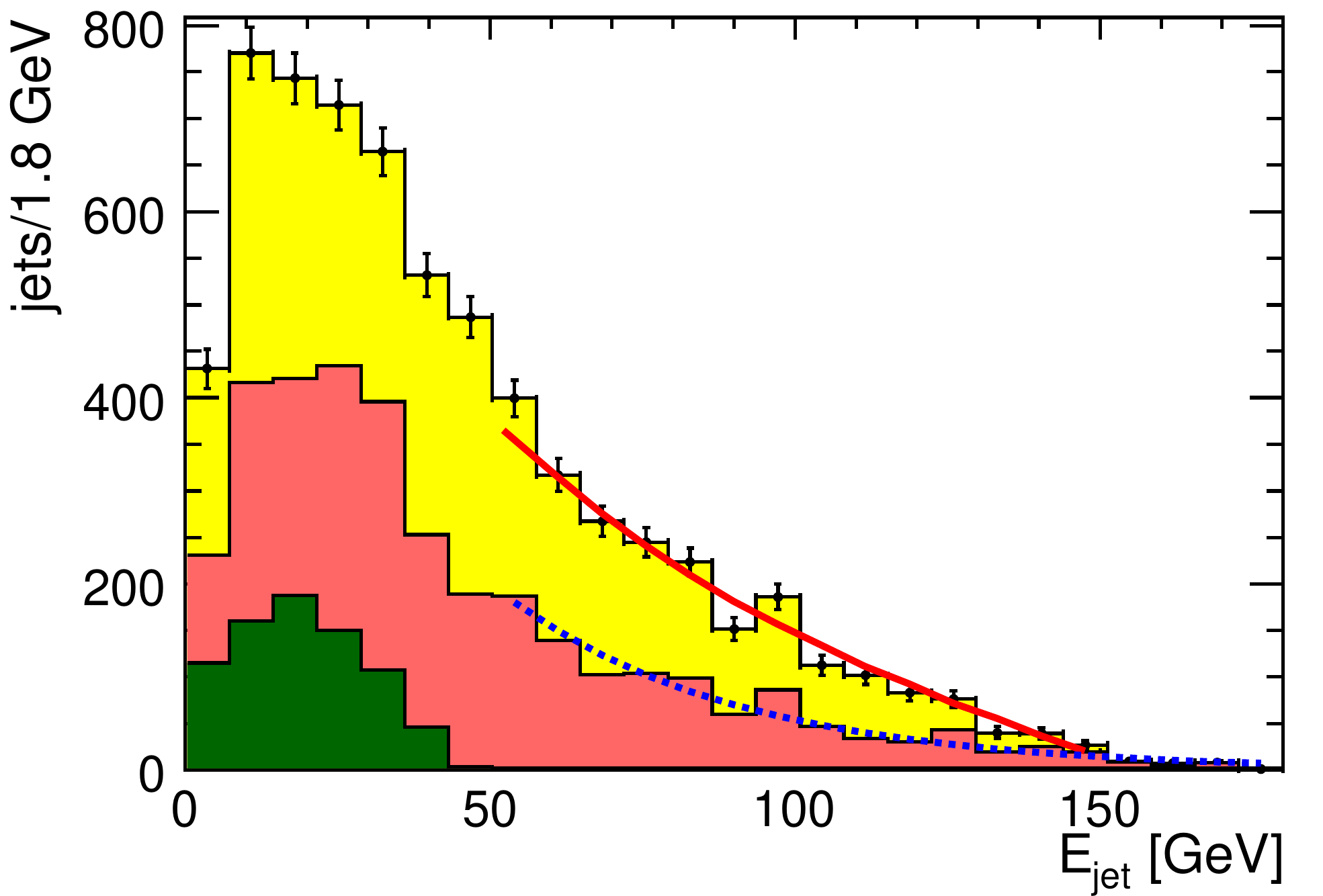}
 \end{center}
  \caption{\label{fig:stc-ilc} Left: \stau{1} spectrum (yellow) and
  background (SM: red, other SUSY: green), with end-point fit. Right: \stau{2} spectrum (yellow) and  background (SM: red, other SUSY: green), with end-point fit. From Ref.~\cite{Bechtle:2009em}. }
\end{figure}

Especially in $\tilde{\tau}$-coannihilation scenarios, a precise determination of the $\tilde{\tau}$ sector is essential in order to be able to predict the expected  
relic density with sufficient precision to test whether the $\XI0{1}$ is indeed the dominant 
Dark Matter constituent. The capabilities for precision measurements in the  $\tilde{\tau}$ sector 
have been studied in full detector simulation~\cite{Bechtle:2009em}. It was shown that the
\stau{1} mass could be determined to 200 MeV,
and the \stau{2} mass to 5 GeV from the endpoint of the $\tau$-jet energy spectrum as illustrated
 in Fig.~\ref{fig:stc-ilc}.

\begin{figure}[htb]
  \begin{center}
\includegraphics[width=0.3\linewidth]{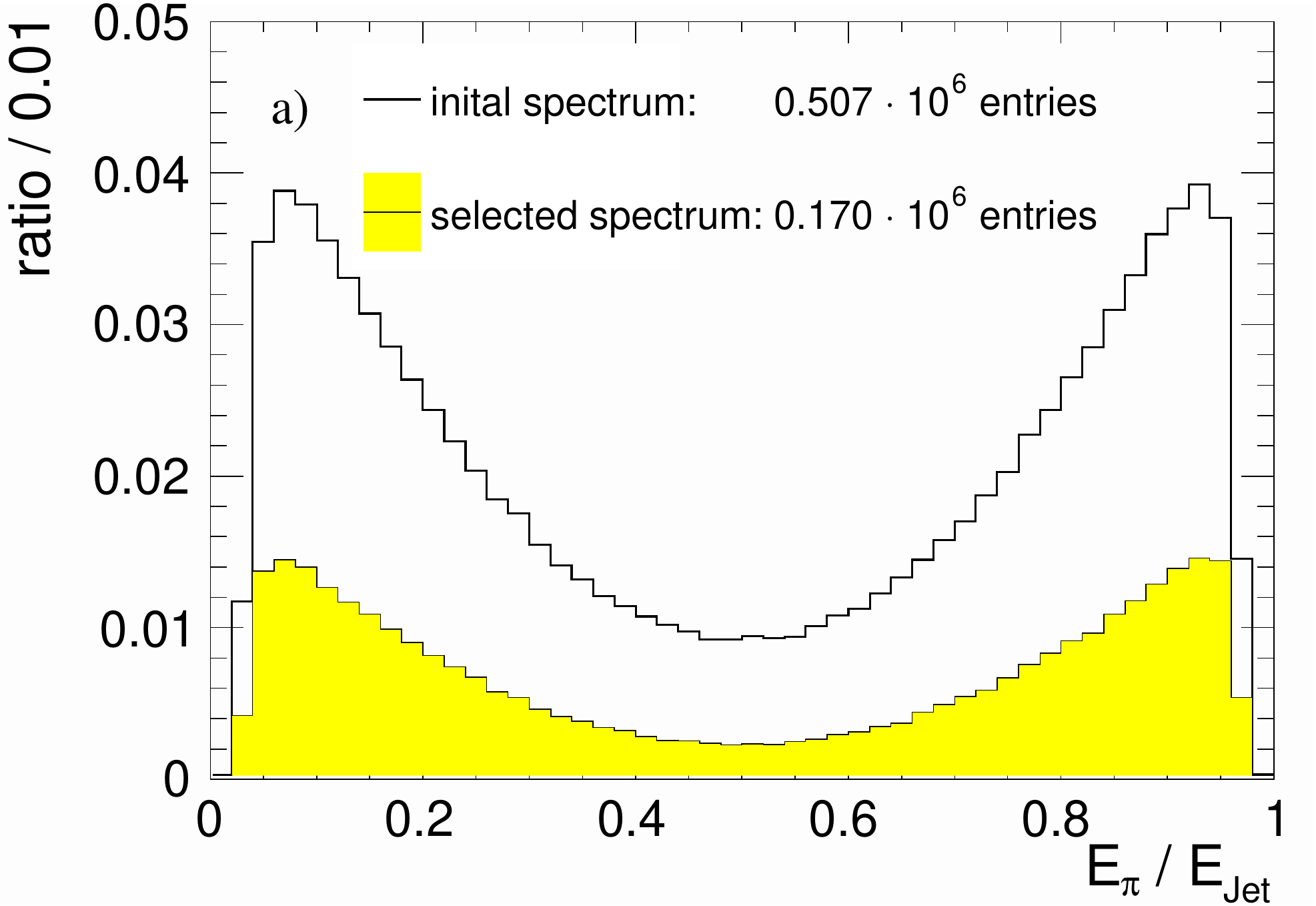}
\hspace{0.1cm}
\includegraphics[width=0.3\linewidth]{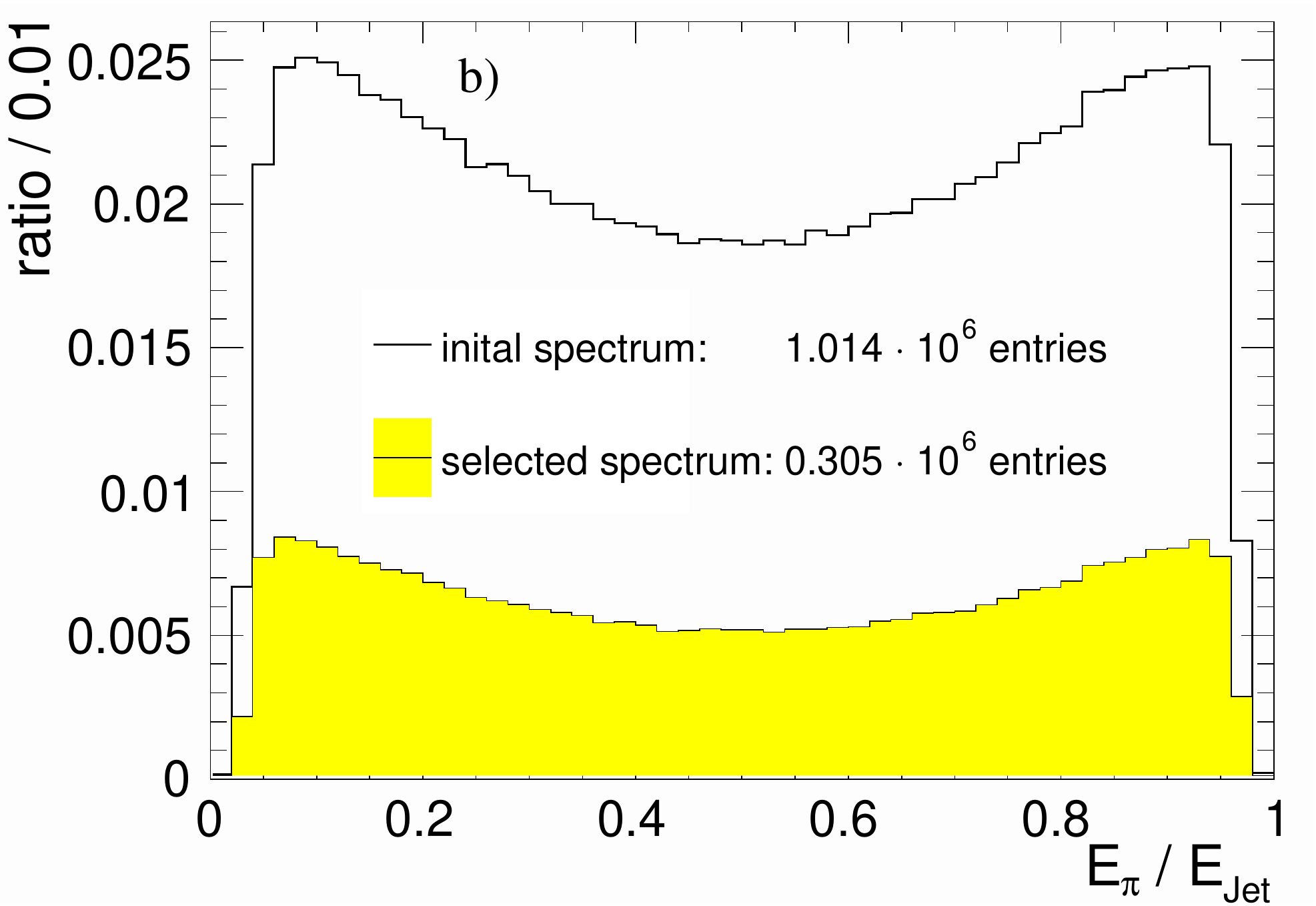}
\hspace{0.1cm}
\includegraphics[width=0.3\linewidth]{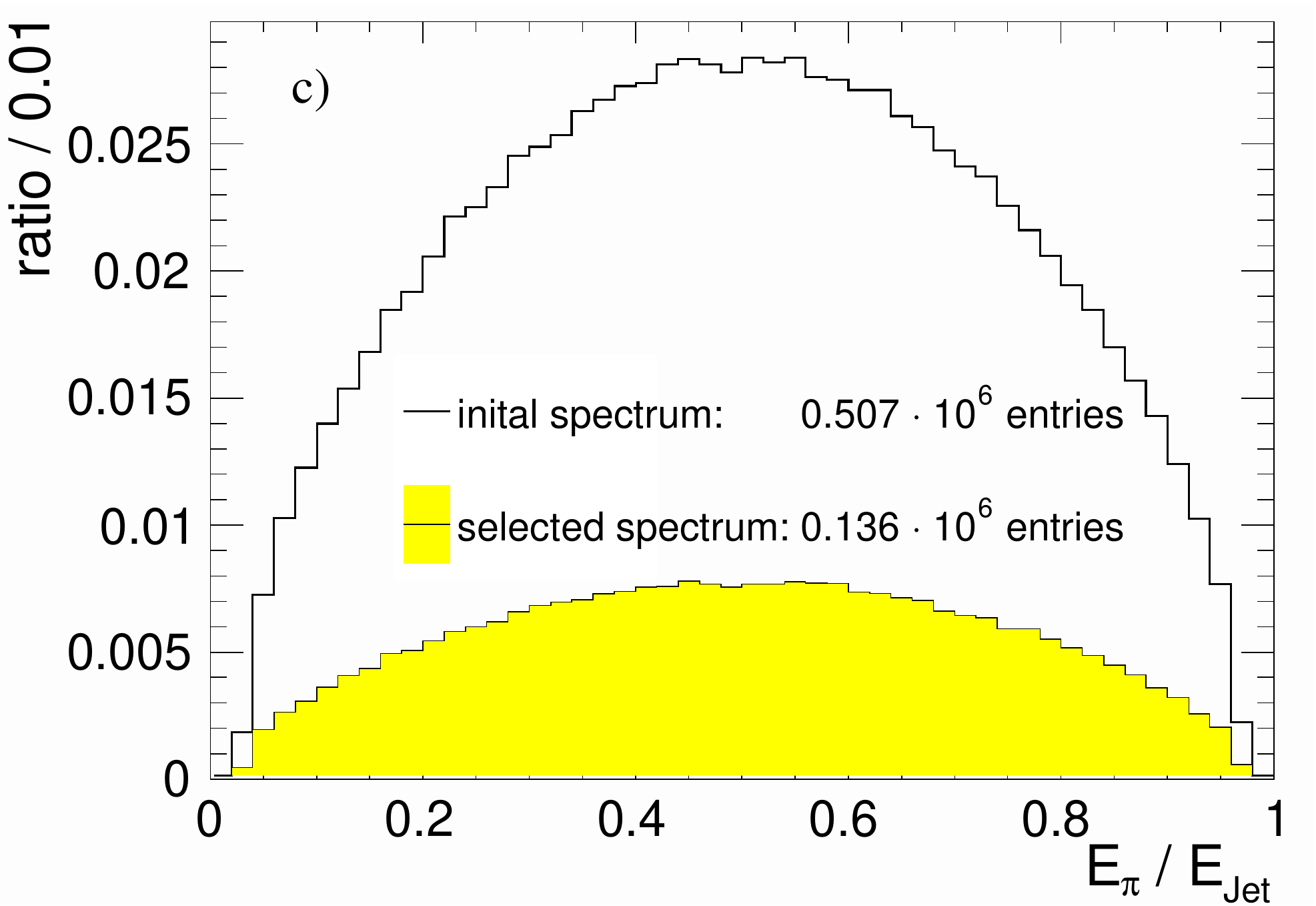}
 \end{center}
  \caption{\label{fig:taupol} Distribution of the observable $R=E_{\pi}/E_{jet}$ before (open histogram) and after event selection (grey (yellow) histogram). a): both $\tau$ leptons are
right-handed. c): the $\tau$ leptons have opposite helicity.
r): both $\tau$ leptons are left-handed.  From~\cite{Bechtle:2009em}.}.
\end{figure}

Production cross section for both these modes can be
determined at the level of 4\%, and the polarization
of $\tau$-leptons from the \stau{1} decay, which gives access to the $\tilde{\tau}$ and $\XI0{1}$ mixing\footnote{Interaction of sfermions and gauginos conserve chirality, while the Yukawa
 interaction of the higgsinos flips chirality.}, 
could be
measured with an accuracy better than 10\%, eg. from $\tau\to\pi^+\nu_{\tau}$ decays.
Fig.~\ref{fig:taupol} illustrates an additional possibility to determine the 
$\tau$-polarization from decays to $\rho$-mesons ($\tau\to\rho^+\nu_{\tau} \to \pi^+ \pi^0\nu_{\tau}$). In this case, the observable $R=E_{\pi}/E_{jet}$ can be used
to measure  the $\tau$-polarization to $\pm5\%$ by a fit of the templates in Fig.~\ref{fig:taupol} to the data.

%By tuning the beam-energy to be between the \stau{1}\stau{2}
%and \stau{2}\stau{2} thresholds the $\tilde{\tau}$ mixing-angle
%can be obtained.

\subsubsection{Final States with Electrons and Missing Four-momentum}
As already illustrated by Fig.~\ref{fig:selectrons}, the selectron pair
production cross section is huge in our scenario due to the $t$-channel
neutralino exchange, allowing a very precise determination
of the masses and polarised cross sections in a short time. 
They give important information on the neutralino mixing, since eg.
in case of light higgsinos the $t$-channel would be strongly supressed by
the small electron Yukawa coupling. In particular, if both beams are given 
right-handed polarizations, only the \eeto\sel{R}\sel{L} process is possible. 
As this reaction proceeds exclusively via neutralino exchange in the $t$-channel, 
it's size gives insight to the neutralino mixing~\cite{MoortgatPick:2005cw}.

\subsubsection{Final States with Muons and Missing Four-momentum}
\begin{figure}[htb]
  \begin{center}
\includegraphics[width=0.5\linewidth]{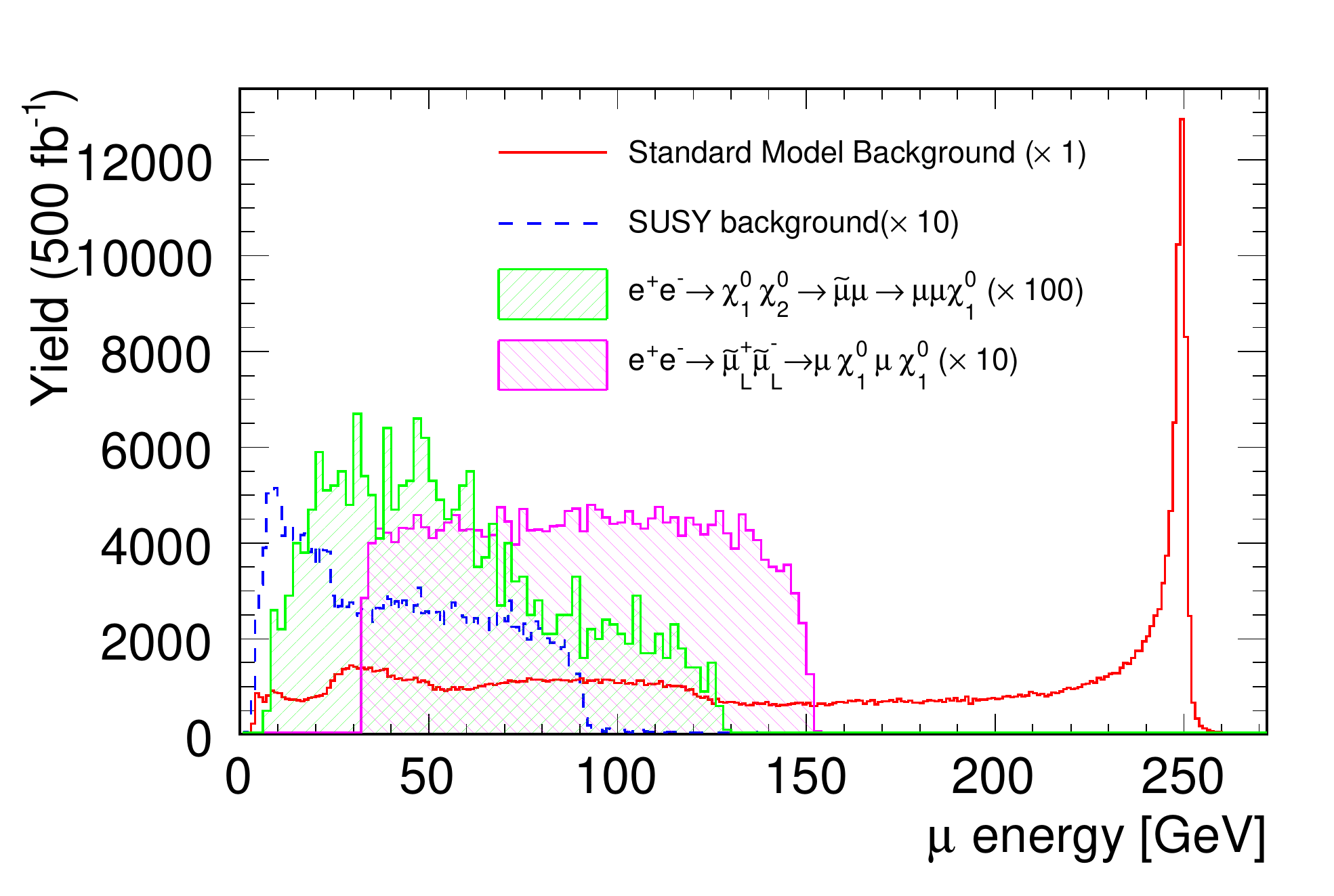}
 \end{center}
  \caption{\label{fig:muene} Inclusive muon energy spectrum from di-muon events. From~\cite{bib:thesis_dascenzo}. }
\end{figure}

Figure~\ref{fig:muene} shows the muon energy distribution for all events with
two muons from SM and SUSY processes, before any selection in full simulation of the ILD detector. The striking peak in the
SM distribution at the beam energy originates from $\eeto \mu^+\mu^-$. The SUSY 
contributions (scaled only by a factor of 10 or 100 to be visible at this fully 
inclusive stage) arise from $\smu{L}\smu{L} \to \mu \mu \XI0{1}\XI0{1}$, 
$\XI0{1}\XI0{2}\to \mu \mu \XI0{1}\XI0{1}$ as well as $\smu{R}\smu{R}$, $\stau{1}\stau{1}$ with $\tau$ decays to muons and others. We will show in the following that all these contributions
can be disentangled and identified.

\begin{figure}[htb]
  \begin{center}
\includegraphics[width=0.45\linewidth]{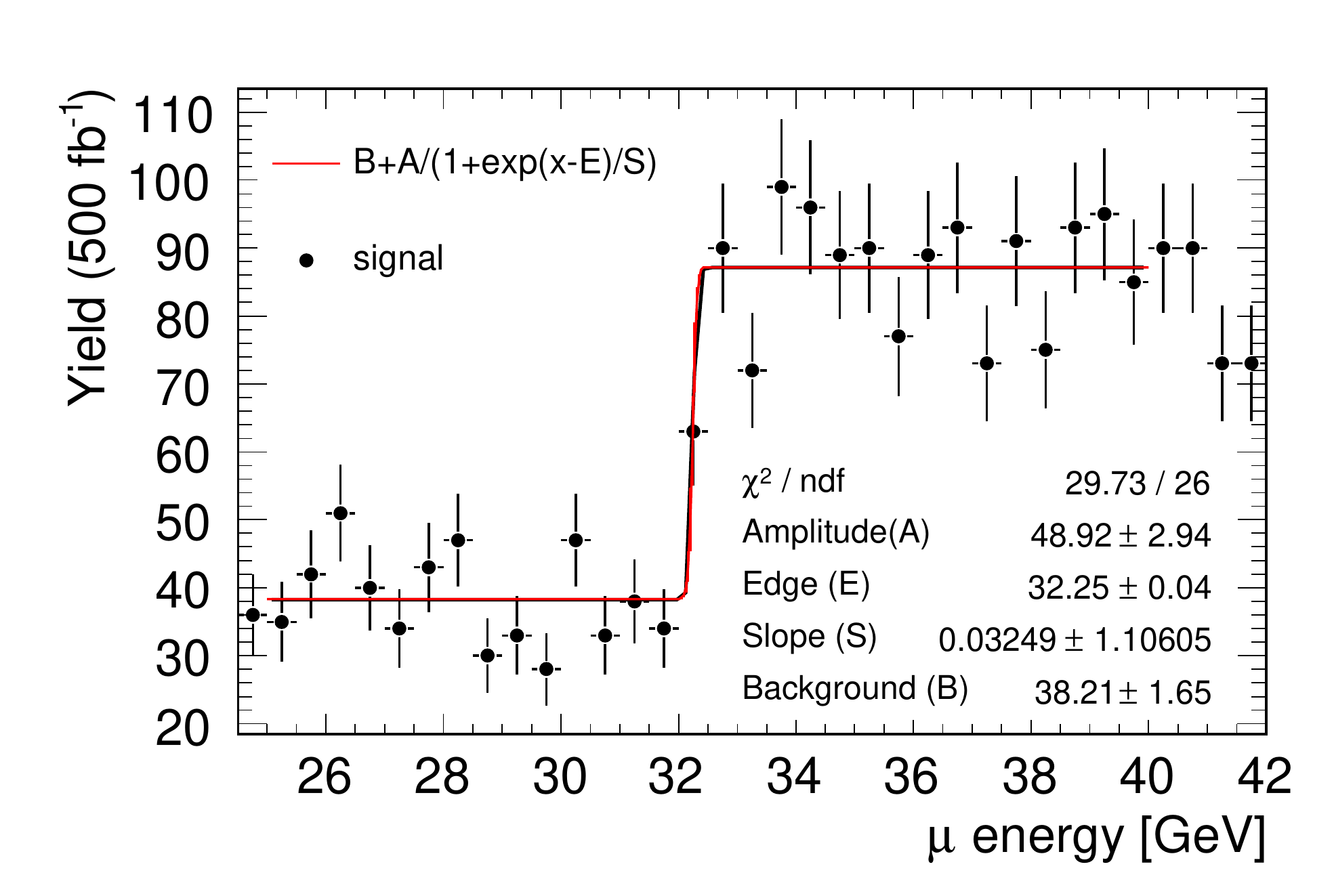}
\hspace{0.1cm}
\includegraphics[width=0.45\linewidth]{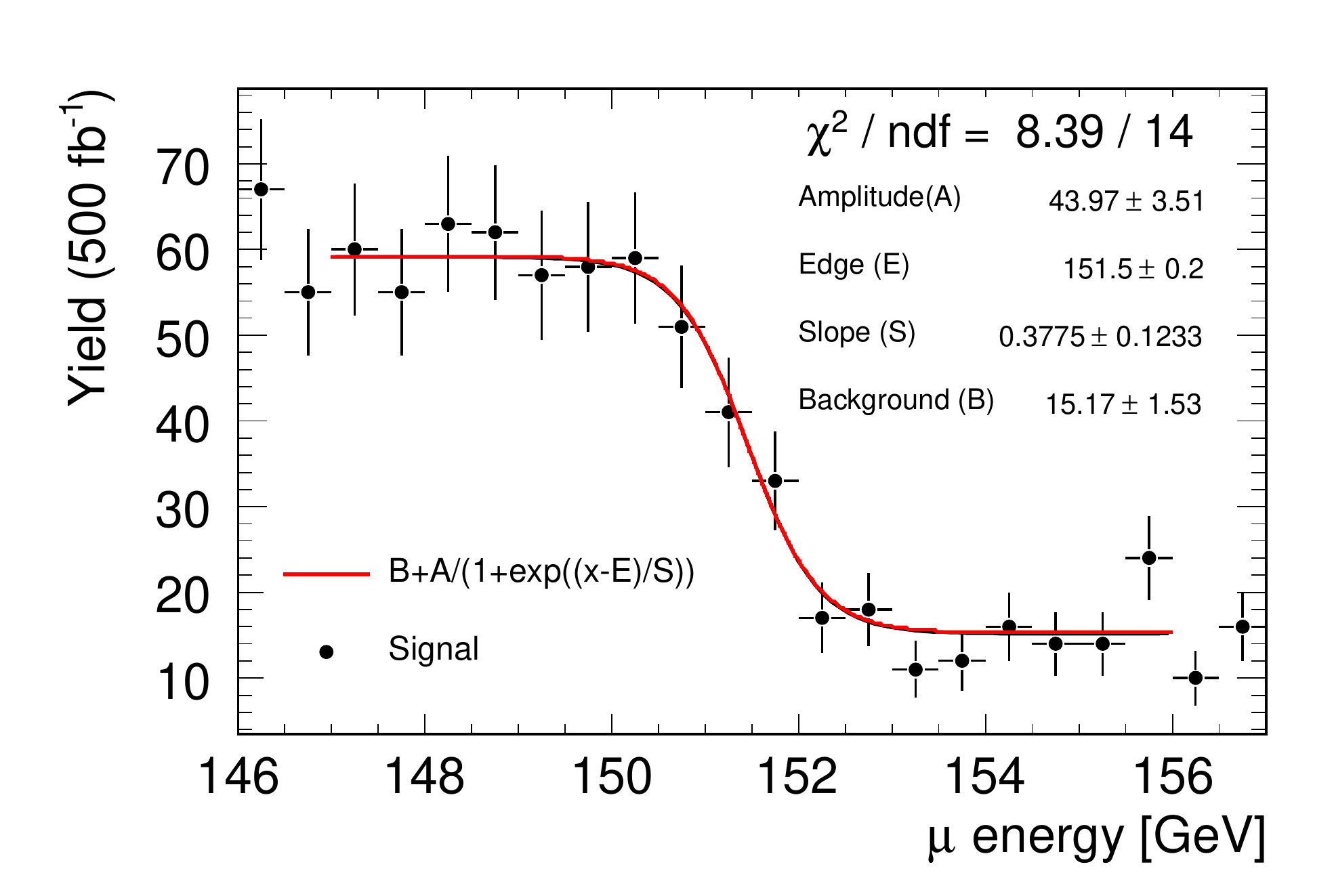}
 \end{center}
  \caption{\label{fig:smuL} Determination of the $\smu{L}$ mass from edges in the muon
  energy spectrum. Left: lower edge; Right: upper edge. From~\cite{bib:thesis_dascenzo}. }
\end{figure}

\begin{figure}[htb]
  \begin{center}
\includegraphics[width=0.45\linewidth]{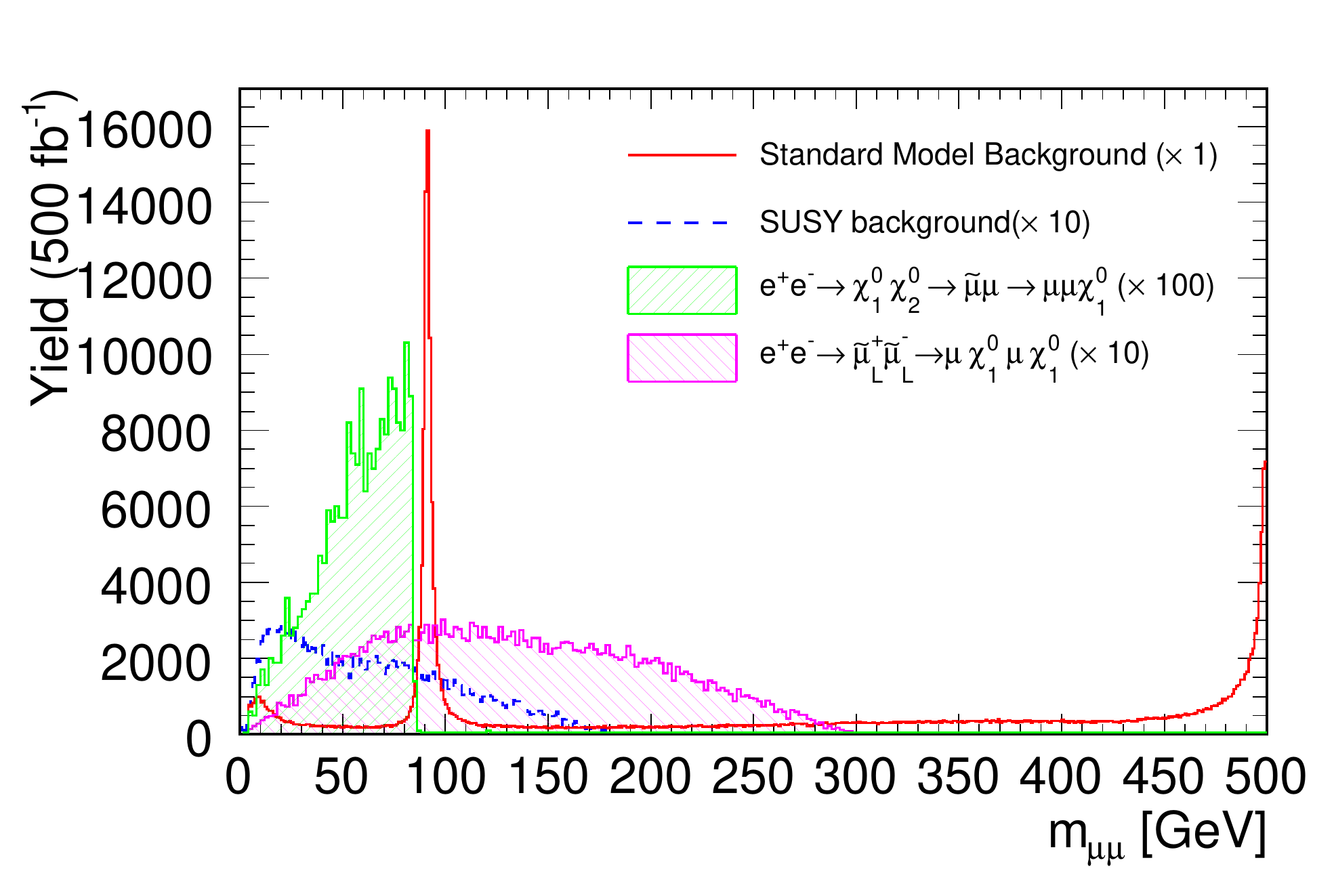}
\hspace{0.1cm}
\includegraphics[width=0.45\linewidth]{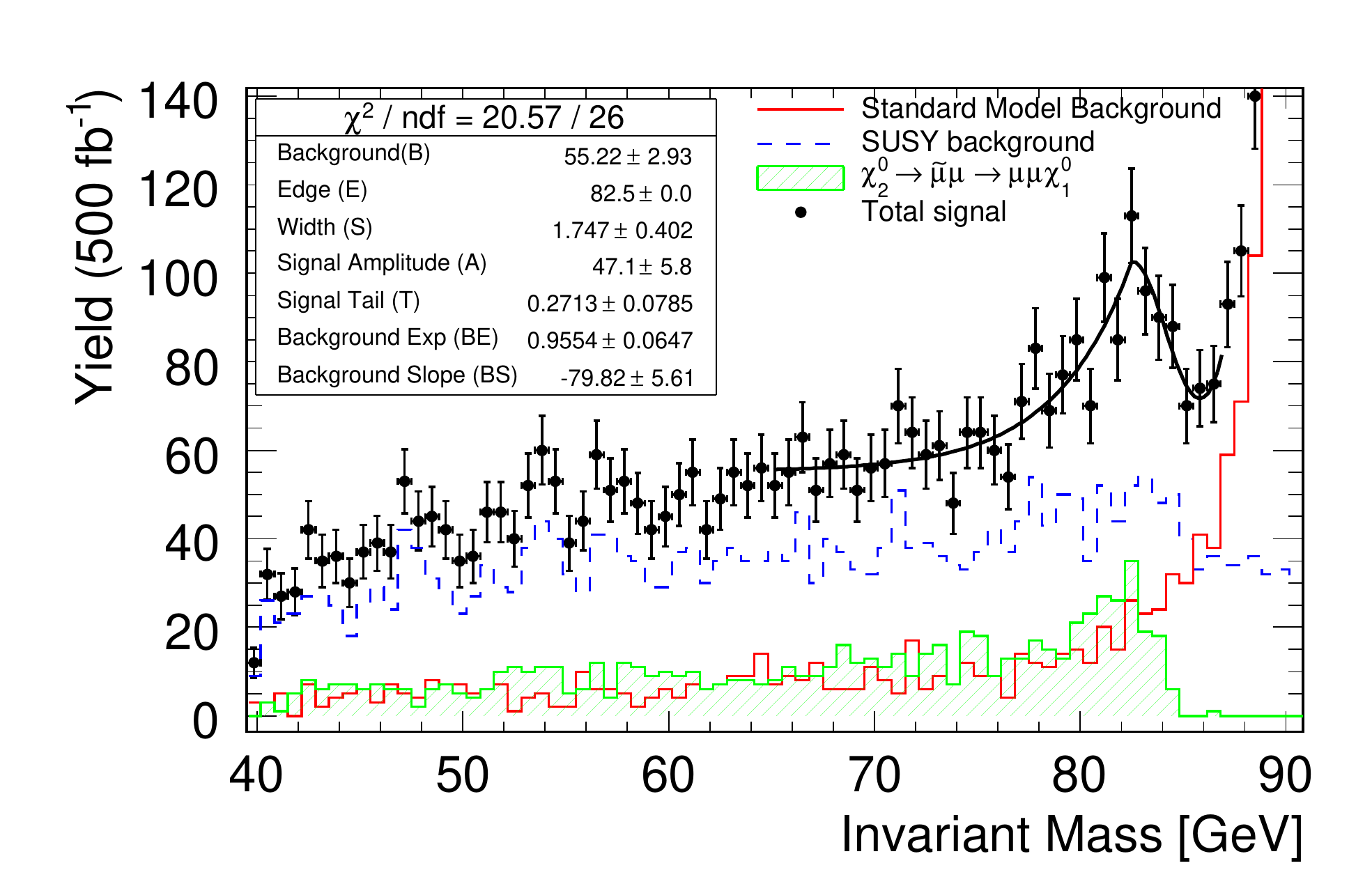}
 \end{center}
  \caption{\label{fig:chi02mu} Determination of the $\XI0{2}$ mass the di-muon invariant mass spectrum Left: full spectrum for inclusive di-muon sample; Right: zoom into signal region after 
  dedicated selection. From~\cite{bib:thesis_dascenzo}. }
\end{figure}

\begin{figure}[htb]
  \begin{center}
\includegraphics[width=0.4\linewidth]{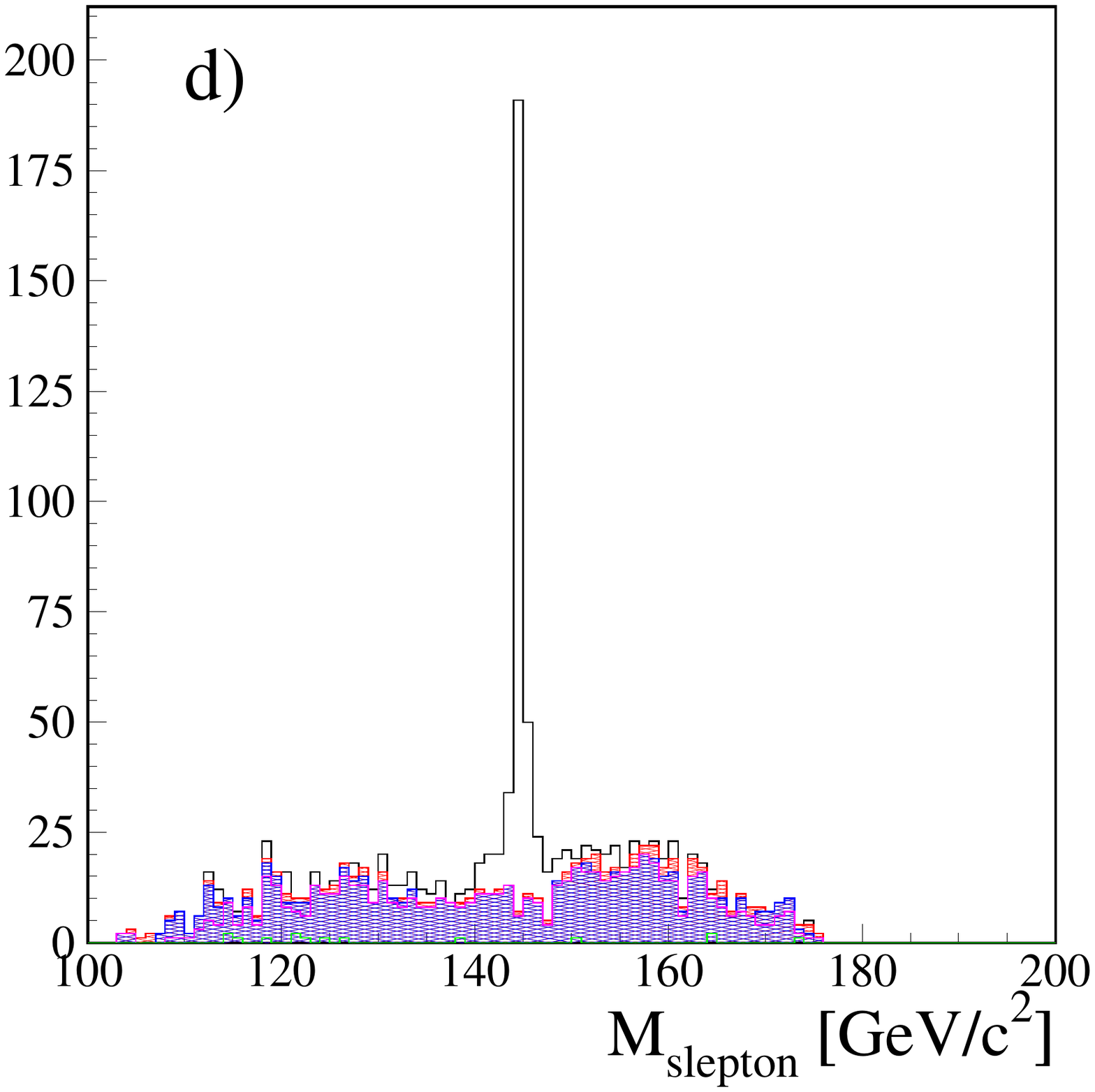}
\hspace{0.1cm}
\includegraphics[width=0.4\linewidth]{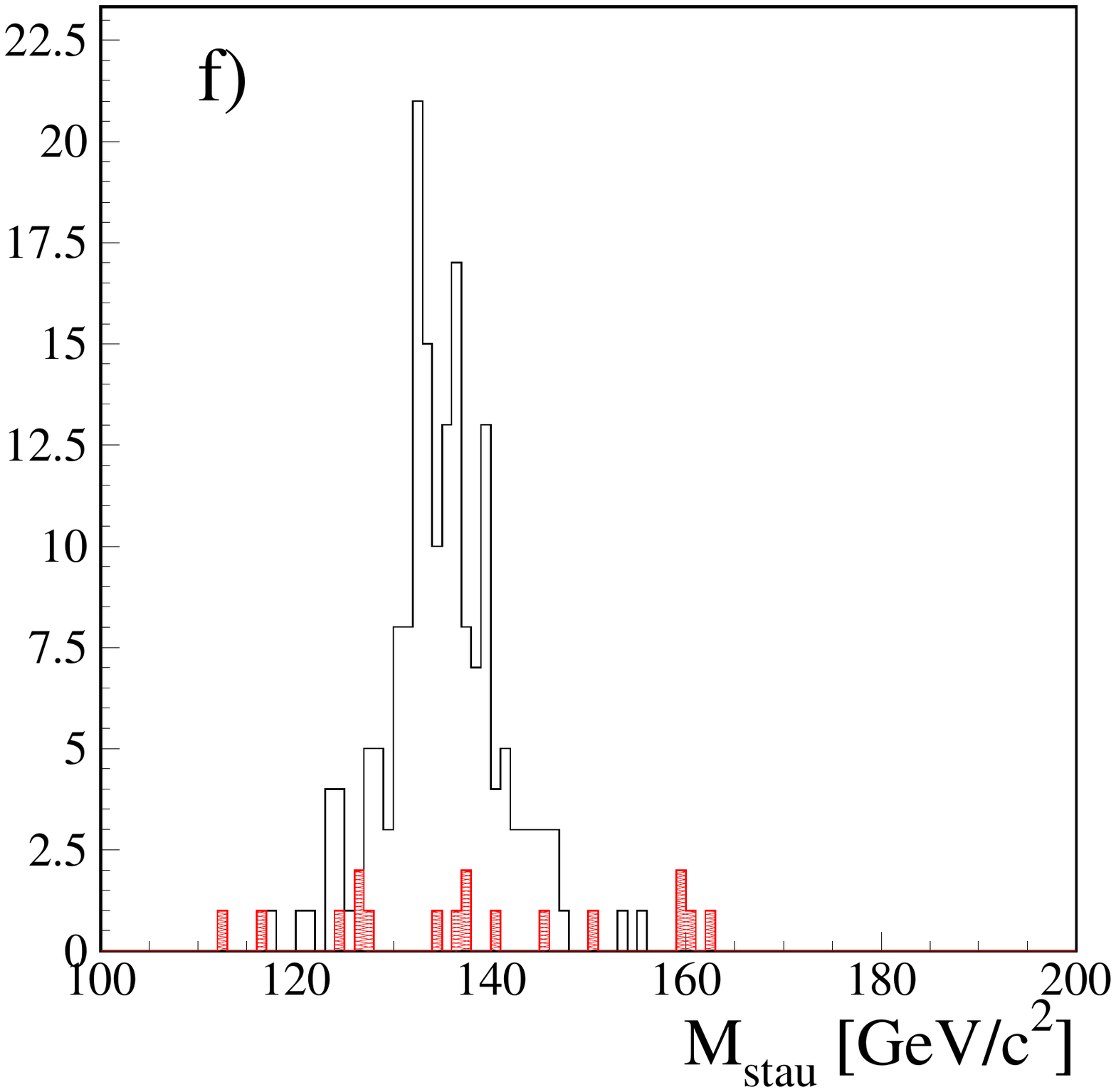}
\end{center}
\caption{\label{fig:kinrec}Reconstruction of slepton masses from $\XI0{2}\XI0{2} \to \tilde{l} l   \tilde{l} l$ in SPS1a, which has a very similar spectrum to our case. Left: reconstructed \smu{R} mass. Right: reconstructed \stau{1} mass. From~\cite{Berggren:2005gp}.}
\end{figure}

For the $\smu{L}$ case, Fig~\ref{fig:smuL} shows zooms into the edge reagions of the muon energy spectrum after a dedicated selection.  From the edge positions, the $\smu{L}$ mass can be determined to $400\,$MeV~\cite{bib:thesis_dascenzo}.

The even smaller contribution from $\XI0{1}\XI0{2}\to \mu \mu \XI0{1}\XI0{1}$ (scaled by 
factor 100 in inclusive plots) can also be identified, eg. in the invariant mass spectrum of the
two muons, as illustrated by Fig~\ref{fig:chi02mu}. From this channel alone, the mass of the
$\XI0{2}$ can be determined to a precision of about $1\,$GeV, depending on the assumed precision
for the mass of $\smu{R}$ and $\XI0{1}$.

A particularly interesting channel is \eeto\XI0{2}\XI0{2} and the \XI0{2} decay to \smu{R}$\mu$ (or equivalently to \sel{R}$e$), even if the branching ratio is at the level of a few percent like in our example point.
These cascade decays can be fully kinematically constrained at the ILC,
and would promise to yield even lower uncertainties on the
\smu{R} and \sel{R} masses than the threshold scans,
of the order of 25 MeV. This is estimated on an earlier study in a scenario with about twice as large branching ratios for the considered decay mode, where a precision of $10\,$MeV~\cite{Berggren:2005gp} was found. The corresponding distribution of the reconstructed \smu{R} mass is shown in the left part of Fig.~\ref{fig:kinrec}, including
all SM and SUSY backgrounds. Even the dominating decays to \stau{1}$\tau$
can be constrained as shown in the right part of Fig.~\ref{fig:kinrec}, and could yield comparable results to a threshold scan. 

%\subsubsection{Electroweakino Production}
%In the bosino sector, 
%further constraints on the neutralino and chargino mixing-matrices
%are possible. All processes, except \eeto\XIPM{2}\XIPM{2}
%and \eeto\XI0{4}\XI0{4} are kinematicly accessible at 500 GeV,
%and all except  \eeto\XI0{3}\XI0{3} have cross-sections above
%1 fb.
%Although the \stau{1} is the NLSP,
%almost all bosinos have sizable BR to other final states
%than the notoriously difficult $\tau$-lepton.
%The use of beam-polarization and tunable $E_{{\mathrm{cms}}}$ will further enhance the
%power of the observations.
%Due to the low background,
%cascade decays can be disentangled.
%A particularly interesting channel is \eeto\XI0{2}\XI0{2}, where
%there is a sizable BR for the decay to \smu{R}$\mu$ or \sel{R}$e$.
%These cascade decays can be fully kinematicaly constrained at the ILC,
%and would promise to yield even lower uncertainties on the
%\smu{R} and \sel{R} masses than the threshold scans,
%of the order of 10 MeV \cite{Berggren:2005gp}.
%%%   see Fig. \ref{fig:stc-ilc}.
%Even the dominating decays to \stau{1}$\tau$
%can be constrained, and could yield comparable results to a
%threshold scan. 

\section{Dark Matter Relic Density} \label{sec:dmrelic}
A final goal would be to perform a closure test on the neutralino Dark Matter hypothesis. This can be achieved by using all available collider observables to determine the SUSY parameters and to predict the relic density based on the assumption that the \XI0{1} is the only contribution to Dark Matter.

\begin{figure}[htb]
\begin{center}
\includegraphics[width=0.4\linewidth]{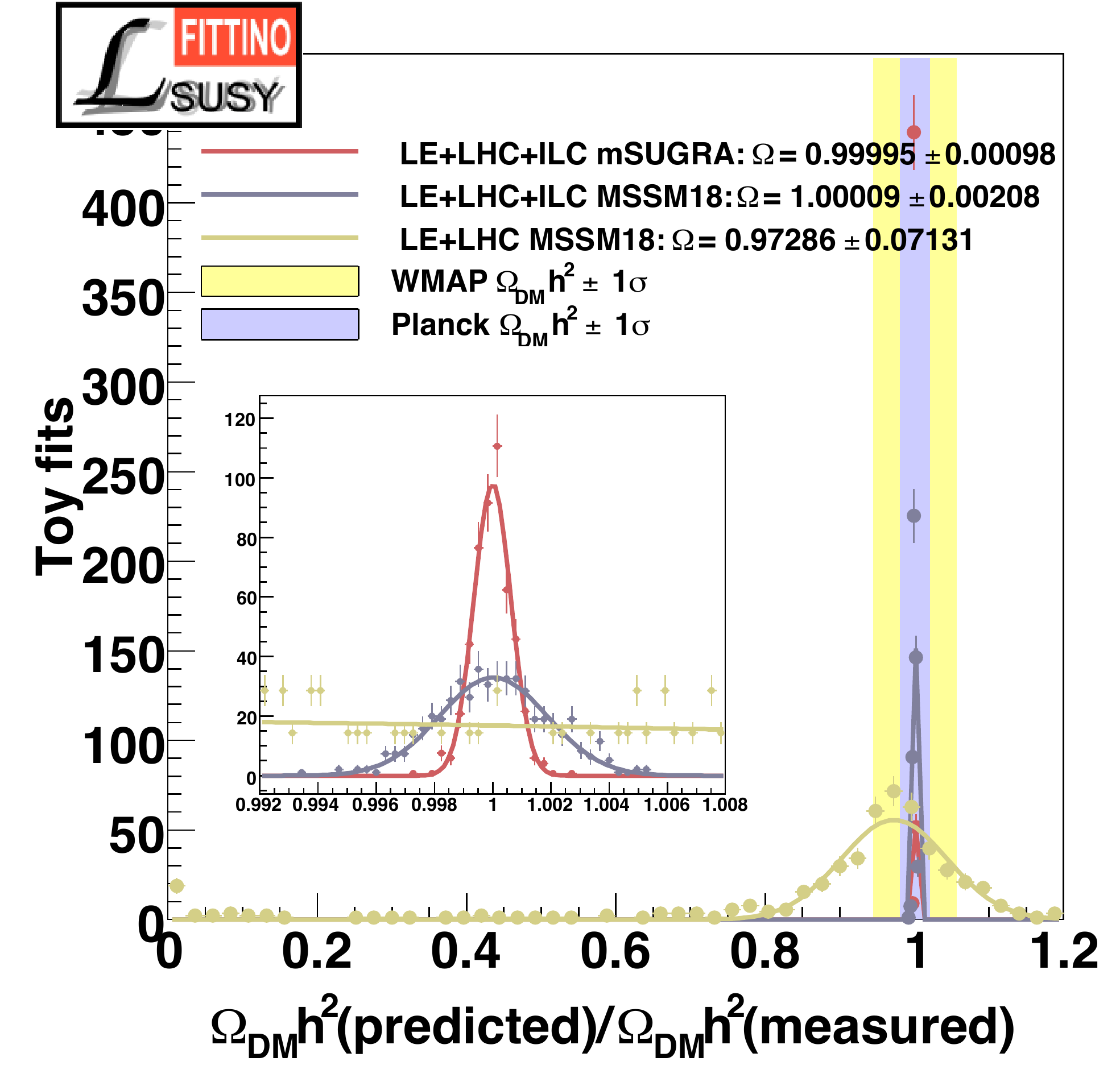}
\caption{Ratio of the predicted value of $\Omega_{\mathrm{pred}}
    h^2$ to the nominal value of $\Omega_{\mathrm{SPS1a}} h^2$ in the
    SPS1a scenario for a variety of Fittino Toy Fits
   without using $\Omega_{\mathrm{CDM}} h^2$ as an observable. From~\cite{Bechtle:2009ty}.
    The anticipate predictions for LHC and ILC measurements are compared to 
       current and projected cosmological observations.} 
\label{fig:omega_DM}
\end{center}
\end{figure}

This has been studied in~\cite{Bechtle:2009ty} for the SPS1a scenario, which is very similar
to our benchmark points apart from the squarks and gluinos. This means
that the projections used for the LHC observables might be too optimistic for the much heavier colored sector in our scenario. This might be partially cancelled by the fact that pre-LHC projections turned out to be rather conservative in many cases. However for the prediction
of the relic density, the colored sector is of less importance, while
it depends crucially on the electroweak sector, and in particular the 
LSP and \stau{1} properties, which are almost identical between SPS1a' and our scenarios.

Figure~\ref{fig:omega_DM} shows the relic density obtained in mSugra and MSSM18 fits to many toy experiment outcomes of LHC and ILC measurements.
It shows that the ILC measurements allow to predict $\Omega_{\mathrm{CDM}} h^2$ in the MSSM18 case almost as precisely
as in mSugra with only $4$ free parameters (and one sign) in stead of $18$ parameters. The LHC alone would leave a comparison with cosmological observations at an inconclusive and thus unsatisfactory level. This example beautifully illustrates the complementarity of the two machines.

\section{Conclusions} \label{sec:summary}
We have presented a series of $\tilde{\tau}$-coannihilation scenarios
based on the pMSSM, which is compatible with all known experimental
constraints. It illustrates that the phenomenology
of full models can be significantly more subtle than suggested by
the simplified model approach. Especially the $\tilde{t}_1$ masses
of this series, ranging from $\sim 300$ to $\sim 700\,$GeV seem to
be excluded by current LHC limits in simplified models. However we
showed that due to many different long decay chains the actual 
analyses are not yet sensitive to these scenarios.

At LHC14, the observability of the considered model points in terms of a deviation from the Standard 
Model depends strongly on the systematic uncertainty on the background. In fully hadronic stop searches 
or stop searches with one lepton, STC4, STC5 and STC6 could be discovered provided that the systematic 
uncertainty on the background does not exceed about $20\%$. Discovery of STC8 as well as electroweakino 
production in any of the scenarios requires systematic uncertainties at the few percent level. Larger 
statistics, like the LHC high-lumi running ($3000\,$fb$^{-1}$) can be exploited only if the systematic 
uncertainties are low enough, roughly in the few percent region. While the stop searches are rather 
robust against pileup, the effect of considering 50 pileup events is clearly visible in the 
electroweakino searches. The simulation results with 140 pileup events support these observations (appendix 
\ref{sec:cut140})

For the scenarios with lighter stop masses, stop pair production amounts up to $\sim 90\%$ of the total 
SUSY cross section. At the 2nd highest considered stop mass, electroweakino production is already 
dominant.  We could not yet investigate how well contributions from these and other open channels (eg. 
sbottom production) can be disentangled from each other and how well properties of individual sparticles 
can be measured.

% JL: I think this is clarified - since all scenarios give about the same sensitivities, there can't be much stop in it...
%In particular it is not yet clear if eg. the electroweakino production
%can be identified in the low stop mass scenarios below the overwhelming
%SUSY background from stop pair production - 

At the ILC, nearly all sleptons and electroweakinos are accessible either in
pair or associated production, several of them would be most likely discoveries
in view of the LHC studies summarised above. We gave a brief summary of some of the
existing studies on spectroscopy in our scenario(s) and also older
studies of points with a very similar electroweak part of the spectrum,
like SPS1a' or SPS1a. In particular for SPS1a it has been shown
in previous studies that ILC precision is mandatory to achieve a
satisfactory precision on the predicted Dark Matter relic density.

\clearpage

\bibliographystyle{utphys}
\bibliography{lit}

\providecommand{\href}[2]{#2}\begingroup\raggedright\begin{thebibliography}{10}

\bibitem{Buchmueller:2009fn}
O.~Buchmueller, R.~Cavanaugh, A.~De~Roeck, J.~Ellis, H.~Flacher, {\em et~al.},
  ``{Likelihood Functions for Supersymmetric Observables in Frequentist
  Analyses of the CMSSM and NUHM1},''
  \href{http://dx.doi.org/10.1140/epjc/s10052-009-1159-z}{{\em Eur.Phys.J.}
  {\bfseries C64} (2009) 391--415},
\href{http://arxiv.org/abs/0907.5568}{{\ttfamily arXiv:0907.5568 [hep-ph]}}.
%%CITATION = ARXIV:0907.5568;%%.

\bibitem{ATLAS-CONF-2013-049}
{\bfseries ATLAS} Collaboration, {ATLAS Collaboration}, ``Search for
  direct-slepton and direct-chargino production in final states with two
  opposite-sign leptons, missing transverse momentum and no jets in
  $20\,$fb$^{-1}$ of pp collisions at $\sqrt{s} = 8\,$tev with the atlas
  detector,'' ATLAS Conference Note ATLAS-CONF-2013-049, 2013.
\newblock \url{http://cds.cern.ch/record/1546777}.

\bibitem{CMS-PAS-SUS-12-022}
{\bfseries CMS} Collaboration, {CMS Collaboration}, ``Search for direct ewk
  production of susy particles in multilepton modes with 8tev data,'' CMS
  Physics Analysis Summary CMS-PAS-SUS-12-022, 2012.
\newblock \url{http://cds.cern.ch/record/1546777}.

\bibitem{Baer:2013ula}
H.~Baer and J.~List, ``{Post LHC8 SUSY benchmark points for ILC physics},''
\href{http://arxiv.org/abs/1307.0782}{{\ttfamily arXiv:1307.0782 [hep-ph]}}.
%%CITATION = ARXIV:1307.0782;%%.

\bibitem{Sjostrand:2006za}
T.~Sjostrand, S.~Mrenna, and P.~Z. Skands, ``{PYTHIA 6.4 Physics and Manual},''
  \href{http://dx.doi.org/10.1088/1126-6708/2006/05/026}{{\em JHEP} {\bfseries
  0605} (2006) 026},
\href{http://arxiv.org/abs/hep-ph/0603175}{{\ttfamily arXiv:hep-ph/0603175
  [hep-ph]}}.
%%CITATION = HEP-PH/0603175;%%.

\bibitem{Beenakker:1996ch}
W.~Beenakker, R.~Hopker, M.~Spira, and P.~Zerwas, ``{Squark and gluino
  production at hadron colliders},''
  \href{http://dx.doi.org/10.1016/S0550-3213(97)80027-2}{{\em Nucl.Phys.}
  {\bfseries B492} (1997) 51--103},
\href{http://arxiv.org/abs/hep-ph/9610490}{{\ttfamily arXiv:hep-ph/9610490
  [hep-ph]}}.
%%CITATION = HEP-PH/9610490;%%.

\bibitem{Beenakker:1997ut}
W.~Beenakker, M.~Kramer, T.~Plehn, M.~Spira, and P.~Zerwas, ``{Stop production
  at hadron colliders},''
  \href{http://dx.doi.org/10.1016/S0550-3213(98)00014-5}{{\em Nucl.Phys.}
  {\bfseries B515} (1998) 3--14},
\href{http://arxiv.org/abs/hep-ph/9710451}{{\ttfamily arXiv:hep-ph/9710451
  [hep-ph]}}.
%%CITATION = HEP-PH/9710451;%%.

\bibitem{ATLAS-CONF-2013-001}
{\bfseries ATLAS} Collaboration, {ATLAS Collaboration}, ``Search for direct
  stop pair production in events with missing transverse momentum and two
  b-jets in 12.8 fb$^-1$ of pp collisions at $\sqrt{s} = 8$~tev with the atlas
  detector,'' ATLAS Conf Note ATLAS-CONF-2013-001, 2013.
\newblock \url{http://cds.cern.ch/record/1503233}.

\bibitem{Graesser:2012qy}
M.~L. Graesser and J.~Shelton, ``{Hunting Asymmetric Stops},''
  \href{http://dx.doi.org/10.1103/PhysRevLett.111.121802}{{\em Phys.Rev.Lett.}
  {\bfseries 111} (2013) 121802},
\href{http://arxiv.org/abs/1212.4495}{{\ttfamily arXiv:1212.4495 [hep-ph]}}.
%%CITATION = ARXIV:1212.4495;%%.

\bibitem{Ovyn:2009tx}
S.~Ovyn, X.~Rouby, and V.~Lemaitre, ``{DELPHES, a framework for fast simulation
  of a generic collider experiment},''
\href{http://arxiv.org/abs/0903.2225}{{\ttfamily arXiv:0903.2225 [hep-ph]}}.
%%CITATION = ARXIV:0903.2225;%%.

\bibitem{Snowmass_MC1}
A.~Avetisyan {\em et~al.}, ``{Snowmass Energy Frontier Simulations for Hadron
  Colliders},'' \href{http://arxiv.org/abs/1307.XXX}{{\ttfamily arXiv:1307.XXX
  [hep-ph]}}.

\bibitem{Snowmass_MC2}
A.~Avetisyan {\em et~al.}, ``{Standard Model Background Generation for Snowmass
  using Madgraph},'' \href{http://arxiv.org/abs/1307.XXX}{{\ttfamily
  arXiv:1307.XXX [hep-ph]}}.

\bibitem{Cacciari:2011ma}
M.~Cacciari, G.~P. Salam, and G.~Soyez, ``{FastJet User Manual},''
  \href{http://dx.doi.org/10.1140/epjc/s10052-012-1896-2}{{\em Eur.Phys.J.}
  {\bfseries C72} (2012) 1896},
\href{http://arxiv.org/abs/1111.6097}{{\ttfamily arXiv:1111.6097 [hep-ph]}}.
%%CITATION = ARXIV:1111.6097;%%.

\bibitem{ATLAS-CONF-2012-165}
{\bfseries ATLAS} Collaboration, {ATLAS Collaboration}, ``Search for scalar
  bottom pair production in final states with missing transverse momentum and
  two b-jets in pp collisions at {\color{red} $\sqrt{s} = 8$~tev} with the
  atlas detector,'' ATLAS Conf Note ATLAS-CONF-2012-165, 2012.
\newblock \url{http://cds.cern.ch/record/1497668}.

\bibitem{Polesello:2009rn}
G.~Polesello and D.~R. Tovey, ``{Supersymmetric particle mass measurement with
  the boost-corrected contransverse mass},''
  \href{http://dx.doi.org/10.1007/JHEP03(2010)030}{{\em JHEP} {\bfseries 1003}
  (2010) 030},
\href{http://arxiv.org/abs/0910.0174}{{\ttfamily arXiv:0910.0174 [hep-ph]}}.
%%CITATION = ARXIV:0910.0174;%%.

\bibitem{CMS-PAS-SUS-13-011}
{\bfseries CMS} Collaboration, {CMS Collaboration}, ``Search for top-squark
  pair production in the single lepton final state in pp collisions at 8 tev,''
  CMS Physics Analysis Summary CMS-PAS-SUS-13-011, 2013.
\newblock \url{http://cdsweb.cern.ch/record/1547550}.

\bibitem{Bai:2012gs}
Y.~Bai, H.-C. Cheng, J.~Gallicchio, and J.~Gu, ``{Stop the Top Background of
  the Stop Search},'' \href{http://dx.doi.org/10.1007/JHEP07(2012)110}{{\em
  JHEP} {\bfseries 1207} (2012) 110},
\href{http://arxiv.org/abs/1203.4813}{{\ttfamily arXiv:1203.4813 [hep-ph]}}.
%%CITATION = ARXIV:1203.4813;%%.

\bibitem{CMS-SUS-12-022}
{\bfseries CMS} Collaboration, {CMS Collaboration}, ``Search for electroweak
  production of charginos, neutralinos, and sleptons using leptonic final
  states in pp collisions at $\sqrt{s} = 8$~tev,'' CMS Physics Analysis Summary
  CMS-PAS-SUS-12-022, 2012.
\newblock \url{http://cdsweb.cern.ch/record/1546777}.

\bibitem{ATLAS-CONF-2013-028}
{\bfseries ATLAS} Collaboration, {ATLAS Collaboration}, ``Search for direct
  slepton and gaugino production in final states with hadronic taus and missing
  transverse momentum with the atlas detector in pp collisions at $\sqrt{s} =
  8$~tev,'' ATLAS Conf Note ATLAS-CONF-2013-028, 2012.
\newblock \url{http://cds.cern.ch/record/1525889}.

\bibitem{bib:thesis_bartels}
C.~Bartels, {\em {WIMP Search and a Cherenkov Detector Prototype for ILC
  Polarimetry}}.
\newblock PhD thesis, University of Hamburg,
  http://www-flc.desy.de/flc/work/group/thesis/doctor.2011.bartels.pdf, 2011.

\bibitem{bib:stefano_ecfa}
S.~Caiazza, ``{Measuring the selectron properties at the ILC},'' in {\em ECFA
  LC2013}.
\newblock 2013.

\bibitem{Bartels:2012rg}
C.~Bartels, O.~Kittel, U.~Langenfeld, and J.~List, ``{Measurement of Radiative
  Neutralino Production},''
\href{http://arxiv.org/abs/1202.6324}{{\ttfamily arXiv:1202.6324 [hep-ph]}}.
%%CITATION = ARXIV:1202.6324;%%.

\bibitem{Brau:2012hv}
J.~E. Brau, R.~M. Godbole, F.~R.~L. Diberder, M.~Thomson, H.~Weerts, {\em
  et~al.}, ``{The Physics Case for an e+e- Linear Collider},''
\href{http://arxiv.org/abs/1210.0202}{{\ttfamily arXiv:1210.0202 [hep-ex]}}.
%%CITATION = ARXIV:1210.0202;%%.

\bibitem{Bechtle:2009em}
P.~Bechtle, M.~Berggren, J.~List, P.~Schade, and O.~Stempel, ``{Prospects for
  the study of the $\tilde{\tau}$-system in SPS1a' at the ILC},''
  \href{http://dx.doi.org/10.1103/PhysRevD.82.055016}{{\em Phys.Rev.}
  {\bfseries D82} (2010) 055016},
\href{http://arxiv.org/abs/0908.0876}{{\ttfamily arXiv:0908.0876 [hep-ex]}}.
%%CITATION = ARXIV:0908.0876;%%.

\bibitem{MoortgatPick:2005cw}
G.~Moortgat-Pick, T.~Abe, G.~Alexander, B.~Ananthanarayan, A.~Babich, {\em
  et~al.}, ``{The Role of polarized positrons and electrons in revealing
  fundamental interactions at the linear collider},''
  \href{http://dx.doi.org/10.1016/j.physrep.2007.12.003}{{\em Phys.Rept.}
  {\bfseries 460} (2008) 131--243},
\href{http://arxiv.org/abs/hep-ph/0507011}{{\ttfamily arXiv:hep-ph/0507011
  [hep-ph]}}.
%%CITATION = HEP-PH/0507011;%%.

\bibitem{bib:thesis_dascenzo}
N.~D'Ascenzo, {\em {Study of the Neutralino Sector and Analysis of the Muon
  Resoponse of a Highly Granular Hadron Calorimeter at the International Linear
  Collider}}.
\newblock PhD thesis, University of Hamburg,
  http://www-library.desy.de/cgi-bin/showprep.pl?desy-thesis-09-004, 2009.

\bibitem{Berggren:2005gp}
M.~Berggren, ``{Reconstructing sleptons in cascade-decays at the linear
  collider},'' in {\em Proceedings of LCWS04}, pp.~907--910.
\newblock Paris, April, 2004.
\newblock
\href{http://arxiv.org/abs/hep-ph/0508247}{{\ttfamily arXiv:hep-ph/0508247
  [hep-ph]}}.
\newblock
%%CITATION = HEP-PH/0508247;%%.

\bibitem{Bechtle:2009ty}
P.~Bechtle, K.~Desch, M.~Uhlenbrock, and P.~Wienemann, ``{Constraining SUSY
  models with Fittino using measurements before, with and beyond the LHC},''
  \href{http://dx.doi.org/10.1140/epjc/s10052-009-1228-3}{{\em Eur.Phys.J.}
  {\bfseries C66} (2010) 215--259},
\href{http://arxiv.org/abs/0907.2589}{{\ttfamily arXiv:0907.2589 [hep-ph]}}.
%%CITATION = ARXIV:0907.2589;%%.

\end{thebibliography}\endgroup

\clearpage

\section{Appendix}\label{sec:appendix}

\subsection{Resolution of the jet momentum}
We investigate the jet momentum for the three different pileup scenarios. Reconstructed jets are matched to generator level jets with the criterium $\Delta R < 0.5$, where the distance parameter $\Delta R = \sqrt{\Delta \eta^2 + \Delta \phi^2}$ is used to match the closed generator level jet. We define the 
resolution as $(p_T^{\rm gen} - p_T^{\rm rec}) / p_T^{\rm gen} $.

\begin{figure}[htbp]
\begin{center}
  \subfigure[]{ \includegraphics[width=0.45\textwidth]{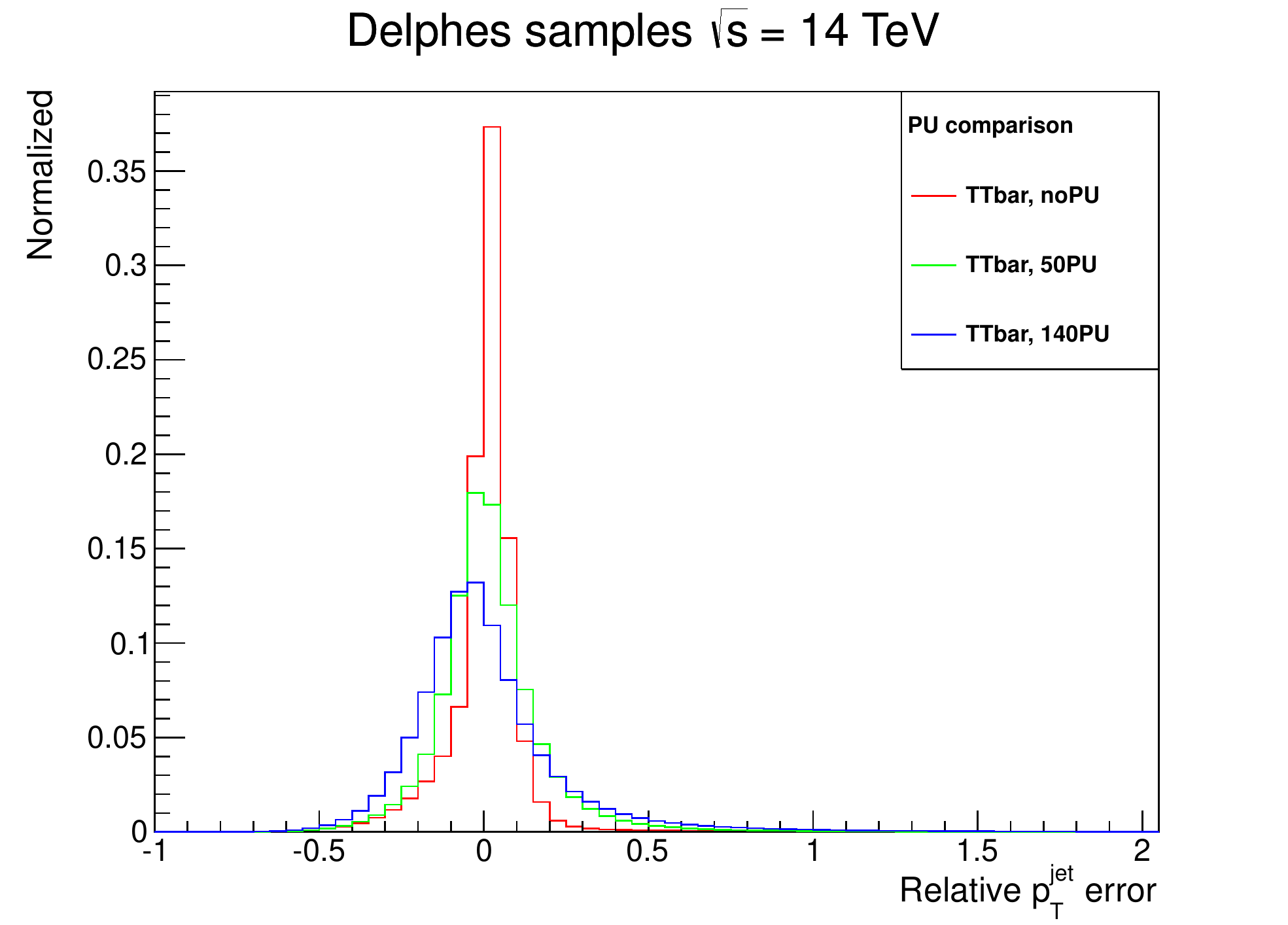} }
  \subfigure[]{ \includegraphics[width=0.45\textwidth]{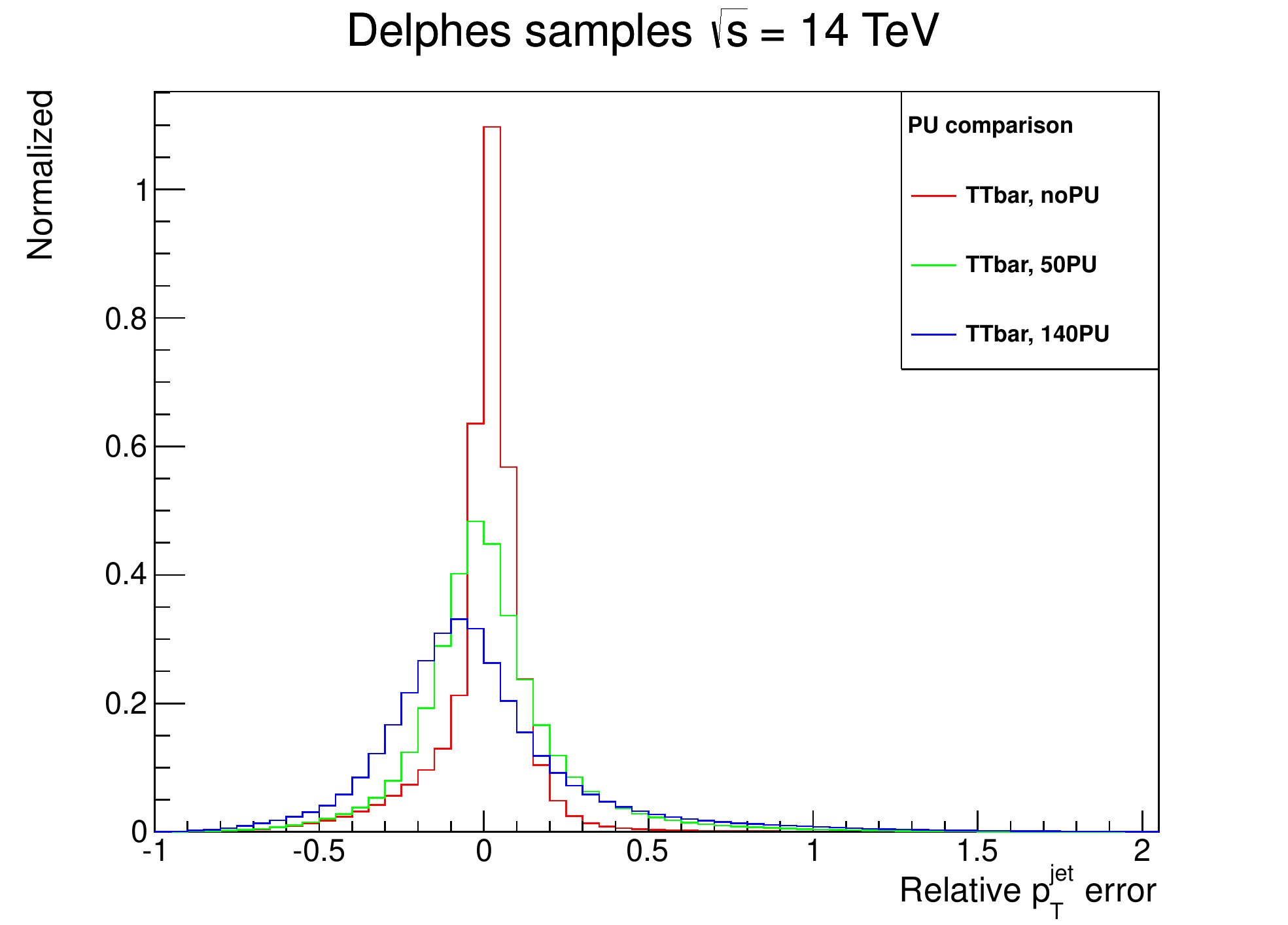} }
\caption{Comparison of the reconstructed and generated $p_T$ for simulated t$\bar{\rm t}$ events with 0, 50 PU and 140 PU events for the hardest (a) and all jets (b). Shown is in both cases the resolution, for the definition see text.}
\label{fig:jet_res}
\end{center}
\end{figure}

\begin{figure}[htbp]
\begin{center}
  \includegraphics[width=0.45\textwidth]{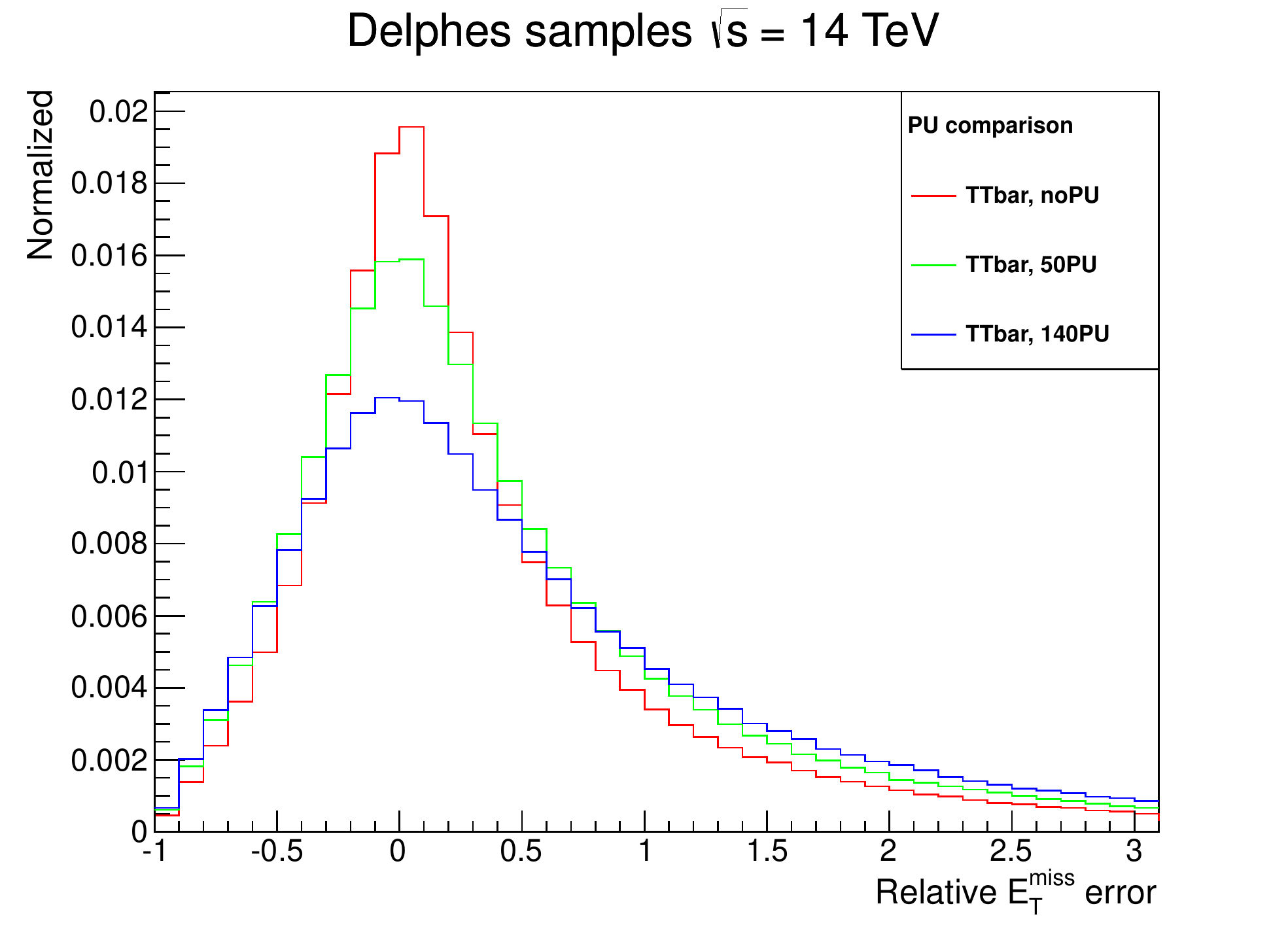}
\caption{Comparison of the reconstructed and generated $E_T^{\rm miss}$ for simulated t$\bar{\rm t}$ events with 0, 50 PU and 140 PU events.}
\label{fig:MET_res}
\end{center}
\end{figure}

Figure~\ref{fig:jet_res}  contains the comparison for the hardest jet and for all jets with the reconstructed momentum $p_T > 30$~GeV. The jet resolution decreases with higher pileup.
Figure~\ref{fig:MET_res} shows a similar plot for the $E_T^{\rm miss}$ resolution. The generator level information contains all objects that are invisible for the detector, like neutrinos or neutralinos. Also the  $E_T^{\rm miss}$ resolution decreases with increasing pileup as expected.

\subsection{Comparison of signal cross sections}

The main production processes at the LHC running at a center-of-mass energy of 14~TeV are summarized in Table~\ref{tab:cs_14}.
The subprocess with the largest cross section in model STC4 is direct stop production. The stops predominantly decay to top quarks and the lightest neutralino (54\%), or to bottom quarks and the lightest chargino (46\%). Here, the chargino decays mainly to tau and neutrino (70 \%), suggesting analyses searching for tops in the final state, either with one or no lepton.

The mass of the stop quarks rises from model STC4 to STC8 subsequently from 293~GeV to 750~GeV, reducing the cross section for stop production significantly, while the production cross section of the electroweak particles stays roughly the same. The latter are very hard to identify at the LHC, as they mainly decay to final states containing tau leptons.

\begin{table}[htdp]
\caption{Overview over the cross sections of the main processes calculated at leading order by Pythia at the LHC with a center-of-mass energy of 14~TeV}
\begin{center}
\begin{tabular}{|c|c|c|c|}
\hline
Model   & Process & Relative cross section & LO cross section from Pythia \\ \hline \hline
                & gg$ \rightarrow \tilde{\rm t_1} \tilde {\bar{\rm t_1}} $                              & 70\%          &       5.2 pb  \\ \hline
                & gg$ \rightarrow \tilde{\rm b_1} \tilde {\bar{\rm b_1}} $                              & 0.1\%                 &       11 fb   \\ \hline
                & q$\bar{\rm q} \rightarrow \tilde{\rm t_1} \tilde {\bar{\rm t_1}}$     & 9.4\%         &       0.70 pb \\ \hline
STC4            & q$\bar{\rm q} \rightarrow \tilde{\chi_1^+} \tilde {\chi_2^0} $      & 8.0\%   &       0.61 pb \\ \hline
                & q$\bar{\rm q} \rightarrow \tilde{\chi_1^-} \tilde {\chi_2^0} $      & 4.5\%   &       0.34 pb \\ \hline
                & q$\bar{\rm q} \rightarrow \tilde{\chi_1^+} \tilde {\chi_1^-} $      & 6.5\%   &       0.49 pb \\ \hline \hline
                & gg$ \rightarrow \tilde{\rm t_1} \tilde {\bar{\rm t_1}}        $               & 31\%          &       0.73 pb \\ \hline
                & gg$ \rightarrow \tilde{\rm b_1} \tilde {\bar{\rm b_1}}        $               & 0.6\%                 &       12 fb   \\ \hline
                & q$\bar{\rm q} \rightarrow \tilde{\rm t_1} \tilde {\bar{\rm t_1}}$     & 6.2\%         &       0.15 pb         \\ \hline
STC5            & q$\bar{\rm q} \rightarrow \tilde{\chi_1^+} \tilde {\chi_2^0} $      & 25\%            &       0.64 pb         \\ \hline
                & q$\bar{\rm q} \rightarrow \tilde{\chi_1^-} \tilde {\chi_2^0} $      & 13\%            &       0.32 pb         \\ \hline
                & q$\bar{\rm q} \rightarrow \tilde{\chi_1^+} \tilde {\chi_1^-} $      & 20\%    &       0.48 pb         \\ \hline \hline
                & gg$ \rightarrow \tilde{\rm t_1} \tilde {\bar{\rm t_1}}        $               & 10\%          &       0.18 pb \\ \hline
                & gg$ \rightarrow \tilde{\rm b_1} \tilde {\bar{\rm b_1}}        $               & 0.6\%                 & 11 fb \\ \hline
                & q$\bar{\rm q} \rightarrow \tilde{\rm t_1} \tilde {\bar{\rm t_1}}$     & 2.9\%         &       0.05 pb         \\ \hline
STC6            & q$\bar{\rm q} \rightarrow \tilde{\chi_1^+} \tilde {\chi_2^0} $      & 36\%            &       0.63 pb         \\ \hline
                & q$\bar{\rm q} \rightarrow \tilde{\chi_1^-} \tilde {\chi_2^0} $      & 19\%            &       0.33 pb         \\ \hline
                & q$\bar{\rm q} \rightarrow \tilde{\chi_1^+} \tilde {\chi_1^-} $      & 27\%    &       0.48 pb         \\ \hline \hline
                & gg$ \rightarrow \tilde{\rm t_1} \tilde {\bar{\rm t_1}}        $               & 1.1\%         &       18 fb   \\ \hline
                & gg$ \rightarrow \tilde{\rm b_1} \tilde {\bar{\rm b_1}}        $               & 0.7\%         &       12 fb   \\ \hline
STC8            & q$\bar{\rm q} \rightarrow \tilde{\chi_1^+} \tilde {\chi_2^0} $      & 40\%            &       0.63 pb         \\ \hline
                & q$\bar{\rm q} \rightarrow \tilde{\chi_1^-} \tilde {\chi_2^0} $      & 21\%            &       0.33 pb         \\ \hline
                & q$\bar{\rm q} \rightarrow \tilde{\chi_1^+} \tilde {\chi_1^-} $      & 31\%    &       0.49 pb         \\ \hline \hline
\end{tabular}
\end{center}
\label{tab:cs_14}
\end{table}

%\newpage
\subsection{Cutflows for the scenario with 140 pileup events and 3000 $fb^{-1}$}\label{sec:cut140}

The following pages show the cutflows for simulations with 140 pileup events. The selections are identical to the 
three cases discussed in section \ref{sec:LHC}. Some of the cuts are adapted to the larger statistics assuming 
3000~fb$^{-1}$ at 14~TeV.

\begin{sidewaystable}[htdp]
\caption{
Cutflow: number of events for the inclusive signal samples and several important backgrounds for the 
full-hadronic stop analyses with 3000~fb$^{-1}$ at 14~TeV with 140 pileup events. The last two lines show the 
significances with an additional systematic background uncertainty of 25\% and, as an optimistic scenario, of 15\%.
}
\label{tab:fullhadronic_result_140PU}
\begin{center}
\begin{tabular}{|c||c|c|c|c||c||c|c|c|c|}
\hline
Description   &diboson  &  ttbar+jets   &boson+jets& single top&sum bgrds  &    STC4  &     STC5  &    STC6  &     STC8 \\ \hline \hline
  
preselection                       & 1108210000 & 2161240000 & 168402000000 &  620665000 & 172292115000 &   38400000   &   11460000   &    7590000   &    6570000   \\ \hline
lepton veto                        &  940242000 & 1487090000 & 159642000000 &  502288000 & 162571620000 &   28679700   &    8447620   &    5632630   &    4994540   \\ \hline
n jets $\ge 3$                    &  201104000 & 1183240000 & 32492600000 &  233184000 & 34110128000 &   16757100   &    3284950   &    1229670   &     691336   \\ \hline
jet1 $> 120$~GeV                   &   74779600 &  546969000 & 8058930000 &   94522500 & 8775201100 &    9572540   &    2458540   &     846769   &     375882   \\ \hline
jet2 $> 70$~GeV                    &   65339800 &  510834000 & 6852930000 &   81913300 & 7511017100 &    8493500   &    2223710   &     749506   &     304717   \\ \hline
jet3 $> 60$~GeV                    &   41908700 &  415604000 & 3625730000 &   50266400 & 4133509100 &    6533640   &    1738650   &     590349   &     218424   \\ \hline
bjets $ge 2$                       &    3372470 &  230206000 &  195456000 &   17357200 &  446391670 &    3424080   &    1012000   &     341458   &     101420   \\ \hline
$H_T > 1000$~GeV                   &     265364 &   12417600 &    8318720 &     631840 &   21633524 &     494592   &     179313   &      83465   &      43579   \\ \hline 
$\Delta \Phi > 0.5$                &     178354 &    8002890 &    5590080 &     405644 &   14176968 &     303096   &     131683   &      65378   &      35453   \\ \hline 
$E_T^{
m miss}/m_{
m eff}>0.2$   &       5599 &     157699 &     115775 &       6005 &     285079 &     138792   &      61382   &      33273   &      19742   \\ \hline \hline
$E_T^{
m miss}>1000$~GeV          &        240 &        540 &       4569 &         27 &       5378 &       5976   &       3769   &       2999   &       2944   \\ \hline
$s/\sqrt{b+(0.25*b)^2}$           &      &      &      &      &      &        4.4 &        2.8 &        2.2 &        2.2 \\ \hline
$s/\sqrt{b+(0.15*b)^2}$           &      &      &      &      &      &        7.4 &        4.7 &        3.7 &        3.6 \\ \hline

\end{tabular}
\end{center}

\end{sidewaystable}%

\begin{sidewaystable}[htdp]

\caption{
Cutflow: number of events for the inclusive signal samples and several important backgrounds for the 
single-lepton stop analyses with 3000~fb$^{-1}$ at 14~TeV with 140 pileup events. The last two lines show the 
significances %without and 
with an additional systematic background uncertainty of 25\% and, as an 
optimistic scenario, of 15\%.
}
\begin{center}
\begin{tabular}{|c||c|c|c|c||c||c|c|c|c|}
\hline
Description   &diboson  &  ttbar+jets   &boson+jets& single top&sum bgrds  &    STC4  &     STC5  &    STC6  &     STC8 \\ \hline \hline
  
preselection                        & 1108210000 & 2161240000 & 168402000000 &  620665000 & 172292115000 &   38400000   &   11460000   &    7590000   &    6570000   \\ \hline
singl. lep. and $\tau$ veto        &   94640200 &  400316000 & 5212300000 &   74380700 & 5781636900 &    4348510   &    1114500   &     618297   &     447906   \\ \hline
n jets $\ge 4$                     &    4732320 &  102106000 &  136368000 &    4329060 &  247535380 &    1562500   &     356513   &     141841   &      52116   \\ \hline
b-tagged jets = 1 or 2              &    1419280 &   81610600 &   31314400 &    3294970 &  117639250 &    1201010   &     270116   &      97023   &      26776   \\ \hline
$E_T^{\rm miss}>500$~GeV           &       4893 &      67767 &      39029 &       2269 &     113958 &      25608   &      12892   &       8471   &       4266   \\ \hline
$\Delta \Phi >0.5$                &       4504 &      59682 &      36012 &       1865 &     102065 &      17616   &      10618   &       7290   &       3876   \\ \hline
$H_T>1500$~GeV                     &        547 &       6403 &       4165 &        150 &      11266 &       4968   &       2802   &       1830   &       1400   \\ \hline
$M_T>120$~GeV                       &         76 &       1103 &        367 &         19 &       1566 &       2760   &       1674   &       1218   &        866   \\ \hline 
\multicolumn{1}{|p{3cm}||}{\centering topness $> 9.5$ and $p_T$ asym.$<-0.2$  }
                       &         48 &        628 &        222 &         13 &        913 &       1632   &       1208    &        881   &        710   \\ \hline \hline
$H_T>1750$~GeV                     &         24 &        313 &        119 &          6 &        463 &       1128   &        868   &        605   &        525   \\ \hline
$s/\sqrt{b+(0.25*b)^2}$           &      &      &      &      &      &        9.6 &        7.4 &        5.1 &        4.5 \\ \hline
$s/\sqrt{b+(0.15*b)^2}$           &      &      &      &      &      &       15.5 &       11.9 &        8.3 &        7.2 \\ \hline

\end{tabular}
\end{center}
\label{tab:single_lepton_result_140PU}
\vspace{1.0cm}

\caption{
Cutflow: number of events for the inclusive signal samples and several important backgrounds for the 
same-sign di-lepton analysis targeting electroweak particles with 3000~$fb^{-1}$ at 14~TeV and with 140
pileup events. The last two lines show the significances 
with an additional systematic 
background uncertainty of 30\% and, as an optimistic scenario, of 20\%.
}

\begin{center}
\begin{tabular}{|c||c|c|c|c||c||c|c|c|c|}
\hline
Description   &diboson  &  ttbar+jets   &boson+jets& single top&sum bgrds  &    STC4  &     STC5  &    STC6  &     STC8 \\ \hline \hline
  
preselection              & 1108210000 & 2161240000 & 168402000000 &  620665000 & 172292115000 &   38400000   &   11460000   &    7590000   &    6570000   \\ \hline
2 lepton req.             &   19295000 &   67454200 &  514871000 &    1627920 &  603248120 &     977736   &     443583   &     329663   &     250383   \\ \hline
$E_T^{miss} >$ 120~GeV    &    1867680 &   12033600 &   20908400 &     270240 &   35079920 &     478392   &     249524   &     180581   &     124423   \\ \hline
same-sign req.            &     112615 &      74246 &      59914 &      20679 &     267455 &      21504   &      28238   &      24820   &      17714   \\ \hline
Z veto                    &      49880 &      73564 &      38490 &      20675 &     182610 &      18792   &      23296   &      19795   &      14597   \\ \hline
b-jet veto                &      42453 &      21839 &      30696 &      10652 &     105641 &      10272   &      10573   &       8850   &       7986   \\ \hline 
$E_T^{miss} >$ 200~GeV    &       8951 &       4132 &       6989 &       1437 &      21511 &       4992   &       5452   &       4927   &       4401   \\ \hline
$E_T^{miss} >$ 400~GeV    &        944 &        218 &        857 &         81 &       2102 &	   1152   &	  1316   &	 1377   &	1170   \\ \hline
$s/\sqrt{b+(0.3*b)^2}$    &	 &	&      &      &      &        1.8 &        2.1 &        2.2 &        1.9 \\ \hline
$s/\sqrt{b+(0.2*b)^2}$    &	 &	&      &      &      &        2.7 &        3.1 &        3.3 &        2.8 \\ \hline
\end{tabular}
\end{center}
\label{tab:ewkino_140PU}
\end{sidewaystable}

\end{document}